\documentclass[fleqn,usenatbib]{mnras}
\usepackage{graphicx}	
\usepackage{amsmath}	
\usepackage[dvipsnames]{xcolor}  
\usepackage{amssymb}	
\usepackage{newtxtext,newtxmath}
\usepackage{macros_vdb} 
\begin{document}
%

\title[Disruption of Dark Matter Substructure]
      {Disruption of Dark Matter Substructure: Fact or Fiction?}

\author[van den Bosch et al.]{%
   Frank~C.~van den Bosch$^1$\thanks{E-mail: frank.vandenbosch@yale.edu},
   Go Ogiya$^2$, 
   Oliver Hahn$^2$,
   Andreas Burkert$^{3,4}$
\vspace*{8pt}
\\
   $^1$Department of Astronomy, Yale University, PO. Box 208101, New Haven, CT 06520-8101\\
   $^2$Laboratoire Lagrange, Observatoire de la C\^ote d'Azur, CNRS, Blvd de l'Observatoire, CS 34229, F-06304  Nice cedex 4, France\\
   $^3$ University Observatory Munich (USM), Scheinerstr. 1, D-81679 Munich, Germany\\
$^4$Associated Scientist, Max-Planck-Institute for Extraterrestrial Physics (MPE), Giessenbachstr. 1, D-85748 Garching, Germany}


\date{}

\pagerange{\pageref{firstpage}--\pageref{lastpage}}
\pubyear{2013}

\maketitle

\label{firstpage}


\begin{abstract}
  Accurately predicting the demographics of dark matter (DM)
  substructure is of paramount importance for many fields of
  astrophysics, including gravitational lensing, galaxy evolution,
  halo occupation modeling, and constraining the nature of dark
  matter. Because of its strongly non-linear nature, DM substructure
  is typically modeled using $N$-body simulations, which reveal that
  large fractions of DM subhaloes undergo complete disruption.  In
  this paper we use both analytical estimates and idealized numerical
  simulations to investigate whether this disruption is mainly
  physical, due to tidal heating and stripping, or numerical (i.e.,
  artificial). We show that, contrary to naive expectation, subhaloes
  that experience a tidal shock $\Delta E$ that exceeds the subhalo's
  binding energy, $|E_\rmb|$, do not undergo disruption, even when
  $\Delta E/|E_\rmb|$ is as large as 100. Along the same line, and
  contrary to existing claims in the literature, instantaneously
  stripping matter from the outskirts of a DM subhalo also does not
  result in its complete disruption, even when the instantaneous
  remnant has positive binding energy. In addition, we show that tidal
  heating due to high-speed (impulsive) encounters with other
  subhaloes (`harassment'), is negligible compared to the tidal
  effects due to the host halo. Hence, we conclude that, in the
  absence of baryonic processes, the complete, physical disruption of
  CDM substructure is extremely rare, and that most disruption in
  numerical simulations therefore must be artificial. We discuss
  various processes that have been associated with numerical
  overmerging, and conclude that inadequate force-softening is the
  most likely culprit.
 \end{abstract} 


\begin{keywords}
methods: analytical ---
methods: numerical ---
galaxies: haloes --- 
cosmology: dark matter ---
gravitation
\end{keywords}


\section{Introduction} 
\label{sec:intro}

In the $\Lambda$ cold dark matter ($\Lambda$CDM) concordance
cosmology, structure develops as primordial density fluctuations grow
to form virialized structures.  Because of the negligible thermal
velocity of CDM, fluctuations survive the early universe on all
scales, and structure develops from the bottom up as small, dense CDM
clumps merge together to build-up large dark matter haloes. During
this hierarchical assembly, the inner regions of early virialized
haloes often survive accretion on to a larger system, thus giving rise
to a population of subhaloes. These are subjected to forces that try
to dissolve them: dynamical friction, tides from the host halo, and
impulsive encounters with other substructure.

Being able to accurately predict the abundance and demographics of
dark matter substructure is of crucial importance for a wide range of
astrophysics. First of all, it is one of the prime discriminators
between different dark matter models; if dark matter is warm (WDM),
rather than cold, free-streaming and enhanced tidal disruption, will
cause a significantly reduced abundance of {\it low mass} subhaloes
\citep[e.g.,][]{Knebe.etal.08, Lovell.etal.14, Colin.etal.15,
  Bose.etal.16}. If dark matter has a significant cross-section for
self-interaction (SIDM), subhaloes are predicted to have
constant-density cores, with significantly lower central densities
than their CDM counterparts \citep[e.g.][]{Burkert.00,
  Vogelsberger.etal.12, Rocha.etal.13}.  The main methods that are
used to test these different dark matter models are gravitational
lensing, either by using time-delays
\citep[e.g.,][]{Keeton.Moustakas.09}, flux-ratio anomalies
\citep[e.g.,][]{Metcalf.Madau.01, Bradac.02, Dalal.Kochanek.02}, or
perturbations of the surface brightness distribution of lensed arcs
and Einstein rings \citep[e.g.][]{Vegetti.Koopmans.09,
  Vegetti.etal.14}.  Alternatively, low mass substructure can reveal
its presence by creating gaps in tidal (stellar) streams
\citep[e.g.][]{Carlberg.09, Sanders.etal.16, Erkal.etal.16}. The
detailed {\it structure} of subhaloes can be tested using kinematic
data of dwarf galaxies \citep[e.g.,][]{Mateo.etal.93, Walker.etal.09,
  Lokas.09}.  The latter has given rise to a potential problem for the
$\Lambda$CDM paradigm in the form of the `too-big-to-fail' problem
\citep{Boylan-Kolchin.etal.10, Boylan-Kolchin.etal.11,
  Garrison-Kimmel.etal.14, Ogiya.Burkert.15}, which has sparked
renewed interest in alternatives such as WDM or SIDM (although
baryonic solution have also been proposed
\citep[e.g.,][]{Zolotov.etal.12, Ogiya.Mori.14, Arraki.etal.14,
  Dutton.etal.16}. Finally, dark matter substructure boosts the
expected dark matter annihilation signal \citep{Bergstrom.etal.99}, an
effect that is typically quantified in terms of a `boost-factor'
\citep[e.g.,][]{Strigari.etal.07, Giocoli.etal.08b, Kuhlen.etal.08,
  Pieri.etal.08, Moline.etal.16}.

In addition to carrying a wealth of information regarding the dark
sector, dark matter substructure is also important for our quest to
understand galaxy formation and large scale structure.  Dark matter
subhaloes are believed to host satellite galaxies and the demographics
of dark matter substructure is therefore directly related to the
(small scale) clustering of galaxies. This idea underlies the popular
technique of subhalo abundance matching
\citep[e.g.,][]{Vale.Ostriker.04, Conroy.etal.06, Guo.etal.10,
  Hearin.etal.13, Moster.etal.13, Behroozi.etal.13c}, in which
galaxies are assigned to dark matter host and sub-haloes in
simulations to create mock galaxy samples. This has become a prime
tool for interpreting galaxy clustering, galaxy-galaxy lensing, and
group multiplicity functions, and is even used to constrain
cosmological parameters \citep{Trujillo-Gomez.etal.11, Hearin.etal.15,
  Hearin.etal.16, Reddick.etal.14, Zentner.etal.14, Zentner.etal.16,
  Lehmann.etal.15}.

The formation and evolution of dark matter substructure is best
studied using numerical $N$-body simulations. Nowadays, large
cosmological simulations routinely resolve an entire hierarchy of
substructure, with haloes hosting subhaloes, which themselves host
sub-subhaloes, etc. \citep[e.g.,][]{Ghigna.etal.98, Tormen.etal.98,
  Diemand.etal.04, Gao.etal.04, Kravtsov.etal.04, Giocoli.etal.08,
  Giocoli.etal.10}. Since simulations are CPU-expensive, and suffer
from limiting mass and force resolution, various authors have
attempted to construct semi-analytical models for the evolution of
dark matter substructure \citep[e.g.,][]{Taylor.Babul.01,
  Penarrubia.Benson.05, vdBosch.etal.05, Zentner.etal.05,
  Kampakoglou.Benson.07, Pullen.etal.14, Jiang.vdBosch.16}. However,
since we lack accurate, analytical descriptions, based on first
principles, for how the mass, structure and orbits of subhaloes evolve
with time, these models typically have several free parameters that
are tuned to reproduce results from numerical simulations.

Prior to 1997, numerical $N$-body simulations suffered from a serious
`overmerging' problem, in that the dark matter haloes revealed little
to no substructure. This was in clear contrast with the wealth of
`substructure' (i.e., satellite galaxies) observed in galaxy groups
and clusters. While some speculated that baryonic physics would
resolve this problem \citep[e.g.,][]{Frenk.etal.88}, others argued
that overmerging might actually be a numerical artifact arising from
insufficient mass and/or force resolution. In particular,
\cite{Carlberg.94} and \cite{vKampen.95, vKampen.00a} identified
particle-subhalo two-body heating as the main cause for overmerging,
while others blamed inadequate force softening
\citep[e.g.,][]{Moore.etal.96a, Klypin.etal.99a}. At the close of the
last millennium, when simulations started to resolve surviving
populations of subhaloes \citep[e.g.,][]{Tormen.etal.97,
  Brainerd.etal.98, Moore.etal.98b}, the discussion as to what might
cause numerical overmerging was quickly eclipsed by a new issue,
namely that CDM simulations seem to predict too {\it much}
substructure \citep{Klypin.etal.99b, Moore.etal.99}. Over the years,
though, it has become clear that this `missing satellite problem' is
mainly a manifestation of poorly understood baryonic physics related
to galaxy formation \citep[see][for a review]{Bullock.BoylanKolchin.17}.

As for overmerging, it has become clear that modern, state-of-the-art
numerical simulations still suffer from numerical overmerging, with
important ramifications for small-scale clustering
\citep[e.g.,][]{Conroy.etal.06, Guo.White.13, Moster.etal.17,
  Campbell.etal.17}, semi-analytical models for galaxy formation
\citep[e.g.,][]{Springel.etal.01, Kang.etal.05, Kitzbichler.White.08},
and other studies that rely on the phase-space distribution of
subhaloes \citep[e.g.,][]{Faltenbacher.Diemand.06, Wu.etal.13,
  Tollet.etal.17}. However, the general consensus seems to be that
this numerical overmerging only affects subhaloes below a mass
resolution limit of 50-100 particles. This notion is based on the fact
that simulations seem to yield consistent, converged results for the
mass function of subhaloes above this mass resolution limit
\citep[e.g.,][]{Springel.etal.08, Onions.etal.12, Knebe.etal.13,
  Cautun.etal.14, vdBosch.Jiang.16, Griffen.etal.16}.

In a recent study, \cite{vdBosch.17} showed that subhalo disruption is
extremely prevalent in modern simulations, with inferred fractional
disruption rates (at $z=0$) of $\sim 13$ percent per Gyr \citep[see
  also][]{Diemand.Moore.Stadel.04}. This implies that $\sim 65$ ($90$)
percent of all subhaloes accreted around $z=1$ ($z=2$) are disrupted
by $z=0$ \citep[][]{Han.etal.16, Jiang.vdBosch.17}. In the simulation
studied by \cite{vdBosch.17}, roughly 20 percent of this disruption
occurs above the mass resolution limit of 50 particles, with a mass
function of disrupting subhaloes that is indistinguishable from that
of the surviving population! What is the dominant cause of this
prevalent disruption of subhaloes in numerical simulations? In
particular, is it artificial (numerical) or real (physical)?  Based on
the fact that subhalo mass functions are deemed converged above a mass
resolution limit of 50-100 particles, it is tempting to conclude that
disruption above this `resolution limit' be physical, rather than
numerical. However, convergence is a necessary, but not a sufficient 
condition, to guarantee that the simulation results are physically 
correct. Furthermore, there is no consensus as to what physical
mechanism dominates, with most studies arguing either for tidal
heating or tidal stripping. We therefore initiated a comprehensive
study aimed at answering these questions.  While researching the
existing literature on this topic, we came across a variety of
conflicting results and claims, and this paper is largely intended to
clarify this confusion, correct a number of misconceptions, and
present a comprehensive overview. As such, this paper serves as a
compendium to two follow-up papers: In van den Bosch et al. (in prep.;
hereafter Paper II) we use a large suite of idealized numerical
simulations and experiments to show that state-of-the-art numerical
$N$-body simulations still suffer from an enormous amount of
overmerging, mainly driven by inadequate force-softening and an
amplification of discreteness noise due to the tidal field. In Ogiya
et al. (in prep.; hereafter Paper III), we use a large suite of
similar idealized simulations, that are properly converged, to probe
large parts of parameter space. The results of these simulations can
be used to calibrate semi-analytical models of subhalo evolution and
to gauge the reliability and accuracy of cosmological simulations. In 
this paper we
\begin{itemize}
\item show that impulsive, tidal heating during peri-centric
  passage of the host halo can inject as much as 100 times the binding
  energy. Yet, even under those extreme conditions, a remnant
  containing $\sim 20$ percent of the mass remains.
\item demonstrate that harassment has a negligible impact on the
  evolution of dark matter substructure.
\item underscore the problems and uncertainties associated with
  analytical treatments of tidal stripping.
\item demonstrate that simply removing matter from the outskirts of
  a dark matter subhalo will never result in its complete disruption,
  even if the remnant has positive binding energy.
\item  correct and elucidate a number of errors and misconceptions
  regarding the numerical, artificial disruption of substructure.
\end{itemize}
We start in \S\ref{sec:tidalradius} with a discussion of the concept
of `tidal radius', including an overview of the various definitions 
that are used in the literature. \S\ref{sec:sims} describes the 
idealized, numerical simulations that we use in this paper for 
comparison with simple analytical estimates of various {\it physical} 
disruption mechanisms (\S\ref{sec:Physical}), and for assessing the 
impact of instantaneously stripping matter off of dark matter haloes
(\S\ref{sec:trunc}). \S\ref{sec:numerics} discusses a variety of {\it
  numerical} processes that might potentially give rise to artificial
disruption of substructure in $N$-body simulations, including two-body
relaxation, impulsive heating due to interactions with overly massive
host halo particles, and inadequate force softening. Finally,
\S\ref{sec:concl} summarizes our findings and briefly discusses some
implications for various areas of astrophysics.


\section{Tidal Radius}
\label{sec:tidalradius}

We start our discussion of the tidal evolution of dark matter
substructure by addressing an important concept, namely the tidal
radius. There are a number of different definitions that are used in the
literature, with varying degrees of approximation. For the sake of
completeness, we briefly overview and compare the most commonly used
definitions.

Consider two point masses, $m$ and $M$, separated by a distance $R$. If
we define the tidal radius, $\rtid$, of $m$ as the radius from $m$ at
which the tidal force from $M$ exceed the self-gravity of $m$, then we 
have that
\begin{equation}\label{rtRoche}
\rtid = R \, \left( {m \over 2\,M} \right)^{1/3}\,.
\end{equation}
This is known as the Roche limit. In general, however, $m$ and $M$
will be in motion with respect to each other, resulting in an
additional centrifugal force.  If we assume that the two bodies are on
a circular orbit of radius $R$, we can account for this using a
derivation based on the restricted three-body problem, resulting in
the Jacobi limit
\begin{equation}\label{rtJacobi}
\rtid = R \, \left( {m \over 3\,M} \right)^{1/3}\,.
\end{equation}
\citep[e.g.,][]{Spitzer.87, Binney.Tremaine.08}.

Going beyond point particles, and taking account of the extended mass
profiles, $m(r)$ and $M(r)$, of the two bodies, the tidal force due 
to  $M$ is equal to the gravitational attraction due to $m$ at a tidal 
radius
\begin{equation}\label{rtTormen}
\rtidA = R \, \left[ {m(\rtidA)/M(R) \over
      2 - {\rmd \ln M \over \rmd \ln R}\vert_R} \right]^{1/3} \,.
\end{equation}
\citep[e.g.,][]{Tormen.etal.98}. Note that $\rtidA$ is only defined
when $\rmd \ln M/\rmd\ln R \leq 2$. This is a consequence of the fact
that (when ignoring angular momentum) the tidal field becomes
compressive in all directions (i.e., no mass will be stripped)
whenever $\rmd\ln\rho/\rmd\ln R \leq -1$ \citep[e.g.,][]{Dekel.etal.03}.

Although Eq.~(\ref{rtTormen}) is used in several studies
\citep[e.g.,][]{Klypin.etal.99a, Hayashi.etal.03, Taffoni.etal.03}, it
ignores the centrifugal force. Assuming that $m$ is on a circular
orbit of radius $R$ around the centre of $M$, this correction results
in
\begin{equation}\label{rt}
  \rtidB  = R \, \left[ {m(\rtidB)/M(R) \over 2 + {\Omega^2 R^3 \over G M(R)} -
      {\rmd \ln M \over \rmd \ln R}\vert_R} \right]^{1/3} =
  \left[ {G m(\rtidB) \over \Omega^2 - {\rmd^2 \Phi_\rmh \over \rmd r^2}} \right]^{1/3}\,.
\end{equation}
where $\Omega$ is the angular speed \citep[e.g.,][]{King.62,
  Tollet.etal.17}.  Using that along a circular orbit $\Omega = V_{\rm
  circ}(R)/R$, we see that Eq.~(\ref{rt}) reduces to
\begin{equation}\label{rtcirc}
\rtidB = R \, \left[ {m(\rtidB)/M(R) \over
      3 - {\rmd \ln M \over \rmd \ln R}\vert_R} \right]^{1/3} \,.
\end{equation}
Hence, for subhaloes on a circular orbit, the tidal field only becomes
compressive in all directions when $\rmd\ln M/\rmd \ln R \geq 3$ ($\rmd
\ln\rho/\rmd\ln R \geq 0$). This implies that subhaloes on a circular
orbit will always be subject to tidal stripping, unless they happen to
find themselves orbiting within the central region of a host halo with 
a constant density core 
\citep[see][for a comprehensive discussion]{Dekel.etal.03}. We
emphasize that the above definition for the tidal radius is strictly
only valid along circular orbits.  However, it is common practice to
also apply it to eccentric orbits, in which case one estimates the
{\it instantaneous} tidal radius by simply using for $R$ and $\Omega$
the corresponding, instantaneous values, i.e., $\Omega = |\vec{V}
\times \vec{R}| / R^2$, where $\vec{V}$ is the instantaneous orbital
velocity of the subhalo \citep[see e.g.,][]{King.62, Taylor.Babul.01,
  Zentner.Bullock.03, Penarrubia.Benson.05}.
\begin{figure*}
\includegraphics[width=\hdsize]{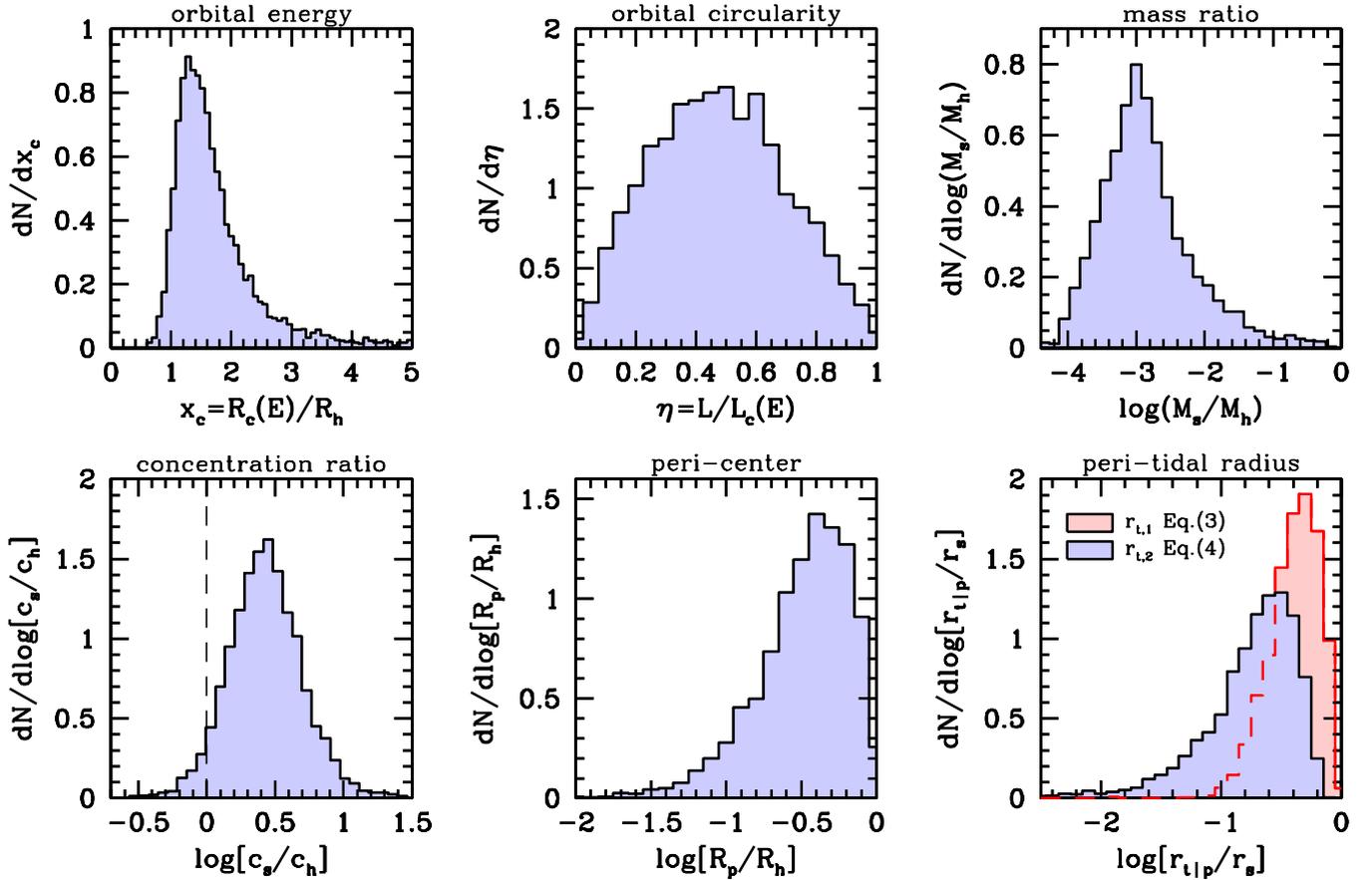}
\centering
\caption{Properties of subhaloes in the Bolshoi simulation at the
  moment they are accreted by their host halo. From left to right and
  top to bottom, the different panels show distributions of the
  orbital energy, characterized by $x_\rmc = R_\rmc(E)/R_\rmh$,
  orbital circularity, $\eta = L/L_\rmc(E)$, the mass ratio of subhalo
  to host halo, $M_\rms/M_\rmh$, the ratio of concentration parameters
  of subhalo and host halo, $c_\rms/c_\rmh$, the orbit's peri-centre
  in units of the virial radius of the host halo, $R_\rmp/R_\rmh$, and
  finally the peri-tidal radius, $\rpt$, (defined as the tidal radius
  at peri-centre), normalized by the size of the subhalo.  For the
  latter we have used the two different definitions of the tidal
  radius, as indicated. See text for a detailed discussion.}
\label{fig:orbit}
\end{figure*}

Finally, for the sake of completeness, we mention that
\cite{Klypin.etal.99a, Klypin.etal.15} have advocated the use of yet
another definition of tidal radius. Motivated by the work of
\cite{Weinberg.94a, Weinberg.94b, Weinberg.97}, which demonstrates that
resonant effects can boost the impact of tidal stripping, Klypin \etal
define their tidal radius as the radius where the frequency of the
tidal force by the host halo is equal to that of the internal motion
within the subhalo. This results in
\begin{equation}\label{rtcirc}
\rtidC = R \, \left[ {m(\rtidC) \over M(R)} \right]^{1/3} \,.
\end{equation}
Following the suggestion by \cite{Klypin.etal.99a}, several studies
have taken the tidal radius to be the minimum of $\rtidA$ and $\rtidC$
\citep[e.g.,][]{Oguri.Lee.04, Arraki.etal.14, Klypin.etal.15}.

Although the tidal radius is a common concept used to discuss tidal
stripping, it is important to realize that all the various definitions
discussed above are only approximate \citep[see
  e.g.,][]{Binney.Tremaine.08, MBW10}. First of all, the
two-dimensional surface along which the radial acceleration of a test
particle vanishes is not spherical, even if $m$ and $M$ are point
masses, and can therefore not be characterized by a single
radius \citep[see][for a detailed discussion]{Tollet.etal.17}. 
Secondly, variance in the orbital motion of the particles
within the subhalo gives rise to scatter in the coriolis forces on
these particles, which in turn introduces some `thickness' to the
shell of particles for which $\ddot{r}=0$ \citep[see][for a detailed
  discussion]{Read.etal.06a}.  Thirdly, the various expressions for
$\rtid$ are all derived under the distant-tide approximation, which
implies that the distance $R$ between $m$ and $M$ is large compared to
the size of $m$.  This condition is not necessarily valid in the case
of subhaloes, since the orbit's peri-centre can be comparable to, or
even smaller than, the size of the subhalo. Finally, the optimal
definition of tidal radius is likely to depend on the orbit in
question. For instance, whereas $\rtidB$ (Eq.~[\ref{rt}]) is expected
to be most applicable to circular orbits, $\rtidA$
(Eq.~[\ref{rtTormen}]) is likely to be more relevant for more radial
orbits. In what follows, we will consider both in order to assess how
the `choice' for the definition of tidal radius impacts simple models
for tidal stripping. As for $\rtidC$ (Eq.~[\ref{rtcirc}]); since this
definition of tidal radius is basically bracketed by the other two,
i.e., one typically has $\rtidB < \rtidC <\rtidA$, we will no longer
consider this definition of the tidal radius in what follows.

\subsection{Properties of Newly Accreted Subhaloes}
\label{sec:prop}

As a subhalo moves towards the central region of its host halo, its
tidal radius typically shrinks\footnote{Since the tidal radius depends
  on the density profile of the host halo, this is not always the case
  though.}, reaching a minimum value during peri-centric
passage. Hence, it is commonly assumed that the amount of material
stripped off a subhalo during a radial orbital period depends on its
tidal radius at peri-centre. In what follows, we refer to this as the
`peri-tidal radius', for which we use the symbol $\rpt$.

What are realistic values for $\rpt$ for dark matter subhaloes during
their first radial orbit? In order to address this question we
consider dark matter subhaloes in the cosmological Bolshoi simulation
\citep[][]{Klypin.etal.11}, which follows the evolution of $2048^3$ dark
matter particles using the Adaptive Refinement Tree (ART) code
\citep[][]{Kravtsov.etal.97} in a flat $\Lambda$CDM model with
parameters $\Omega_{\rm m,0} = 1 - \Omega_{\Lambda,0} = 0.27$,
$\Omega_{\rm b,0} = 0.0469$, $h = H_0/(100 \kmsmpc) = 0.7$, $\sigma_8
= 0.82$ and $n_\rms = 0.95$ (hereafter `Bolshoi cosmology').  The box
size of the Bolshoi simulation is $L_{\rm box} = 250 h^{-1} \Mpc$,
resulting in a particle mass of $m_\rmp = 1.35 \times 10^8 \Msunh$.

We use the publicly available halo
catalogs\footnote{http://hipacc.ucsc.edu/Bolshoi/MergerTrees.html}
obtained using the phase-space halo finder \Rockstar
\citep[][]{Behroozi.etal.13a}, which uses adaptive, hierarchical
refinement of friends-of-friends groups in six phase-space dimensions
and one time dimension.  \Rockstar haloes are defined as spheres with
an average density equal to $\Delta_{\rm vir}(z) \rho_{\rm
  crit}(z)$. Here $\rho_{\rm crit}(z) = 3 H^2(z)/8 \pi G$ is the
critical density for closure at redshift $z$, and $\Delta_{\rm
  vir}(z)$ is given by the fitting function of \cite{Bryan.Norman.98}.

Following \cite[][hereafter `vdB17']{vdBosch.17} we use the 19
simulation outputs at $z \leq 0.0605$, from which we select a random
subset of 5,000 newly accreted subhaloes (labeled `A' subhaloes in
vdB17) by identifying those subhaloes in the halo catalogs that in the
previous output were still considered host haloes and which had never
before been subhaloes. In order to focus on well resolved systems,
each subhalo is requested to have at least 250 particles and to be
accreted in a host halo with mass $M_\rmh \geq 3 \times
10^{13}\Msunh$. For each of the subhaloes thus selected we register
\begin{itemize}

\item $x_\rmc \equiv R_\rmc(E)/R_\rmh$, the radius of the circular
  orbit corresponding to the orbital energy, $E$, expressed in terms of
  the virial radius of the host halo, $R_\rmh$.

\item $\eta \equiv L/L_\rmc(E)$, the orbital circularity, defined as
  the ratio of the orbital angular momentum, $L$, and the angular
  momentum $L_\rmc(E)$ corresponding to a circular orbit of energy
  $E$. Radial and circular orbits have $\eta = 0$ and 1, respectively.

\item $M_\rms/M_\rmh$, the ratio of subhalo mass to host halo mass.

\item $c_\rms$ the concentration parameter of the subhalo.

\item $c_\rmh$ the concentration parameter of the host halo.

\end{itemize}
The concentration parameters are defined as the ratios of the virial
radius and the scale parameter of the NFW profile (see below) that
best fits the density distribution of the (sub)halo \citep[see][for
  details on how these quantities are computed in
  \Rockstar]{Behroozi.etal.13a}.
  
For each subhalo thus selected we then compute the orbit's apo-centre,
$R_\rma$, and peri-centre, $R_\rmp$, which are given by the two roots
for
\begin{equation}\label{apoperi}
  {1 \over R^2} + {2 [\Phi_\rmh(R) - E] \over L^2} = 0\,,
\end{equation}
as well as the radial, orbital period, which is given by
\begin{equation}\label{Trad}
T_\rmr = 2 \int_{R_\rma}^{R_\rmp} {\rmd R \over \sqrt{2 [E -
      \Phi_\rmh(R)] - L^2/R^2}} \,,
\end{equation}
and the increase in azimuthal angle during a radial period
\begin{equation}\label{deltapsi}
\Delta  \Psi = 2 L \int_{R_\rma}^{R_\rmp} {\rmd R \over R^2
   \sqrt{2 [E - \Phi_\rmh(R)] - L^2/R^2}} \,,
\end{equation}
\citep[][]{Binney.Tremaine.08}.  Throughout we assume that the host
halo is an NFW halo \citep[][]{Navarro.etal.97}, for which
\begin{equation}\label{psiNFW}
\Phi_\rmh(R) = -V^2_\rmh \, {\ln(1+c_\rmh x) \over f(c_\rmh) \, x}\,,
\end{equation}
where $V_\rmh = \sqrt{G M_\rmh/R_\rmh}$ is the host halo's virial
velocity, $x=R/R_\rmh$, and $f(x) = \ln(1+x) - x/(1+x)$.  Finally, we
compute the subhalo's peri-tidal radius, $\rpt$, in units of the
subhalo's virial radius, $r_\rms$, using both definitions of the tidal
radius; $\rtidA$ and $\rtidB$. Note that all these quantities are
computed using the instantaneous (i.e., at the epoch of accretion)
properties of the host halo and subhalo. We thus ignore the fact that
$\Phi_\rmh(R)$, is time-dependent, and may change appreciably between
accretion and first peri-centric passage.

The upper left and middle panels of Fig.~\ref{fig:orbit} show the
distributions of $x_\rmc$ and $\eta$, characterizing the orbits of
newly accreted subhaloes. The median $x_\rmc$ is $1.26$, and the
distribution is skewed towards higher values, indicating that at
accretion the orbits of subhaloes are only weakly bound (cf.,
vdB17). The $\eta$ distribution is fairly symmetric around $\eta \sim
0.5$, indicating that purely radial and circular orbits are rare
\citep[cf.,][]{Tormen.97, Zentner.etal.05, Wetzel.11, Jiang.etal.15}.
The mass ratio, $M_\rms/M_\rmh$, is peaked around $10^{-3}$ (upper
right-hand panel)\footnote{The drop-off at $M_\rms/M_\rmh < 10^{-3}$
  is an artifact of our selection criteria and the limiting mass
  resolution of the simulation \citep[see][]{vdBosch.etal.16}.}, while
the concentration ratio follows a roughly log-normal distribution
centered around $c_\rms/c_\rmh \simeq 2$. Hence, subhaloes are, on
average, more concentrated than their host haloes, which is a natural
outcome of the concentration-mass relation of CDM haloes. The lower,
middle panel depicts the ratio of the subhalo's peri-centre $R_\rmp$
to the host halo's virial ratio, $R_\rmh$.  The median $R_\rmp/R_\rmh
= 0.37$, while 7.3\% (0.9\%) of subhaloes have a first peri-centric
passage $R_\rmp < 0.1 R_\rmh$ ($0.03 \, R_\rmh$). These numbers are in
good agreement with \cite{Wetzel.11} (see his Appendix~A).  Finally,
the lower-right panel plots the distributions of the peri-tidal
radius, $\rpt$, in units of the subhalo's original (at accretion)
virial radius, $r_\rms$. Red and blue histograms correspond to the two
different definitions of the tidal radius, $\rtidA$ and $\rtidB$,
respectively.  Note that $\rtidA$, which ignores the centrifugal
force, is significantly larger than $\rtidB$; the median $\rtidA$ is
$0.43 r_\rms$, with only 0.7\% of subhaloes having $\rtidA/r_\rms <
0.1$. In comparison, the median $\rtidB$ is $0.21 r_\rms$, with 21.5\%
(2.0\%) of subhaloes having $\rtidB/r_\rms < 0.1$ ($< 0.01$). Thus we
see that the definition of tidal radius can have a big impact on the
expected stripping rate (see \S\ref{sec:tidalstripping} below).


\section{Numerical Simulations}
\label{sec:sims}

Although this paper mainly focuses on simple (semi)-analytical
treatments of subhalo evolution, we also present some simulation
results for comparison. This section briefly describes these numerical
simulations, and we refer the interested reader to Papers~II and~III
for more details (and many more simulation results).  Our (idealized)
simulations follow the evolution of a single N-body subhalo orbiting
in the fixed, analytical potential of a host halo. Our goal is to
investigate how much mass is stripped from the subhalo during its
first radial orbital period, and to compare the results to
expectations based on the various processes discussed in
\S\ref{sec:Physical}.

Each halo (host and sub) is assumed to (initially) be spherical, and
to have a NFW density profile
\begin{equation}
  \rho(r) = \rho_0 \, \left({r \over r_0}\right)^{-1} \,
  \left(1 + {r\over r_0}\right)^{-2}\,,  
\end{equation}
where $r_0$ is the characteristic scale radius. If we define the
virial radius as enclosing an average density that is 97 times the
critical density (which is valid for haloes at $z=0$), the 
crossing time for such a halo is
\begin{equation}\label{tcross}
t_{\rm cross} \equiv {\rvir \over \Vvir} = 2.006 \Gyr
\end{equation}
where we have adopted a Hubble constant of $H_0 = 70\kmsmpc$.

Throughout we assume that, prior to stripping, the halo has an
isotropic velocity distribution, such that its distribution function
(DF) depends only on energy, i.e., $f = f(E)$.  We use the method of
\cite{Widrow.00} to sample particles from the DF using the standard
acceptance-rejection technique \citep{Press.etal.92,
  Kuijken.Dubinski.94, Drakos.etal.17}, and truncate the halo at its
virial radius. Since the DF that we use to generate the ICs is
computed using the \cite{Eddington.16} inversion equation, which
assumes that the halo extends to infinity, our initial system is not
going to be in perfect equilibrium. However, as we demonstrate and
discuss in Paper~II, this has no significant impact on our results.

Throughout we adopt model units in which the gravitational constant,
$G$, the scale radius, $r_0$, and the initial (virial) mass of the
subhalo, $M_\rms$, are all unity. We restrict ourselves to subhaloes
with a concentration parameter $c_\rms=10$, for which $t_{\rm cross} =
c_\rms^{3/2} = 31.6$. Based on Eq.~(\ref{tcross}) we thus have that a
time interval of $\Delta t = 1$ (model units) corresponds to $63.4
\Myr$.

The simulations are carried out using a treecode designed for graphic
processing unit (GPU) clusters. This code, which is optimized for
speed, uses a second-order Runge-Kutta integrator and is described in
\citet{Ogiya.etal.13}. We have compared the results of this simulation
code with those of the treecode described in \S\ref{sec:numsim} below,
with uses a second-order leap-frog integrator, and find the results to
be in excellent agreement.
\begin{table}\label{tab:EvolvSims}
\caption{Simulation Results}
\begin{center}
\begin{tabular}{lcccccc}
\hline\hline
ID & $\eta$ & $f_{\rm bound}$ & $f_{\rm bound}$ & $f_{\rm bound}$ & $f_{\rm bound}$ & $f_{\rm bound}$ \\
 & & $0.2T_\rmr$ & $0.4T_\rmr$ & $0.6T_\rmr$ & $0.8T_\rmr$ & $T_\rmr$ \\
(1) & (2) & (3) & (4) & (5) & (6) & (7) \\
\hline
 S0 & 0.0 & 0.982 & 0.909 & 0.120 & 0.111 & 0.110 \\
 S1 & 0.1 & 0.982 & 0.908 & 0.262 & 0.244 & 0.230 \\
 S2 & 0.2 & 0.982 & 0.905 & 0.367 & 0.347 & 0.323 \\
 S3 & 0.3 & 0.981 & 0.901 & 0.429 & 0.409 & 0.378 \\
 S4 & 0.4 & 0.981 & 0.894 & 0.479 & 0.454 & 0.419 \\
 S5 & 0.5 & 0.980 & 0.887 & 0.522 & 0.489 & 0.453 \\
 S6 & 0.6 & 0.979 & 0.877 & 0.559 & 0.520 & 0.484 \\
 S7 & 0.7 & 0.977 & 0.867 & 0.594 & 0.550 & 0.514 \\
 S8 & 0.8 & 0.975 & 0.858 & 0.631 & 0.580 & 0.543 \\
 S9 & 0.9 & 0.969 & 0.849 & 0.672 & 0.612 & 0.573 \\
S10 & 1.0 & 0.942 & 0.837 & 0.732 & 0.664 & 0.613 \\
\hline\hline
\end{tabular}
\end{center}
\medskip
\begin{minipage}{\hssize}
  The bound fractions at five different epochs, $t/T_\rmr = 0.2, 0.4,
  ..., 1.0$ (Columns 3-7) inferred from our idealized simulations of
  subhaloes orbiting in the fixed potential of an NFW host. Each
  simulation has $M_\rmh/M_\rms = 1000$, $c_\rmh = 5$, $c_\rms = 10$
  and $x_\rmc = 1.0$, and only differ in their value for the orbital
  circularity, $\eta$ (Column 2). In each simulation the subhalo,
  which is simulated with $N=10^7$ particles, starts out at it's own
  apo-centre (cf. Fig.~\ref{fig:simres}).  
\end{minipage}
\end{table}

We integrate the subhalo in the external tidal field of an NFW host
halo whose virial mass (radius) is 1000 (10) times larger than that of
the subhalo. For such a large mass ratio, dynamical friction, which is
not accounted for in our idealized simulations, can be safely ignored.
In the $\Lambda$CDM cosmology, to good approximation, concentration
scales with halo mass as $c \propto M^{-0.1}$
\citep[e.g.,][]{Dutton.Maccio.14}. Hence, for a mass ratio of 1000,
the ratio in concentration parameters is roughly 2, in agreement with
the results shown in Fig.~\ref{fig:orbit}, and we therefore adopt a
concentration for the host halo of $c_\rmh = 5$. Each subhalo is
modeled using $N=10^7$ particles and a softening length of
$\varepsilon = 0.003$.  All simulations adopt a tree opening angle of
$\theta=0.7$, and a fixed time step of $\Delta t = 0.02$
(corresponding to $\sim 1.3 \Myr$). The latter implies that we resolve
the {\it minimal} orbital time of the system, defined as $\tau_{\rm
  min} \simeq [3 \pi / G \bar{\rho}(<\varepsilon)]^{1/2}$, with
$\bar{\rho}(<\varepsilon)$ the average density enclosed by the
softening length, with 30 time steps.  As discussed in great detail in
Paper II, for these parameters we obtain results that are both
converged (i.e., stable to an increase in $N$) and reproducible (i.e.,
different random realizations yield indistinguishable results).
\begin{figure*}
\includegraphics[width=\hdsize]{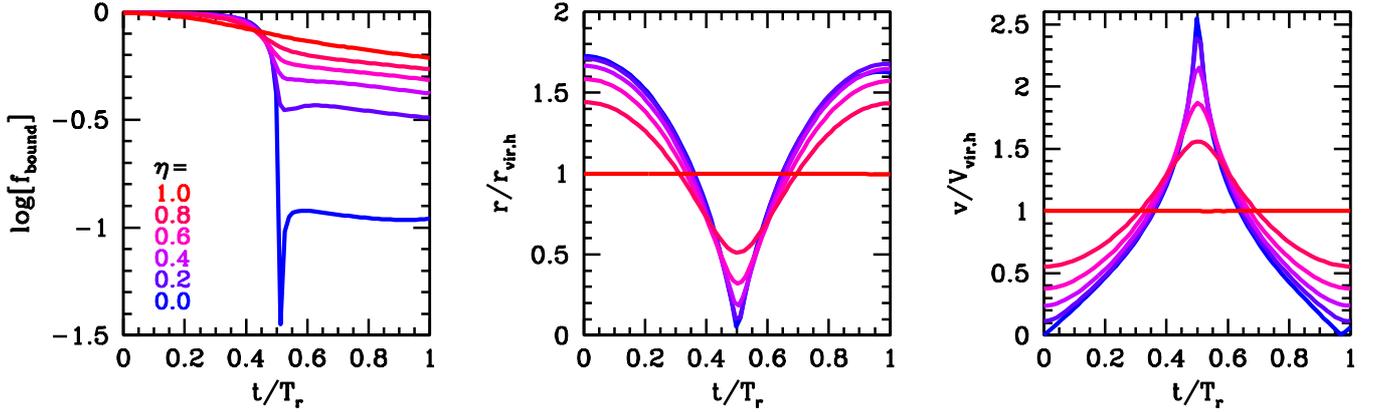}
\centering
\caption{Results for a subset of the idealized numerical simulations
  described in \S\ref{sec:sims} and listed in Table~1. From left to
  right, the panels show the time-evolution (where $t/T_\rmr$ is the
  time normalized by the radial period) of the logarithm of the bound
  mass fraction, $\log[f_{\rm bound}]$, the orbital radius in units of
  the virial radius of the host, $r/r_{\rm vir,h}$, and the orbital
  speed, in units of the virial velocity of the host, $v/V_{\rm
    vir,h}$. Different colors corresponds to different orbital
  circularities, as indicated. Note how the subhalo looses between
  $\sim 40$ ($\eta=1$) and $\sim 90$ ($\eta=0$) percent of its mass
  during its first radial period.}
\label{fig:simres}
\end{figure*}
We initially position the subhalo at the apo-centre of an orbit with
$x_\rmc = 1$ and $\eta = 0.0, 0.1,...,1.0$, and integrate the system
for a total of $10 \Gyr$ (corresponding to $\sim 7890$ time steps).
This is slightly longer than a radial orbital period, which varies
from $T_\rmr = 9.18$ Gyr for $\eta=0.0$ to $T_\rmr = 9.53$ Gyr for
$\eta=1.0$.  We use an extremely robust, iterative approach to
determine the bound fraction, $f_{\rm bound}$, as function of time,
which is described in detail in App.~\ref{App:fbound} (see also
Paper~II).

\subsection{Results}
\label{sec:results}

Fig.~\ref{fig:simres} shows some of the simulation results. The
left-hand panel shows the logarithm of the bound fraction as a function
of time, normalized by the radial period, $T_\rmr$, while the middle
and right-hand panels show the radius and velocity of the subhalo with
respect to the host halo, again as functions of $t/T_\rmr$. Different
colors correspond to different values of the orbital circularity,
$\eta$, as indicated, and for clarity we only show the results for a
subset of 6 simulations. Table~1 lists the bound fractions at 5
different epochs for the full set of 11 simulations.  Note how, for
fixed orbital energy, subhaloes on more radial orbits are stripped of
a larger fraction of their mass, per orbital period.  For the purely
radial orbit ($\eta=0$), as much as 89 percent of the original subhalo
mass is stripped away during its first radial orbit.

During peri-centric passage, the bound mass fraction can fluctuate
rapidly, especially for the more radial orbits \citep[see
  also][]{Diemand.etal.07, Han.etal.12}. This is a manifestation of
the re-binding of some particles as a consequence of the subhalo 
undergoing re-virialization. The tidal shock causes a stretching of 
the subhalo in the direction towards the centre of the host halo, 
and a compression in the direction perpendicular to the orbital plane. 
Hence, immediately following the tidal shock the subhalo finds itself 
out of virial equilibrium. While some matter has received a large 
enough boost in kinetic energy that it leaves the subhalo indefinitely, 
some of the particles that are unbound immediately following the tidal 
shock get rebound as the subhalo's potential re-adjusts as part of 
the re-virialization process.

What is most relevant for the discussion in this paper, is that none
of the simulated subhaloes undergo complete disruption. All orbits
simulated have $x_\rmc = 1$, which is at the low end of the
distribution (cf. upper-left panel of Fig.~\ref{fig:orbit}). Hence,
these are among the most bound orbits among newly accreted
subhaloes. And the simulations suggest that at most about 90 percent
of the subhalo mass will be stripped during the first radial period
(which lasts between 9 and 9.5 Gyr). It is intriguing to contrast this
with results from the cosmological Bolshoi simulation, in which a
significant fraction of subhaloes are disrupted during, or shortly
after, their first peri-centric passage \citep[][]{vdBosch.17}. This
suggests that much of this disruption is likely to be a numerical
artifact. We will elaborate on this in the following sections, as well
as in Papers II and III.


\section{Physical Disruption Processes}
\label{sec:Physical}

As discussed in \S\ref{sec:intro}, dark matter subhaloes are subjected
to tidal forces that strip the subhaloes of their mass, and which may
ultimately result in their complete disruption. Depending on the rate
with which the tidal forces change, the subhalo will respond
differently.  In the limit where the external tidal field changes
slowly (i.e., along close-to-circular orbits), the subhalo will loose
mass from beyond some limiting tidal radius. We refer to this as
`tidal stripping'. If the tidal field changes rapidly, particles
in the subhalo experience an impulsive kick, resulting in a
net heating of the subhalo.  We refer to this as tidal heating or
tidal shocking. Throughout we distinguish between tidal heating due to
the host halo (associated with the fast peri-centric passage along
highly eccentric orbits), and due to high-speed encounters with other
subhaloes, to which we refer as `harassment'.  In this section we
describe each of these processes in turn, and use simple analytical
estimates together with the simulation results presented in the
previous section to compare their relative importance.

Another process that is relevant for the evolution of dark matter
substructure is dynamical friction, which transfers orbital energy and
angular momentum of the subhalo to the particles of the host halo,
thereby causing the subhalo to `sink' towards the centre of the host,
where the tidal forces are stronger. Dynamical friction is only
expected to be significant for the most massive subhaloes, with a mass
that exceeds a few percent of that of the host halo
\citep[e.g.,][]{Binney.Tremaine.08, MBW10}, and various studies have
pointed out that indeed dynamical friction makes a negligible
contribution to the evolution of {\it the ensemble of} subhaloes
\citep[e.g.,][]{Zhao.04, Penarrubia.Benson.05, Ogiya.Burkert.16}. In
this paper we mainly focus on less massive subhaloes, which vastly
outnumber their more massive counter-parts, and therefore ignore
dynamical friction in what follows.

\subsection{Impulsive Heating due to the Host Halo}
\label{sec:TidalHeat}

During the high-speed peri-centric passage, the rapid, impulsive
change in the external potential causes a transfer of orbital energy
to subhalo internal energy. The resulting increase in subhalo internal
energy, $\Delta E$, can be computed using the impulse approximation
\citep[][]{Spitzer.58}, and in what follows we closely follow the
treatment of \citet[][hereafter GHO99]{Gnedin.etal.99}.

Consider a subhalo orbiting a spherical, NFW host halo of mass
$M_\rmh$ and concentration parameter $c_\rmh$. Let $\vec{R}_{\rm
  orb}(t)$ be the trajectory of the centre of the subhalo with respect
to that of the host halo, and $r$ be the distance of a subhalo
particle from the centre of the subhalo. The position vector of that
particle is then indicated by $\vec{R} = \vec{R}_{\rm orb} +
\vec{r}$. The subhalo's orbit is confined to a plane and, in the
absence of dynamical friction, characterized by two independent
parameters; the orbital energy $E$ and the orbital angular momentum
$L$, or equivalently, the parameters $x_\rmc = R_\rmc(E)/R_\rmh$, and
$\eta \equiv L/L_c(E)$ (see \S\ref{sec:prop}).  Following GHO99, we
assume that the orbit is in the $X$-$Y$ plane, so that $\vec{R}_{\rm
  orb}(t) = \cos\theta(t) \, \hat{X} + \sin\theta(t) \, \hat{Y}$,
where $\theta=0$ for $R_{\rm orb} = R_\rmp$ (i.e., at peri-centric
passage). Then, the impulsive change in the velocity of a subhalo
particle located at $\vec{r} = (x,y,z)$ can be written as
\begin{equation}\label{dvshock}
\Delta\vec{v} =  V_\rmh \, {\calQ \over R_\rmh} \,
\left\{ (B_1-B_3) x , (B_2 - B_3) y, -B_3 z \right\}\,,
\end{equation}
where
\begin{equation}
\calQ = {1 \over \eta\,x_\rmc^{1/2}} \,
  \left[ {f(c_\rmh) \over f(x_\rmc c_\rmh)} \right]^{1/2}\,,
\end{equation}
and\footnote{Note that the integrals for $B_k$ (k=1,2,3) defined
  here differ from those in GHO by a factor $c_\rmh$.}
\begin{equation}\label{Bone}
  B_1 = \int_{-\theta_\rmm}^{\theta_\rmm}
  {3 \mu_0(\zeta) - \mu_1(\zeta) \over \zeta} \, \cos^2\theta \, \rmd\theta\,,
\end{equation}
\begin{equation}\label{Btwo}
  B_2 = \int_{-\theta_\rmm}^{\theta_\rmm} 
  {3 \mu_0(\zeta) - \mu_1(\zeta) \over \zeta} \, \sin^2\theta \, \rmd\theta\,,
\end{equation}
\begin{equation}\label{Bthree}
  B_3 = \int_{-\theta_\rmm}^{\theta_\rmm}
  {\mu_0(\zeta) \over \zeta} \, \rmd\theta\,.
\end{equation}
Here $\zeta = \zeta(\theta) \equiv R_{\rm orb}/R_\rmh$ is the orbital
radius of the subhalo in units of the host halo's virial radius,
$\theta_\rmm = \Delta\Psi/2$ (see Eq.~[\ref{deltapsi}]), and the
functions $\mu_0(x)$ and $\mu_1(x)$ for an NFW host halo are given by
\begin{equation}\label{muzero}
  \mu_0(x) = {1 \over f(c_\rmh)} \, \left[ \ln(1+c_\rmh x) - {c_\rmh x \over
      1 + c_\rmh x } \right] \,,
\end{equation}
and
\begin{equation}\label{muone}
  \mu_1(x) = {1 \over f(c_\rmh)} \, \left[ {(c_\rmh x)^2 \over
      (1 + c_\rmh x)^2 } \right] \,.
\end{equation}
Note that the above integrations are along one orbit $R_{\rm
  orb}(\theta)$ of the subhalo, running from apo-centre to apo-centre,
as parameterized by the position angle $\theta$. We compute $R_{\rm
  orb}(\theta)$ by integrating the subhalo's orbit in the host halo
potential using a fifth-order Cash-Karp Runge-Kutta method with
adaptive step-size control \citep[e.g.][]{Press.etal.92}.  When
computing the integrals for $B_1$, $B_2$ and $B_3$, we use bi-cubic
spline interpolation of the resulting $R_{\rm orb}(\theta)$.

Eq.~(\ref{dvshock}) implies that subhalo particles at subhalo-centric
radius $r$ experience an average change in specific (kinetic) energy,
per radial period, equal to
\begin{equation}\label{TidalShock}
  \langle \Delta E \rangle(r) = {1 \over 2} \langle \Delta\vec{v} \cdot
  \Delta\vec{v} \rangle(r) = {1 \over 6} \, \calQ^2 \,
  V^2_\rms \, \chi_{\rm orb} \, {r^2 \over r^2_\rms}\,.
\end{equation}
Here $\chi_{\rm orb} \equiv (B_1-B_3)^2 + (B_2-B_3)^2 + B_3^2$, and we
have used that, by virtue of the definition of `virial radius',
$V_\rmh/R_\rmh = V_\rms/r_\rms$.  In addition, we have assumed an
isotropic velocity distribution, so that $\langle \vec{v} \cdot
\Delta\vec{v} \rangle = 0$.
\begin{figure*}
\includegraphics[width=\hdsize]{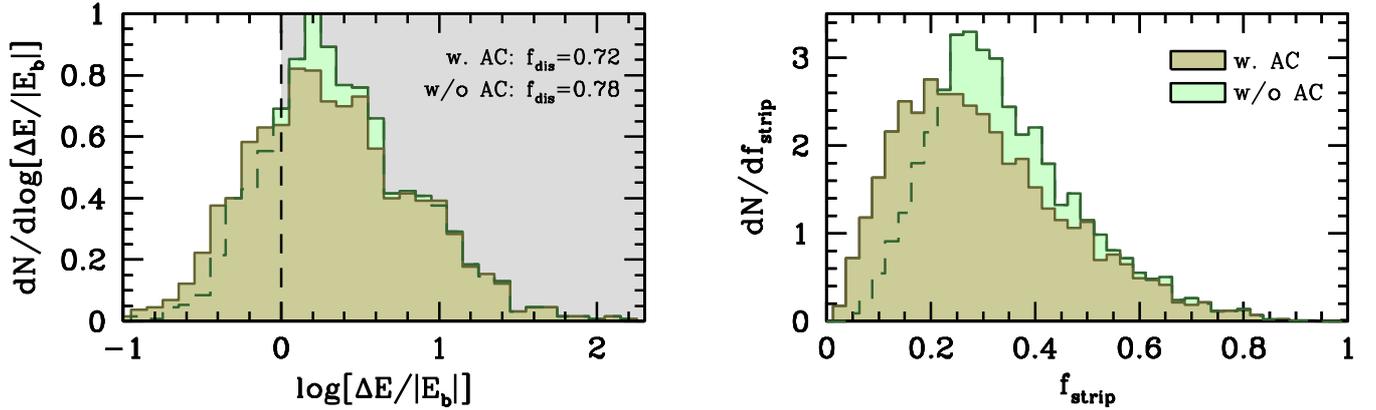}
\centering
\caption{{\it Left-hand panel:} The distribution of $\log[\Delta
    E/|E_\rmb|]$ for the same set of newly accreted subhaloes as in
  Fig.~\ref{fig:orbit}. Here $\Delta E$ is the energy injected into
  the subhalo during its first radial period, due to the impulsive
  shock associated with peri-centric passage (Eq.~[\ref{dEtotorb}]),
  while $E_\rmb$ is the subhalo's binding energy
  (Eq.~[\ref{Etotsub}]). Ochre and green histograms indicate the
  results obtained with and without the adiabatic correction
  (Eq.~[\ref{ACorr}]), respectively, which has little impact.  The
  grey-shaded region indicates where $\Delta E > |E_\rmb|$, and thus
  where the energy injected due to the impulsive shock exceeds the
  subhalo's original binding energy. The fractions of subhaloes that
  meet this criterion are indicated. {\it Right-hand panel:} the
  corresponding fraction of mass lost by the subhalo as a consequence
  of the tidal shock, calculated under the instantaneous mass loss
  approximation as described in the text.}
\label{fig:heating}
\end{figure*}

The above derivation is based on the impulse approximation, wherein
the subhalo particles are assumed to move little during the tidal
shock. This is, however, only appropriate for the least bound
particles, in the outer regions of the subhalo. The more bound
particles will have dynamical times that can be substantially shorter
than the duration of the tidal shock, $\tau_{\rm sh} \simeq
R_\rmp/V_\rmp$, where $V_\rmp = \sqrt{2[E-\Phi_\rmh(R_\rmp)]}$ is the
subhalo speed at peri-centre. For such particles, the peri-centric
passage of the host halo is adiabatic, rather than impulsive, and the
conservation of adiabatic invariants prevents these particles from
gaining any energy \citep[][]{Spitzer.87, Weinberg.94a, Weinberg.94b}.
\cite{Gnedin.Ostriker.99} show that one can correct for this
`adiabatic shielding' by simply multiplying $\langle \Delta E
\rangle(r)$ of Eq.~(\ref{TidalShock}) with an adiabatic correction of
the form
\begin{equation}\label{ACorr}
A_{\rm corr}(r) = \left[1 + \omega^2(r) \, \tau^2_{\rm sh}\right]^{-3/2}\,.
\end{equation}
Here $\omega(r)$ is the orbital frequency of subhalo particles at
radius $r$. Throughout we adopt $\omega(r) = \sigma_r(r)/r$, where
$\sigma_r(r)$ is the radial velocity dispersion at subhalo-centric
radius $r$, obtained by solving the Jeans equation for a spherical,
isotropic NFW subhalo \citep[e.g., Eq.~(11) in][]{vdBosch.etal.04}.

Finally, we can use the above to compute the total energy injected
into the subhalo per radial orbit using
\begin{eqnarray}\label{dEtotorb}
  \Delta E & = & 4 \pi \int_0^{r_\rms}
  \rho_{\rm s}(r) \,\ \langle \Delta E \rangle(r) \, A_{\rm corr}(r) \,
  r^2 \, \rmd r \nonumber \\
  & = & {1 \over 2} M_\rms V^2_\rms \,
  {\chi_{\rm orb} \, \calQ^2 \over 3 \, c^2_\rms f(c_\rms)} \, \calC_{\rm corr}\,,
\end{eqnarray}
where
\begin{equation}\label{Ccorr}
\calC_{\rm corr} = \int_0^{c_\rms} {x^3 A_{\rm corr}(x r_0) \over (1+x)^2} \rmd x\,,
\end{equation}
with $r_0 = r_\rms/c_\rms$ the scale radius of the subhalo.

\subsubsection{Application to Bolshoi Subhaloes}
\label{sec:applB}

In order to assess the potential impact of tidal heating we compute
$\Delta E / |E_\rmb|$ for each of the 5,000 newly accreted subhaloes
from the Bolshoi simulation (see \S\ref{sec:prop}). Here $E_\rmb$ is
the total binding energy of the initial subhalo (i.e., prior to the
tidal shock). Since we consider the initial subhalo to be truncated at
its virial radius, we have that
\begin{equation}\label{Etotsub}
E_\rmb = -{1 \over 2} \, M_\rms \, V^2_\rms \, f_\rmE(c_\rms) \,,
\end{equation}
where
\begin{equation}\label{fE}
f_E(c) = {c \over 2 \, f^2(c)} \, 
 \left[1 - {1 \over (1+c)^2} - {2 \ln(1+c) \over (1+c)} \right]\,,
\end{equation}
\citep{Mo.etal.98}. The ochre histogram in the left-hand panel of
Fig.~\ref{fig:heating} shows the distribution of $\Delta E / |E_\rmb|$
values for the newly accreted subhaloes in the Bolshoi simulation,
computed using Eqs.~(\ref{dEtotorb})-(\ref{fE}).  The median is $1.9$,
indicating that a typical subhalo, during its first-pericentric
passage, experiences a tidal, impulsive shock that injects an amount
of energy that is roughly twice its original binding energy.  If we
were to associate $\Delta E > |E_\rmb|$ with the complete disruption
of the subhalo, as is done for example in \cite{Gonzales.etal.94}, 
then the inference would be that 72 percent of subhaloes do not survive 
their first peri-centric passage.  For comparison, the green histogram 
shows the distribution obtained when ignoring the adiabatic correction 
(i.e., by setting $A_{\rm corr}=1$). This boosts the distributions to 
slightly larger values of $\Delta E / |E_\rmb|$, and increases the 
`disruption fraction' to $0.78$.
\begin{figure*}
\includegraphics[width=\hdsize]{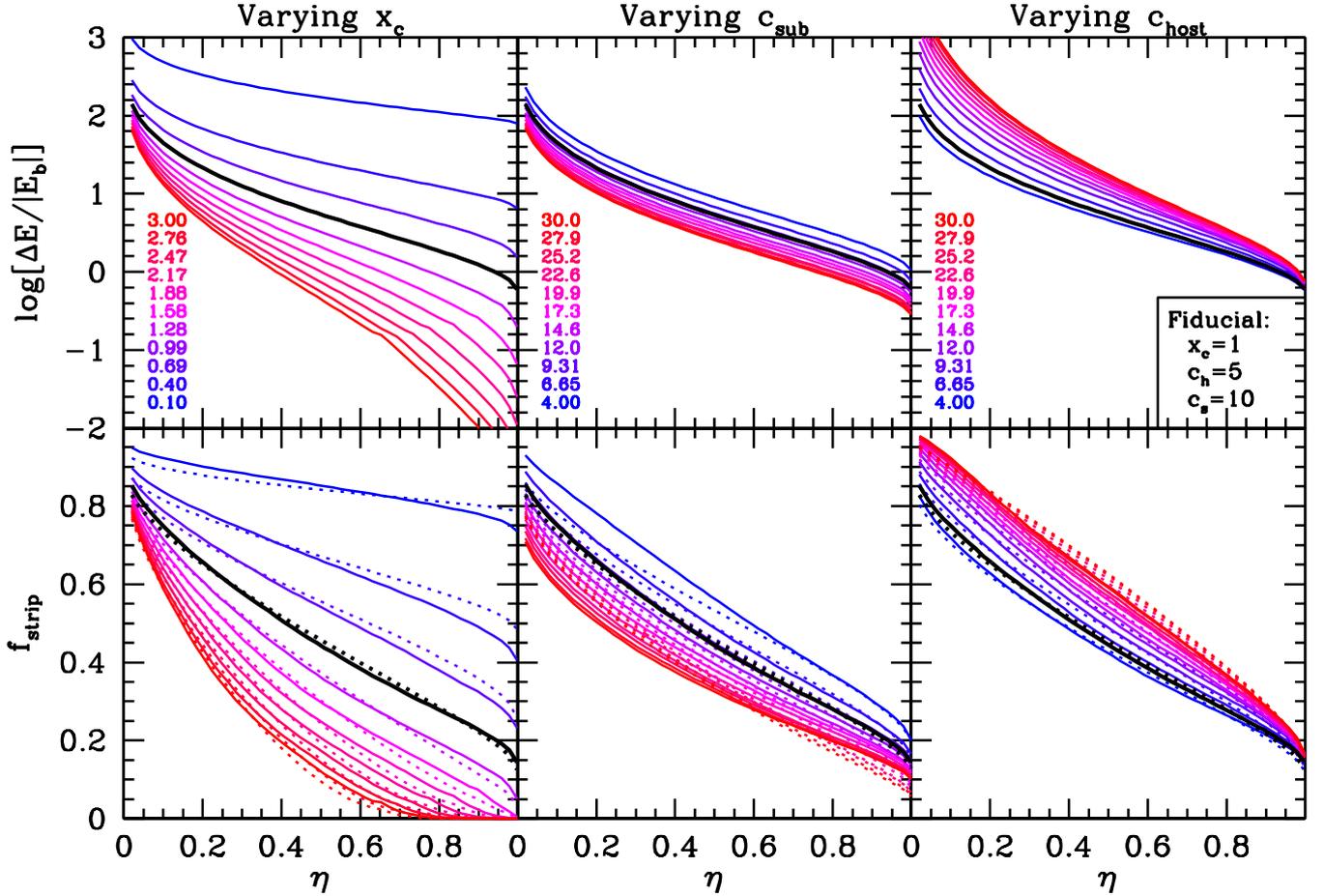}
\centering
\caption{The logarithm of the ratio $\Delta E/|E_\rmb|$ (upper panels)
  and the corresponding stripped mass fraction, $f_{\rm strip}$,
  (lower panels) as functions of the orbital circularity, $\eta$. All
  results assume a host halo to sub-halo mass ratio of 1000. Solid and
  dotted curved correspond to $\Delta E$ having been calculated with
  and without adiabatic correction, respectively.  Results are shown
  for different values of the orbital energy, as parameterized via
  $x_\rmc$ (left-hand column), the concentration of the subhalo,
  $c_\rms$, (middle column), and the concentration of the host halo,
  $c_\rmh$,(right-hand column); corresponding values are
  indicated. The black line corresponds to the fiducial model, which
  has $x_\rmc = 1$, $c_\rms = 10$, and $c_\rmh = 5$. Note that $\Delta
  E/|E_\rmb| = 1$ corresponds to $f_{\rm strip} \sim 0.2$; hence even
  if the energy injected impulsively exceeds the original subhalo
  binding energy, only a modest fraction of subhalo mass will be
  stripped.}
\label{fig:params}
\end{figure*}

There are two important points to be made here. First of all, the
overall impact of adiabatic shielding is fairly modest, and one does
not make a large error if one simply ignores the adiabatic
correction. Secondly, if indeed subhaloes for which $\Delta E >
|E_\rmb|$ were to experience complete disruption, the inference would
be that the vast majority of subhaloes do not survive first
peri-centric passage. However, this interpretation of $\Delta
E/|E_\rmb|$ is extremely naive, and dramatically overestimates the
efficiency of subhalo disruption.  The reason is that $\Delta E$ does
not specify how that energy is distributed over the particles. That
this is important is evident from the following {\it reductio ad
  absurdum}; if all the energy were given to only a single particle,
the system clearly would not disrupt, even if $\Delta E \gg
|E_\rmb|$. Rather, only that single particle would escape, and leave
the remainder of the halo virtually intact. As is evident from
Eq.~(\ref{TidalShock}), the average change in kinetic energy due to an
impulsive tidal shock scales with $r^2$. Hence, particles at the
outskirts of the subhalo, which need less energy to escape, receive
far more energy than the more bound particles in the central regions.

In order to properly assess the impact of $\Delta E$ on the system as
a whole, one needs to compare, for each individual particle $i$, the
particle's individual $(\Delta E)_i$ to its individual binding energy,
$|E_i|$.  One can then express the impact of the tidal shock in terms
of the fraction of particles with $(\Delta E)_i > |E_i|$. We will
refer to this fraction as $f_{\rm strip}$. Note that this
`instantaneous mass loss approximation'
\citep[cf.,][]{Aguilar.White.85} ignores potential re-binding and
unbinding of particles during the re-virialization process immediately
following the tidal shock.  However, these processes are neither
calculable, nor do they appear to have a significant impact, as
demonstrated in \S\ref{sec:compsim} below.

In order to compute $f_{\rm strip}$ for individual subhaloes, we
proceed as follows. We construct an $N$-body realization, consisting
of $N_\rmp=50,000$ particles, of each individual subhalo, using the
method described in \S\ref{sec:sims} above. We do not evolve these
realizations with an actual simulation code; rather for each particle
$i$ we analytically compute its binding energy, $E_i = v^2_i/2 +
\Phi_\rms(r_i)$, and
\begin{equation}\label{dEpart}
(\Delta E)_i = A_{\rm corr}(r_i) \left[
    \vec{v}_i \cdot \Delta\vec{v}_i + {1 \over 2} (\Delta v_i)^2\right]
\end{equation}
where $\Delta\vec{v}$ is given by Eq.~(\ref{dvshock}), and the factor
$A_{\rm corr}$ (Eq.~[\ref{ACorr}]) accounts for the adiabatic
shielding of the most bound particles. Finally, we compute the fraction
$f_{\rm strip}$, of particles for which $(\Delta E)_i > |E_i|$.

Before showing the results for the subhaloes in the Bolshoi
simulation, we first illustrate how $f_{\rm strip}$ relates to $\Delta
E / |E_\rmb|$, by computing both values for a variety of orbits and
concentration parameters of both host- and sub-halo.
Fig.~\ref{fig:params} shows $\Delta E / |E_\rmb|$ (upper panels) and
$f_{\rm strip}$ (lower panels) as functions of the orbital
circularity, $\eta$. The left-hand panels show the dependence on the
orbital energy, as parameterized by $x_\rmc = R_\rmc(E)/R_\rmh$, while
the middle and right-hand panels show the dependences on the
concentrations of the subhalo, $c_\rms$, and the host halo, $c_\rmh$,
respectively. In all cases we adopt a host halo-to-subhalo mass ratio
of $M_\rmh/M_\rms = 1000$. As expected, the impact of tidal heating
increases if the orbit is more radial (smaller $\eta$) and/or more
bound (smaller $x_\rmc$), if the subhalo is less concentrated (smaller
$c_\rms$), or if the host halo is more concentrated (larger $c_\rmh$).
It is not uncommon for $\Delta E$ to exceed $|E_\rmb|$ by more than an
order of magnitude, especially when $\eta \lta 0.5$. However, the
stripped mass fraction rarely exceeds 80 percent. We find a tight
relation between $\Delta E / |E_\rmb|$ and $f_{\rm strip}$, which is
reasonably well fit by
\begin{equation}\label{fitfunc}
f_{\rm strip} = {\rm exp}\left[-1.65 \, (\Delta E/|E_\rmb|)^{-0.44}\right]\,.
\end{equation}
The stripped mass fractions predicted from $\Delta E/|E_\rmb|$ using
Eq.~(\ref{fitfunc}) are indicated in the lower panels of
Fig.~\ref{fig:params} as dotted curves, and (except for some of the
more extreme concentration parameters) typically agree with the values
inferred using the Monte Carlo method described above to better than a
few percent.  Using this equation, which holds for NFW subhaloes
covering a wide range of concentration parameters and orbital
parameters, we see that an impulsive tidal shock with $\Delta
E/|E_\rmb| = 1$ ($10$) only unbinds roughly 20 (55) percent of the
subhalo mass.

The right-hand panel of Fig.~\ref{fig:heating} shows the distribution
of $f_{\rm strip}$, computed using the Monte-Carlo method described
above, for the newly accreted subhaloes in the Bolshoi simulation.
Ochre and green histograms correspond to the bound fractions obtained
with and without the adiabatic correction, respectively, which has
little impact. Under the instantaneous mass loss approximation tidal
heating due to the first peri-centric passage is expected to strip on
average about 20-30 percent of the mass of a subhalo. Note that there
are no subhaloes for which $f_{\rm strip} > 0.9$.
\begin{figure}
\includegraphics[width=\hssize]{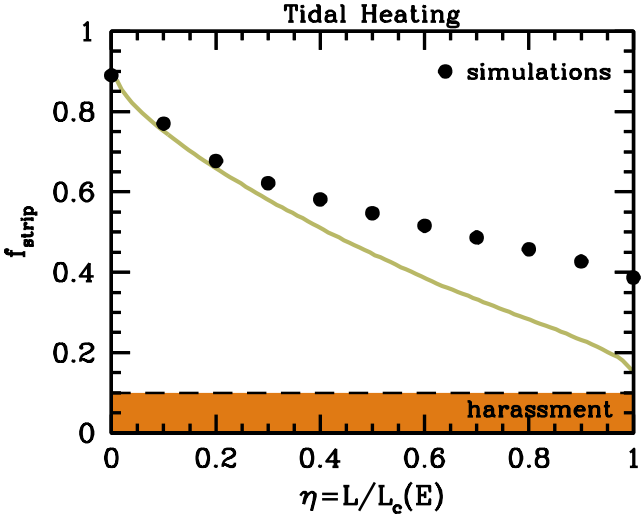}
\centering
\caption{The subhalo mass fraction that is stripped off {\it during
    the first radial period}, $f_{\rm strip}$, as function of the
  orbital circularity, $\eta$. The subhalo is modeled as a spherical
  NFW halo with concentration parameter $c_\rms = 10$, orbiting in a
  host halo of mass $M_\rmh = 1000 M_\rms$ and concentration parameter
  $c_\rmh = 5$, on an orbit with $x_\rmc = 1$.  Solid dots indicate
  the results from our numerical simulations (see \S\ref{sec:sims}
  and Table~1). The solid, ochre curve indicates the fraction of
  subhalo particles that receive an impulsive, tidal shock, $\Delta
  E$, that is larger than its binding energy, computed using the
  method described in \S\ref{sec:applB}. Note the good agreement
  with the simulation results for the more radial orbits with $\eta
  \lta 0.2$.}
\label{fig:comp_heating}
\end{figure}

\subsubsection{Comparison with Idealized Simulations}
\label{sec:compsim}

We now compare our semi-analytical predictions to the results of our
idealized numerical simulations described in \S\ref{sec:sims}. We thus
consider subhaloes with mass $M_\rms$ and concentration parameter
$c_\rms = 10$ orbiting in a host halo of mass $M_\rmh = 1000 M_\rms$
and concentration parameter $c_\rmh = 5$. We adopt an orbital energy
characterized by $x_\rmc=1$, and vary the orbital circularity $\eta$.
The solid, ochre curve in Fig.~\ref{fig:comp_heating} shows the mass
fraction that is stripped due to tidal heating during the subhalo's
first radial orbital period, computed under the instantaneous mass
loss approximation as the fraction of particles with $(\Delta E)_i >
|E_i|$. For comparison, the solid dots show the stripped mass fraction
in the simulations after one radial period.

For $\eta \lta 0.2$, the semi-analytical predictions are in excellent
agreement with the simulation results, indicating that the evolution
of the bound fraction for subhaloes on relatively radial orbits is
governed by tidal shocking, and that the impulse approximation
combined with the instantaneous mass loss approximation accurately
describes the resulting mass evolution of subhaloes. For less radial
orbits, tidal heating by itself underpredicts the stripped mass
fraction. This is expected, since the changes in the tidal field
become less `impulsive' and the impact of tidal stripping is expected
to become more important.

In Paper III in this series (Ogiya et al. in prep.) we compare our
analytical estimates presented here to a large suite of (properly
converged) simulations, covering a much larger range in orbital
energies, orbital angular momentum and halo concentrations.

\subsection{Impulsive Heating due to Subhalo-Subhalo Encounters}
\label{sec:SubSubHeat}

In addition to experiencing tidal shocks during peri-centric passages,
subhaloes also undergo tidal heating due to high-speed (impulsive)
encounters with other subhaloes. The cumulative impact of many such
high-speed impulsive encounters is sometimes called `harassment'
\citep{Moore.etal.96b}.
\begin{figure*}
\includegraphics[width=\hdsize]{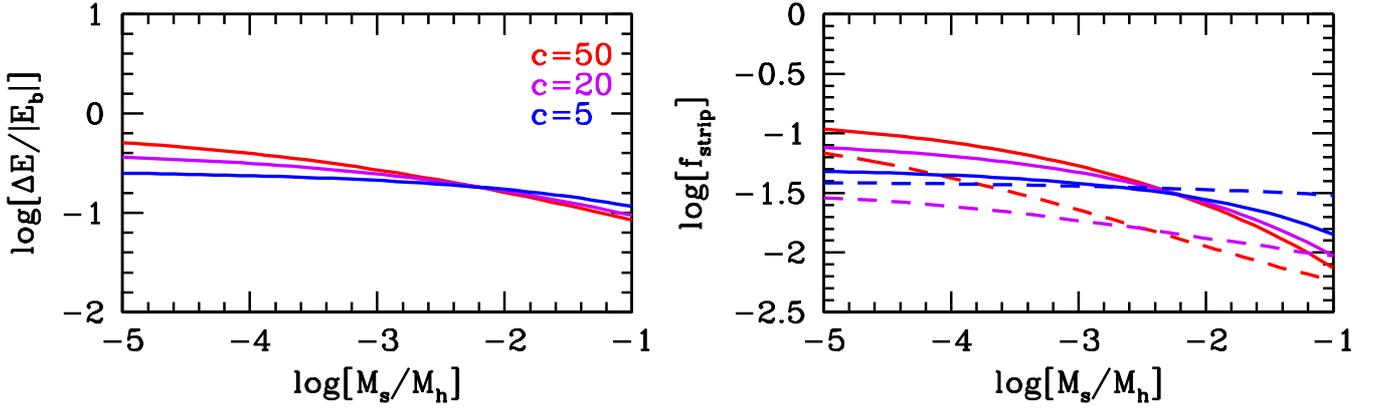}
\centering
\caption{{\it Left-hand panel:} The tidal heating per crossing time
  due to impulsive subhalo-subhalo encounters (harassment). Different
  curves plot the logarithm of $(\Delta E)_{\rm cross}/|E_\rmb|$
  (Eq.~[\ref{harassment}]) as function of the logarithm of the ratio
  of the mass of the target subhalo, $M_\rms$, to the mass of the host
  halo, $M_\rmh$, for three different values of the assumed
  concentration of dark matter subhaloes, as indicated.  {\it
    Right-hand panel:} The logarithm of the corresponding stripped
  mass fraction, $f_{\rm strip}$, computed under two extreme
  conditions; where $\Delta E$ is dominated by a single encounter
  (solid lines) and by many, independent encounters (dashed
  lines). Note how tidal harassment due to subhalo-subhalo encounters
  is expected to only strip between 1 and 10 percent of the subhalo's
  mass per crossing time. This is significantly less than the mass
  fraction stripped due to the first peri-centric passage of the host
  halo (cf.  Figs.~\ref{fig:params} and~\ref{fig:heating}).}
\label{fig:harassment}
\end{figure*}

In this section we use a crude, analytical estimate of how the tidal
heating due to harassment compares to that due to a peri-centric
passage. Consider a target subhalo (hereafter the subject) of mass
$M_\rms$ and size $r_\rms$ on a radial orbit ($\eta=0$) in a host halo
of mass $M_\rmh$ and size $R_\rmh$. Assume for simplicity that the
host halo has a uniform distribution of subhaloes with mass function
$\rmd n/\rmd m$, which are orbiting the host with an orbital velocity
dispersion $\sigma_\rmh \simeq (G M_\rmh/2 R_\rmh)^{1/2}$ (a good
approximation for the global 1D velocity dispersion of an isotropic
NFW halo).

In order to calculate the total tidal heating $\Delta E$ of our target
subhalo per radial period due to harassment, we first consider a
single encounter with a perturbing subhalo of mass $m$ and size
$r_{\rm m}$. Let $b$ and $V$ be the impact parameter and relative
velocity of the encounter, which we assume to follow a straight path
$\vec{R}_{\rm orb}(t) = (b,Vt,0)$ with respect to the centre of the
subject mass. As shown by GHO99, in the distant tide approximation ($b
\gg r_\rms$), the velocity kick of a particle located at $\vec{r} =
(x,y,z)$ is given by
\begin{eqnarray}
\lefteqn{\Delta \vec{v} = {2 G m \over b^2 V} }  \\
 & & \left\{ (3J_0 - J_1 - I_0) \, x,
  (2I_0 - I_1 - 3J_0 + J_1) \, y, -I_0 \,z\right\}\, \nonumber
\end{eqnarray}
where
\begin{equation}
I_k(b) = \int_1^{\infty} \mu_k(b\zeta) {\rmd \zeta \over \zeta^2
    (\zeta^2-1)^{1/2}}\,
\end{equation}
and
\begin{equation}
J_k(b) = \int_1^{\infty} \mu_k(b\zeta) {\rmd \zeta \over \zeta^4
    (\zeta^2-1)^{1/2}}\,
\end{equation}
with $\mu_0(x)$ and $\mu_1(x)$ given by Eqs.~(\ref{muzero})
and~(\ref{muone}), respectively. Integrating over the spherically
symmetric subject mass yields a total tidal heating of
\begin{equation}\label{dEbm}
\Delta E_{\rm dt}(b,m) = {4 G^2 m^2 \over b^4 V^2} \, M_\rms \, 
{\langle r^2\rangle_\rms \over 3} \, \chi_{\rm st}(b)\,.
\end{equation}
where the subscript indicates that this is only valid under the
distant tide approximation,
\begin{equation}\label{chist}
\chi_{\rm st} = {1 \over 2} 
[(3J_0 - J_1 - I_0)^2 +  (2I_0 - I_1 - 3J_0 + J_1)^2 + I_0^2]\,,
\end{equation}
and
\begin{equation}
\langle r^2 \rangle_\rms = {4 \pi \over M_\rms} \int_0^{r_\rms} \rho_\rms(r) 
\, r^4 \, \rmd r\,,
\end{equation}

For a head-on collision (i.e., $b=0$), the cylindrical symmetry
assures that the velocity kick for a particle only depends on, and is
pointing in the direction of, the cylindrical radius $R = \sqrt{x^2 +
  z^2}$. In particular,
\begin{equation}
\Delta v_R = 2 \int_0^{\infty} {G m(r) \over r^2} \, {R \over r} \rmd t
  = {2 G m \over V} \, {I_0(R) \over R}
\end{equation}
which implies a tidal heating
\begin{equation}\label{dEb0m}
\Delta E_{\rm ho}(m) = {4 G^2 m^2 \over V^2} \, \pi \, 
\int_0^{r_\rms} I_0^2(R) \, \Sigma_\rms(R) \, {\rmd R \over R}\,
\end{equation}
with $\Sigma_\rms(R)$ the projected surface density profile of the
subject mass, and the subscript indicates that this is the heating for
a head-on collision. Throughout we assume that subhaloes have NFW
density profiles\footnote{We acknowledge that this is not a
  particularly accurate assumption \citep[e.g.,][]{Hayashi.etal.03},
  but the detailed density profile of subhaloes has little impact on
  what is anyways a fairly crude estimate.}, for which
\begin{equation}
  \Sigma_\rms(R) = {M_\rms \over 2 \pi r_0^2} {1 \over f(c_\rms)} g(R/r_0)
\end{equation}
with $r_0 = r_\rms/c_\rms$ the scale radius, and
\begin{equation}\label{gNFW}
g(x) = \left\{
  \begin{array}{ll}
{1 \over x^2 -1} \left[1 - {2 \over \sqrt{1-x^2}} {\rm arctanh}\sqrt{1-x \over 1+x} \right] & \mbox{$x<1$} \\
{1 \over 3} & \mbox{$x=1$} \\
{1 \over x^2 -1} \left[1 - {2 \over \sqrt{x^2-1}} {\rm arctan}\sqrt{x-1 \over x+1} \right] & \mbox{$x>1$} \\
  \end{array}\right.   
\end{equation}

We may assume that $\Delta E(b,m)$ for any impact parameter $b$ is a
smooth interpolation between $\Delta E_{\rm dt}(b,m)$, which only
holds for $b \gg r_\rms$ and $\Delta E_{\rm ho}(m)$, which holds for
$b=0$. In fact, we won't make a big error if we simply set $\Delta
E(b,m)$ to be the minimum of $\Delta E_{\rm dt}(b,m)$ and $\Delta
E_{\rm ho}(m)$, i.e.,
\begin{equation}\label{dEcomb}
  \langle \Delta E \rangle(b,m) = {4 G^2 m^2 M_\rms \over 3 V^2} \,
  \langle r^2\rangle_\rms \, \left\{
  \begin{array}{ll}
    {\chi_{\rm st}(b) \over b^4} & \mbox{if $b > b_0$} \\
    {\chi_{\rm st}(b_0) \over b_0^4} & \mbox{if $b\leq b_0$} 
  \end{array}
  \right.
\end{equation}
Here $b_0$ is defined by $\Delta E_{\rm dt}(b_0,m) = \Delta E_{\rm
  ho}(m)$; i.e., $b_0$ is the impact parameter for which the heating
computed using Eq.~(\ref{dEbm}) is equal to that of a head-on collision
(Eq.~[\ref{dEb0m}]), and can be solved for using a simple root-finder.

Under the assumption that the subhaloes are uniformly distributed
within the host halo, the total heating experienced by our target
subhalo during one passage through the host halo is given by
\begin{equation}
(\Delta E)_{\rm cross} = 2 \pi \int_0^{M_\rmh} \rmd m \, {\rmd n \over \rmd m} \int_0^{R_\rmh}
  \Delta E(b,m) \, {b \, \rmd b \over \pi R^2_\rmh} 
\end{equation}
Substituting Eq.~(\ref{dEcomb}) and using that $V^2 \simeq 2
\sigma^2_\rmh$, yields
\begin{eqnarray}\label{harassment}
  \lefteqn{ {(\Delta E)_{\rm cross} \over |E_\rmb|} = {16 \over 3} \,
\left({M_\rmh \over M_\rms}\right)^{2/3} \, {\langle r^2\rangle_\rms \over
  f_\rmE(c_\rms)}} \nonumber \\
  & &  \int_0^1 \rmd\psi \, \psi \, {\rmd n \over \rmd\ln\psi} \,
\left[ {\chi_{\rm st}(b_0) \over 2 b_0^2}
  + \int_{b_0}^{R_\rmh} {\chi_{\rm st}(b) \over b^3} \, \rmd b\right]\,.
\end{eqnarray}
Here $\psi = m/M_\rmh$, and we have made the additional assumption
that $V_\rmh/V_\rms = (M_\rmh/M_\rms)^{1/3}$.
\begin{figure*}
\includegraphics[width=\hdsize]{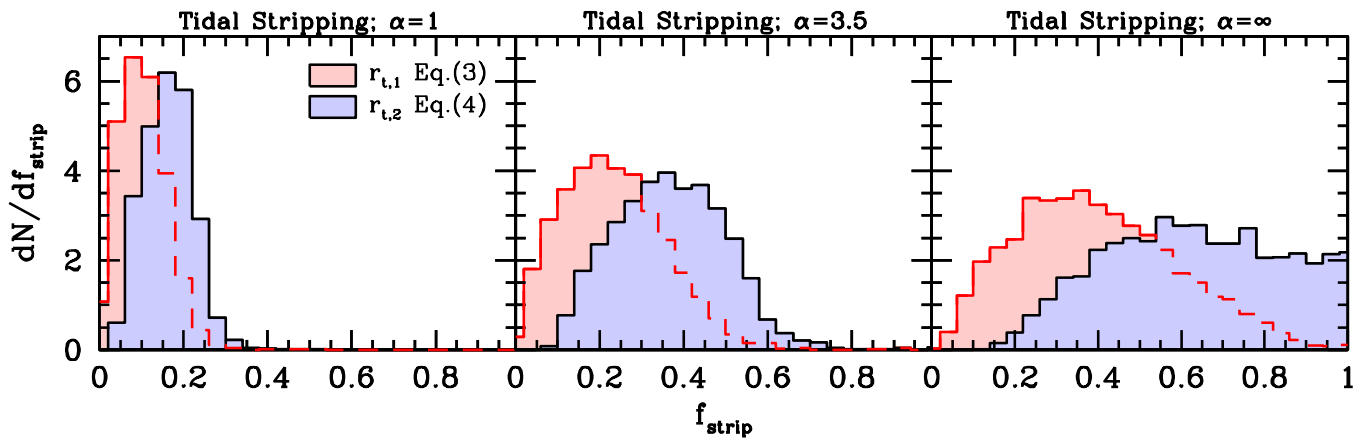}
\centering
\caption{Distributions of the mass fraction $f_{\rm strip}$ that is
  stripped off during a subhalo's first radial orbital period due to
  tidal (non-impulsive) stripping for the same set of newly accreted
  subhaloes as in Fig.~\ref{fig:orbit}. Red and blue histograms are
  based on the different definitions of the tidal radius, as
  indicated, while different panels correspond to different stripping
  rates, as parameterized by $\alpha$ (cf. Eq.~[\ref{striprate}]). The
  case were $\alpha=\infty$ (right-hand panel) corresponds to
  instantaneous stripping, for which $f_{\rm strip} =
  M_\rms(>\rpt)/M_\rms$.}
\label{fig:tidstrip}
\end{figure*}

The left-hand panel of Fig.~\ref{fig:harassment} shows $(\Delta
E)_{\rm cross}/|E_\rmb|$ as function of the mass ratio
$M_\rmh/M_\rms$, computed with Eq.~(\ref{harassment}) using the
universal fitting function for the (evolved) subhalo mass function of
\cite{Jiang.vdBosch.16}, and adopting a typical subhalo mass fraction
of 10 percent. For simplicity, we assume that all subhaloes (subject
as well as perturbers) have the same NFW concentration parameter, $c$.
Results are shown for three values of $c$, as indicated. Note that the
expected amount of tidal heating due to harassment experienced by a
subhalo is typically between 10 and 50 percent of its binding energy,
per crossing time. Typically, as one might expect, more massive
subhaloes experience less harassment, but the mass dependence is
fairly weak.

As discussed in \S\ref{sec:applB}, it is difficult to gauge the impact
of $\Delta E$ on a subhalo without addressing how this energy is
distributed over the $N$ constituent particles. In the case of
harassment this is difficult to address with accuracy. The reason is
that particles in the subject mass move in between separate collisions
with individual perturbing subhaloes. However, we may get some insight
using two limiting cases. If $\Delta E$ is dominated by the impulsive
shock from a single encounter (with a massive subhalo and/or a small
impact parameter), then we are in the limit where the energy gain for
particle $i$ scales as $(\Delta E)_i \propto r_i^2$ and we can use
Eq.~(\ref{fitfunc}) to estimate the stripped mass fraction. The
results are indicated as solid lines in the right-hand panel of
Fig.~\ref{fig:harassment}. If, on the other hand, $\Delta E$ is
dominated by the cumulative effect of many encounters (with small
subhaloes, covering the entire range of impact parameters), then we
are in the regime where $(\Delta E)_i = (\Delta E)_{\rm cross}/N$;
i.e., the energy gain is independent of location within the subhalo
(see \S\ref{sec:PStidal}).  In this limit we can compute the stripped
mass fraction using a similar Monte-Carlo approach as in
\S\ref{sec:applB} and simply counting the fraction of particles for
which $(\Delta E)_i > E_i$. This results in the stripped mass
fractions indicated by dashed lines in the right-hand
panel. Interestingly, the two extreme cases yield stripped mass
fractions that are relatively similar, spanning the range $0.01 \lta
f_{\rm strip} \lta 0.1$. This range is indicated as an orange-shaded
region in Fig.~\ref{fig:comp_heating}.

Although the above estimate for the tidal heating due to harassment is
fairly crude, it is clear that the impact of harassment is sub-dominant
to that due to the subhalo's first peri-centric passage. We therefore
conclude that {\it the tidal heating due to subhalo-subhalo encounters
  can safely be neglected when trying to assess the survivability of
  dark matter substructure.}

This conclusion is at odds with a study by \citet{Knebe.etal.06}, who
claim that harassment may contribute as much as 40 percent to the
total mass loss rate of a subhalo. However, their estimate is based on
comparing the force on the subhalo due to the host halo to that due to
all other subhalos, without taking account of the actual tides
(relevant for tidal stripping) or the relative velocities (relevant
for tidal heating).  Although we believe our estimate to be more
accurate, a more detailed study is required to settle this
disagreement, which is beyond the scope of this paper.

\subsection{Tidal Stripping}
\label{sec:tidalstripping}

In the previous two subsections we focused on tidal shocking, which
manifests itself whenever the tidal field changes rapidly. We now turn
to the impact of a slowly varying tidal field. This is the situation
one encounters for subhaloes on close to circular orbits. In this
limit one expects the material outside of the tidal radius to be
stripped off.  Modeling this tidal stripping, however, is far from
trivial. First of all, as discussed in \S\ref{sec:tidalradius}, the
tidal radius is a poorly defined concept, with significantly different
definitions in use. Secondly, it is unclear a priori at what {\it
  rate} the material that is located outside of the (instantaneous)
tidal radius is going to be stripped off.

Effectively, all semi-analytical models for the tidal evolution of
subhaloes adopt a tidal stripping rate given by
\begin{equation}\label{striprate}
{\rmd m \over \rmd t} = {m(>\rtid) \over \tau_{\rm orb}/\alpha}\,.
\end{equation}
Here $\rtid$ is the {\it instantaneous} tidal radius, $\tau_{\rm
  orb}=2 \pi / \omega$ with $\omega$ the {\it instantaneous} angular
velocity of the subhalo, and $\alpha$ is a `free parameter' that
differs substantially from author to author.  Whereas
\cite{Taylor.Babul.01, Taylor.Babul.04}, \cite{Taffoni.etal.03} and
\cite{Zentner.Bullock.03} all adopt $\alpha = 1$,
\cite{Zentner.etal.05} and \cite{Pullen.etal.14} tune $\alpha$ by
matching their predicted subhalo mass function to simulation
results. This yields $\alpha = 3.5$ and $2.5$,
respectively\footnote{As indicated in footnote 13 of
  \cite{Diemand.etal.07}, there is a factor $2 \pi$ missing in the
  formulation of \cite{Zentner.etal.05}.}. For comparison, Diemand
\etal (2007) find that the mass evolution of subhaloes in their `Via
Lactea' simulation is best modeled using Eq.~(\ref{striprate}) with
$\alpha \sim 6$. Finally, some authors have assumed that the stripping
is instantaneous \citep[e.g.,][]{Oguri.Lee.04, Penarrubia.Benson.05},
which effectively implies $\alpha = \infty$.  Since the tidal radius
is typically minimal at the orbit's peri-centre, this implies that 
per radial orbital period, all the mass outside of the peri-tidal 
radius, $\rpt$, will be stripped. This is the same assumption that 
underlies the orbit-averaged models of \cite{vdBosch.etal.05} and
\cite{Jiang.vdBosch.16, Jiang.vdBosch.17}, which therefore effectively
have adopted Eq.~(\ref{striprate}) with $\alpha = \infty$.
\begin{figure*}
\includegraphics[width=\hdsize]{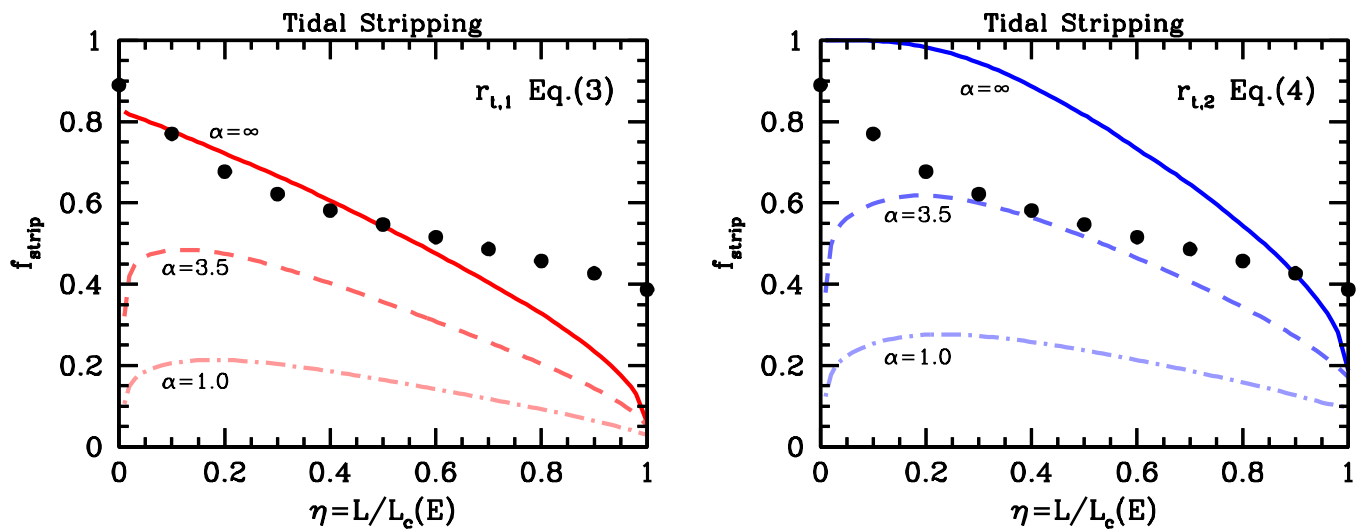}
\centering
\caption{Same as Fig.~\ref{fig:comp_heating}, except that this time
  the curves reflect the predicted mass fractions that are stripped
  off subhaloes during their first radial orbit due to (non-impulsive)
  tidal stripping. The two panels differ in the adopted definition for
  the tidal radius, as indicated, while solid, dashed and dot-dashed
  curves correspond to assumed stripping rates with $\alpha=\infty$,
  $3.5$, and $1.0$, respectively.  As in Fig.~\ref{fig:comp_heating},
  the solid dots indicate the results from the idealized numerical
  simulations (see \S\ref{sec:sims} and Table~1). See text for a
  detailed discussion.}
\label{fig:comp_stripping}
\end{figure*}

In this section we investigate what all these different assumptions
imply for the amount of mass that is stripped from a typical dark
matter subhalo during its first orbit after infall.  Practically, the
way we compute this is as follows. Using orbit-integration, we
discretize the subhalo's orbit in $2000$ equal time steps $\Delta
t$. Let $M_n$, $r_{{\rm t},n}$, and $\omega_n$ correspond to the
instantaneous bound subhalo mass, the instantaneous tidal radius, and
the instantaneous angular velocity of the subhalo at time step $n$. If
we ignore for the moment that the subhalo will respond to tidal
stripping by re-virialization, and simply assume that the mass
distribution of the bound material remains identical to the initial
mass distribution (prior to any stripping), then
\begin{equation}\label{alphadef}
M_{n+1} = M_n - \alpha \, \dot{M}_n \, \Delta t
\end{equation}
where
\begin{eqnarray}
  \dot{M}_n =  \left\{ \begin{array}{ll}
{ \left[ M_{n} - M(<r_{{\rm t},n}) \right] \over (2\pi/\omega_n)} & \mbox{if $M_{n} > M(<r_{{\rm t},n})$} \\
                                      0 & \mbox{otherwise}
    \end{array} \right.
\end{eqnarray}

Using this method, and the properties and orbits of the 5,000
subhaloes at accretion from the Bolshoi simulation depicted in
Fig.~\ref{fig:orbit}, we obtain the stripped mass fractions shown in
Fig.~\ref{fig:tidstrip}. The three panels differ in the value of the
parameter $\alpha$, as indicated, while the red and blue histograms
correspond to the results obtained using $\rtidA$
(Eq.~[\ref{rtTormen}]) and $\rtidB$ (Eq.~[\ref{rt}]), respectively.
Note that $\alpha=\infty$ (right-hand panel) corresponds to
instantaneous stripping, in which case the stripped mass fraction is
simply given by $f_{\rm strip} = M_\rms(>\rpt)/M_\rms$.

In general, since $\rtidB < \rtidA$, accounting for the centrifugal
force in the definition of the tidal radius results in predictions for
the stripped mass fraction that are significantly larger. Also,
$f_{\rm strip}$ depends strongly on the value of $\alpha$: for
$\alpha=1$, the median $f_{\rm strip}$ for subhaloes during their
first radial orbit is only $0.10$ ($0.16$) for $\rtid=\rtidA$
($\rtidB$).  For $\alpha=\infty$ this increases to $0.38$
($0.63$). Note that the assumption of instantaneous stripping combined
with $\rtid=\rtidB$ results in an appreciable fraction of subhaloes
with extremely large stripped mass fractions; among our sample of
5,000 newly accreted subhaloes 3.4 (1.2) percent have $f_{\rm strip} >
0.99$ ($> 0.999$). Clearly, depending on how one defines the tidal
radius, and on what one assumes regarding the rate at which material
beyond the tidal radius is stripped, the predicted impact of tidal
stripping varies dramatically.

This is also evident from Fig.~\ref{fig:comp_stripping}, which
compares our semi-analytical predictions for $f_{\rm strip}$ (for
the first radial orbital period), to the results from our idealized
numerical simulations. The two panels differ in the definition of
the tidal radius, as indicated, while different curves in each panel
correspond to different values of $\alpha$.

Clearly, no single combination of tidal radius definition and value
for $\alpha$ results in predictions for $f_{\rm strip}$ that are in
good agreement with the simulation results over the entire range in
$\eta$. It is not surprising that tidal stripping is unable to capture
the tidal evolution of subhaloes on highly radial (i.e., low $\eta$)
orbits, as their evolution is clearly governed by impulsive, tidal
heating (cf., Fig.~\ref{fig:comp_heating}). However, it may seem
surprising that the tidal stripping model also fails to describe the
results along close-to-circular orbits. After all, this is the regime
where the tidal field varies slowly, and thus where the tidal
stripping model is expected to be most accurate. However, there are
several problems with the tidal stripping model that are ultimately
responsible for its failure.

First of all, the notion that only the matter that is {\it initially}
located outside of $\rtid$ will be stripped off is clearly
incorrect. After all, some of the particles inside of the tidal radius
will be on orbits whose apo-centre $r_{\rm apo} > \rtid$; Hence, if
the tidal radius doesn't evolve much with time (i.e., if the orbit is
close to circular), one expects those particles to be stripped as
well. And since the dynamical time of particles in the subhalo is
comparable to the orbital time of the subhalo within its host, the
time-scale over which this stripping occurs should be comparable to
the radial orbital period $T_\rmr$. Using the same Monte-Carlo
realization used to compute the stripped mass fraction due to tidal
heating (see \S\ref{sec:applB}), we compute the apo-centre for each of
the 50,000 subhalo particles.  For $\eta=1$ (i.e., circular orbit) we
find that 41 percent of all subhalo particles have an apo-centre
$r_\rma > \rtidB$. This is in excellent agreement with the stripped
mass fraction in the simulation, which has $f_{\rm strip} = 0.394$
(after one radial period) for $\eta=1$ (cf. Table~1), and suggests
that along orbits that are close to circular, one may estimate $f_{\rm
  strip}$ as the fraction of particles with $r_\rma > \rtidB$. Note,
though, that this approximation rapidly deteriorates for decreasing
$\eta$\footnote{For $\eta=0.9$ it already overpredicts $f_{\rm strip}$
  by more than 30 percent}. Furthermore, there is an additional
problem with the stripping model that is even more difficult to
overcome.  After the outer layers of a subhalo are (instantaneously)
removed, the remaining remnant is no longer in virial equilibrium.  As
we show in App.~\ref{App:NFWener}, the remnant has a virial ratio
$K/|W| > 0.5$.  Hence, the only way the system can re-virialize, is by
converting kinetic to potential energy. In doing so, the system `puffs
up'; i.e., its characteristic radius increases while its overall
density decreases.  This in turn results in a decrease of the tidal
radius, and hence in additional mass loss
\citep[][]{Kampakoglou.Benson.07}. As a result, the mass loss is a
continuous process, even in the static field experienced along a
circular orbit. This is evident from the left-hand panel of Fig. 2
(red curve), which shows the evolution of the bound fraction for the
duration of one radial period, $T_\rmr = 9.53 \Gyr$. In Paper~II we
show that the mass loss continuous well beyond that, and that
subhaloes on circular orbits continue to loose mass even after 60 Gyrs
of evolution.


\section{Can Instantaneous Stripping result in Disruption?}
\label{sec:trunc}

In an influential paper, \cite{Hayashi.etal.03} pointed out that
instantaneously removing all material from a halo down to some truncation 
radius, $\rtrunc$, may leave a remnant with {\it positive} binding energy, 
as long as $\rtrunc < r_{\rm crit}$.  For a spherical, isotropic NFW
halo, $r_{\rm crit}$ is 0.77 times the NFW scale radius. This is
demonstrated in App.~\ref{App:NFWener}, were it is also shown how the
critical radius, $r_{\rm crit}$, depends on the halo's
anisotropy. Typically, $r_{\rm crit}$ becomes larger for haloes that
are more radially anisotropic, and in general $r_{\rm crit}>0$, unless
haloes are strongly tangentially anisotropic.

This seems to suggest, as was advocated by Hayashi et al., that a
subhalo whose tidal radius is smaller than $r_{\rm crit}$ would
`spontaneously' disrupt, even without the need to invoke tidal shock
heating.  And since a typical NFW halo has roughly between 5 and 10
percent of its mass enclosed within this critical radius, this
suggests that a typical subhalo should physically disrupt once it has
lost between 90 and 95 percent of its original (i.e., at accretion)
mass. Interestingly, this is not inconsistent with the results from
numerical simulations, which indeed reveal very few subhaloes whose
mass is less than 5 percent of that at accretion
\citep[e.g.,][]{Jiang.vdBosch.16, Han.etal.16, vdBosch.17}, and
various authors have implemented a treatment of subhalo disruption
based on this notion \citep[e.g.,][]{Zentner.Bullock.03,
  Taylor.Babul.04, Klypin.etal.15}.

However, as we now demonstrate, this notion is seriously flawed, and,
if simulated with sufficient resolution, the instantaneous removal of
outer layers always leaves a bound remnant, even when these outer
layers contain more than 99.9 percent of the total virial mass. The
reason is once again that the {\it total} binding energy is not
sufficiently informative. What matters is the distribution of binding
energies of the constituent particles, not its sum.  Note that we
already encountered an example of subhaloes with positive binding
energy, namely subhaloes immediately following an impulsive shock with
$\Delta E/|E_\rmb| > 1$. As we saw in \S\ref{sec:TidalHeat} neither of
these disrupted, even when $\Delta E > 100 |E_\rmb|$. In order to test
how subhaloes respond to instantaneous mass stripping down to radii
below $r_{\rm crit}$, we now use numerical $N$-body simulations to
evolve instantaneously stripped, isotropic NFW haloes and track their
bound mass fraction over time.

\subsection{Numerical Simulations}
\label{sec:numsim}

We simulate the evolution of isolated, spherical NFW host haloes that
are instantaneously truncated.  We use the same method as in
\S\ref{sec:sims} to draw the initial positions and velocities of the
particles: we assume that, prior to truncation, the halo has an
isotropic velocity distribution, and we truncate the halo at a radius
$\rtrunc$, which is a free parameter, while initializing the
velocities as if the halo extends out to infinity (this is the
expected result if the stripping is instantaneous). All our
simulations have $N=10^5$ particles {\it inside the truncation
  radius}, and are carried out using a modified version of the
hierarchical N-body code {\tt treecode}, written by Joshua Barnes with
some modifications due to John Dubinski. {\tt treecode} uses a
\cite{Barnes.Hut.86} octree to compute accelerations based on a
multipole expansion up to quadrupole order, and uses a straightforward
second order leap-frog integration scheme to solve the equations of
motion. Since we use fixed time steps, our integration scheme is fully
symplectic. Forces between particles are softened using a simple
Plummer softening.

As in \S\ref{sec:sims}, we adopt model units in which the
gravitational constant, $G$, the scale radius, $r_0$, and the virial
mass of the halo, $\Mvir$, are all unity. Using numerical simulations
of NFW haloes in isolation (see Paper~II), we infer an optimal
softening length of $\varepsilon = 0.05$ (in model units) for
simulations with $N=10^5$ particles inside the virial radius. In order
to gauge the sensitivity to the specific choice of the softening
length, we run two different sets of simulations: in set {\tt TruncA}
we adopt a softening length of $\varepsilon = 0.05$, independent of
$\rtrunc$, while in set {\tt TruncB}, we scale the softening length
linearly with the size of the simulated system according to
$\varepsilon=0.05\,(\rtrunc/\rvir)$; such a scaling is consistent with
simple expectations for the optimal softening length
\citep[e.g.,][]{Power.etal.03, vKampen.00a}.  We run each truncated
halo in isolation for $40,000$ time steps of $\Delta t = 0.02$,
corresponding to a total integration time of $\sim 50 \Gyr$.  Finally,
we adopt a tree opening angle of $\theta=0.7$, and we have verified
that our results are extremely stable to changes in $\Delta t$ and/or
$\theta$. The parameters of the simulations in {\tt TruncA} and {\tt
  TruncB} are listed in Table~2.
\begin{table}\label{tab:StaticSims}
\caption{Parameters of Simulations of Instantaneous Truncation}
\begin{center}
\begin{tabular}{cccccc}
\hline\hline
ID & $\rtrunc/r_0$ & $f(\rtrunc)$ & $\varepsilon/r_0$ & $f_{\rm b,rem}$ & $f_{\rm b,rem}$ \\
 & & & & $[t=0]$ & $[50\,{\rm Gyr}]$ \\
(1) & (2) & (3) & (4) & (5) & (6) \\
\hline
A01 & 7.11 & 0.817 & 0.05 & 0.98 & 0.98 \\
A02 & 6.46 & 0.768 & 0.05 & 0.98 & 0.98 \\
A03 & 5.81 & 0.715 & 0.05 & 0.98 & 0.97 \\
A04 & 5.16 & 0.659 & 0.05 & 0.98 & 0.97 \\
A05 & 4.51 & 0.596 & 0.05 & 0.97 & 0.97 \\
A06 & 3.85 & 0.528 & 0.05 & 0.97 & 0.96 \\
A07 & 3.19 & 0.451 & 0.05 & 0.96 & 0.95 \\
A08 & 2.52 & 0.365 & 0.05 & 0.94 & 0.93 \\
A09 & 1.83 & 0.265 & 0.05 & 0.91 & 0.90 \\
A10 & 1.48 & 0.209 & 0.05 & 0.89 & 0.88 \\
A11 & 1.10 & 0.147 & 0.05 & 0.83 & 0.82 \\
A12 & 0.68 & 0.076 & 0.05 & 0.71 & 0.71 \\
A13 & 0.48 & 0.046 & 0.05 & 0.61 & 0.62 \\
A14 & 0.35 & 0.027 & 0.05 & 0.46 & 0.49 \\
A15 & 0.20 & 0.011 & 0.05 & 0.23 & 0.31 \\
A16 & 0.10 & 0.003 & 0.05 & 0.15 & 0.18 \\
A17 & 0.08 & 0.001 & 0.05 & 0.00 & disrupt \\
\hline
B01 & 2.52 & 0.365 & 0.0125 & 0.94 & 0.93 \\
B02 & 1.83 & 0.265 & 0.0092 & 0.91 & 0.90 \\
B03 & 1.48 & 0.209 & 0.0074 & 0.89 & 0.88 \\
B04 & 1.10 & 0.147 & 0.0051 & 0.83 & 0.82 \\
B05 & 0.68 & 0.076 & 0.0034 & 0.73 & 0.72 \\
B06 & 0.35 & 0.027 & 0.0017 & 0.55 & 0.57 \\
B07 & 0.20 & 0.011 & 0.0010 & 0.43 & 0.46 \\
B08 & 0.10 & 0.003 & 0.0005 & 0.34 & 0.37 \\
B09 & 0.08 & 0.001 & 0.0004 & 0.31 & 0.35 \\
\hline\hline
\end{tabular}
\end{center}
\medskip
\begin{minipage}{\hssize}
  Simulations of instantaneously truncated NFW haloes. Listed are the
  simulation ID (column 1), the ratio of the truncation radius,
  $\rtrunc$, to the scale radius $r_0$ (column 2), the fraction
  $f(\rtrunc) \equiv M(\rtrunc)/M_{\rm vir}$ of the virial mass
  located inside the truncation radius (column 3), the softening
  length in units of the scale radius (column 4), and the bound
  fractions of the remnants, $f_{\rm b,rem}$, at $t=0$ (column 5) and
  $t = 50\Gyr$ (column 6).  Simulations A01-A17 belong to {\tt
    TruncA}, and all adopt a softening length of $\varepsilon = 0.05$,
  while the {\tt TruncB} simulations B01-B09 adopt a softening length
  that scales linearly with the truncation radius according to
  $\varepsilon = 0.05 (\rtrunc/r_{\rm vir})$.
\end{minipage}
\end{table}

\subsection{Results}
\label{sec:resB}

We use the method described in App.~\ref{App:fbound} to investigate
how the bound-mass of the truncated halo evolves over time. We
distinguish two bound fractions:
\begin{equation}\label{fbrem}
f_{\rm b,rem}(t) \equiv {M(t) \over M(\rtrunc)} = {N_{\rm bound} \over
  N_{\rm tot}}\,,
\end{equation}
and
\begin{equation}\label{fbtot}
f_{\rm b,tot}(t) \equiv {M(t) \over \Mvir} =
  f_{\rm b,rem}(t) \times {M(\rtrunc) \over \Mvir},.
\end{equation}
Here $M(t)$ is the bound mass of the halo at time $t$, $M(r)$ is
the {\it initial} (prior to truncation) mass of the subhalo inside
radius $r$, $\Mvir = M(\rvir)$ is the subhalo's initial virial mass,
$N_{\rm bound}$ is the number of bound particles at time $t$, and
$N_{\rm tot}=10^5$ is the total number of particles used in the
simulation.  Thus, $f_{\rm b,rem}$ is the fraction of the mass
initially located inside the truncation radius that remains bound,
while $f_{\rm b, tot}$ is the fraction of mass that remains bound
compared to the original virial mass of the halo in question.
\begin{figure}
\includegraphics[width=\hssize]{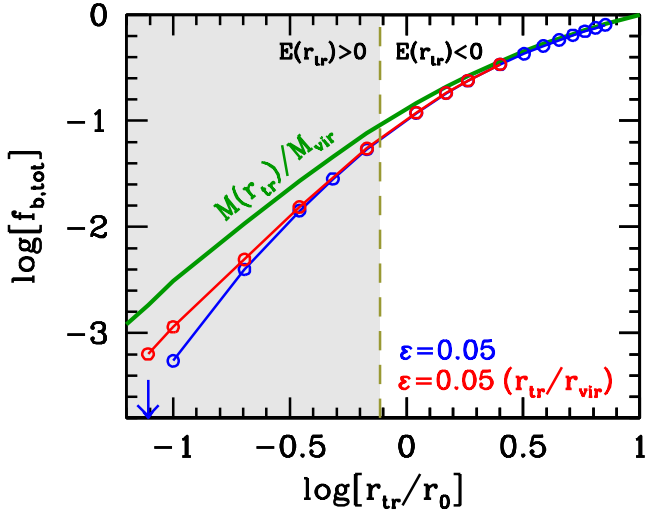}
\centering
\caption{The bound fraction, $f_{\rm b,tot}$, for NFW haloes with
  concentration parameter $c=10$ as a function of their truncation
  radius, $\rtrunc$, expressed in units of the scale radius $r_0$. The
  green, solid line indicates $M(\rtrunc)/M_{\rm vir}$, and reflects
  the initial mass fraction enclosed by $\rtrunc$. The open circles
  are the mass fractions that are found to be bound in numerical
  simulations with $N=10^5$ particles after 50 Gyr of evolution. Blue
  and red circles correspond to simulations from {\tt TruncA} and {\tt
    TruncB}, respectively, that only differ in their choice of
  softening length, as indicated. The dashed, vertical line
  corresponds to $\rtrunc = 0.77 r_0$ where the binding energy of the
  remnant transits from being negative to being positive. The blue,
  downward arrow indicates that the {\tt TruncA} simulation with
  $\rtrunc = 0.078 r_0$ disrupts (i.e., has $\fbound \rightarrow
  0$).}
\label{fig:static}
\end{figure}

\begin{figure}
\includegraphics[width=\hssize]{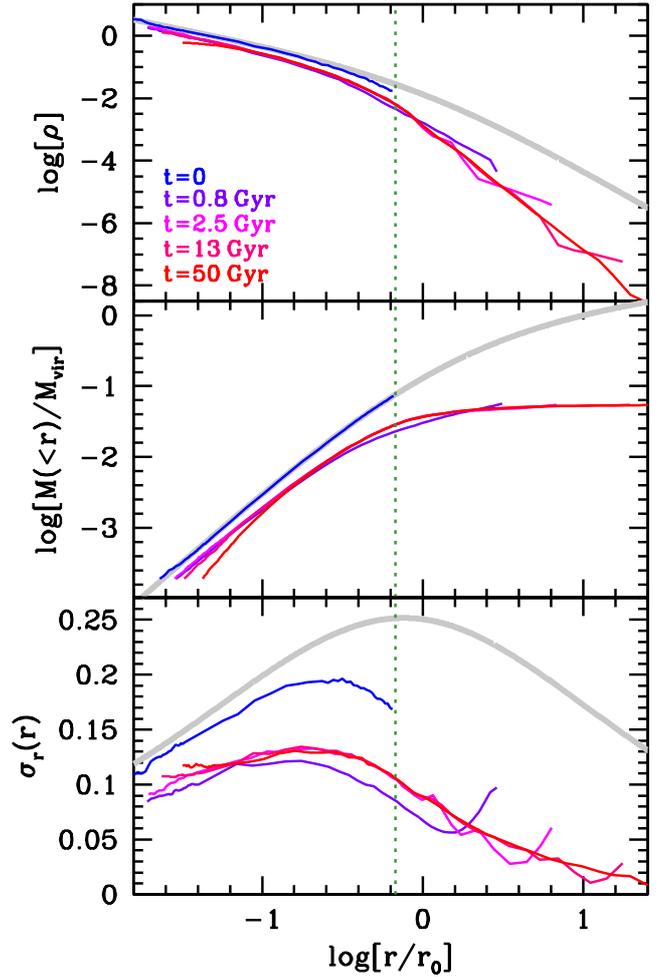}
\centering
\caption{The density profiles (upper panel), enclosed mass profiles
  (middle panel), and radial velocity dispersion profiles (lower
  panel) at 5 different evolutionary stages for an NFW halo that is
  instantaneously truncated (at $t=0$) at $\rtrunc = 0.68 r_0$
  (simulation B05 in Table~2). Note that the profiles are for the
  bound particles only.}
\label{fig:statevol}
\end{figure}

Fig.~\ref{fig:static} summarizes the simulation results (see also
Table~2). The open circles indicate $f_{\rm b,tot}$ at the end of each
simulation (i.e., after $50\Gyr$ of evolution).  Blue and red circles
correspond to simulations from {\tt TruncA} and {\tt TruncB},
respectively, and the results are shown as a function of the
truncation radius, $\rtrunc$, in units of the halo's (original) scale
radius, $r_0$. The green, solid line indicates $M(\rtrunc)/\Mvir$, and
reflects the initial mass fraction enclosed by $\rtrunc$. The dashed,
vertical line marks $\rtrunc = 0.77 r_0$, which is the critical value
for the truncation radius, to the left of which the remnants
(immediately after the instantaneous truncation) have positive binding
energy. Clearly, there is no indication of anything special happening
around this scale. In fact, in all simulations we find that part of
the remnant remains bound, and quickly resettles into a new,
virialized halo. As an example, Fig.~\ref{fig:statevol} shows the
evolution in the density profile (upper panel), enclosed mass profile
(middle panel) and radial velocity dispersion profile (lower panel)
for a NFW halo that is instantaneously truncated at $\rtrunc = 0.68
r_0$. The inner region quickly (within $\sim 2\Gyr$) re-virializes,
while it takes more than a Hubble time for the weakly bound particles
to build a virialized structure that extends to $>10 r_0$ (i.e.,
beyond the original virial radius of the subhalo).

In general, a smaller truncation radius implies a larger fraction of
particles inside this truncation radius that are unbound (and escape),
but in each case a bound subset remains. The only exception is
simulation A17 from {\tt TruncA}, which has $\rtrunc = 0.078 r_0$ and
which experiences complete disruption immediately after the onset of
the simulation (indicated by a blue, downward pointing arrow in
Fig.~\ref{fig:static}).  We point out, though, that in this case
$\rtrunc$ is only $\sim 1.5$ times the softening length; clearly, the
potential is `over-softened' and has little to no resemblance to the
central region of a NFW profile. The same simulation in {\tt TruncB}
(simulation B09), which uses a much smaller softening length (see
Table~2) results in $\sim 35$ percent of the initial particles of the
remnant surviving as a self-bound entity.

In all simulations we find that the bound fraction does not evolve
significantly with time: the fraction of particles that remain bound
together is basically the same after more than 50 Gyr of evolution as
it is at $t=0$ (cf. columns (5) and (6) of Table~2). Only when $\rtrunc
\lta r_0/2$ do we find that the bound fraction evolves somewhat
(typically $f_{\rm b,rem}$ {\it increases} by a few percent), during
the first $\sim 0.2 \Gyr$ after the onset of the simulation. This is a
manifestation of (violent) relaxation in response to the instantaneous
removal of the halo's outer layers; as shown in App.~\ref{App:NFWener}
the remnant can be far from virial equilibrium.

To summarize, unless dark matter haloes are strongly tangentially
anisotropic, instantaneously stripping matter from the outskirts may
leave a remnant with total, positive binding energy. However, contrary
to naive expectations \citep[e.g.,][]{Hayashi.etal.03,
  Taylor.Babul.04, Klypin.etal.15}, a system with positive binding
energy does not necessarily disrupt entirely. After all, the system
will typically have a broad distribution of binding energies, and a
significant fraction of the particles can still be bound, even if the
total binding energy is positive. In the $N$-body experiments
described above, we only find that a truncated, isotropic NFW halo
disrupts if the truncation radius is less than, or comparable to, the
softening length used. Although we have only tested this for isotropic
NFW haloes, we emphasize that (the central parts) of NFW haloes in
cosmological simulations are indeed close to isotropic
\citep[e.g.,][]{Navarro.etal.10}. We therefore conclude that
instantaneous truncation of a NFW halo does not lead to halo
disruption. Artificial disruption in simulations may occur, however,
if the system is simulated with insufficient force resolution (see
\S\ref{sec:softening} and Paper~II).


\section{Numerical Disruption Processes}
\label{sec:numerics}

As discussed in \S\ref{sec:intro}, it wasn't until 1997 that numerical
$N$-body simulations started to resolve surviving populations of dark
matter substructure. Prior to this, the simulations suffered from a
serious overmerging problem, mainly due to inadequate force softening
\citep[][]{Moore.etal.96a}. However, numerical overmerging (i.e., the
artificial disruption of substructure in numerical simulations) is
still present in modern state-of-the-art simulations. This is the
reason why methods to populate dark-matter-only simulations with
galaxies, such as semi-analytical models for galaxy formation
\citep[e.g.,][]{Springel.etal.01, Kang.etal.05, Kitzbichler.White.08}
subhalo abundance matching models \citep[e.g.,][]{Conroy.etal.06,
  Guo.White.13, Campbell.etal.17} and empirical models
\citep[e.g.,][]{Moster.etal.10, Moster.etal.17, Tollet.etal.17} often
include `orphan' galaxies, i.e., mock galaxies without an associated
subhalo in the simulation. Although it is therefore recognized that 
simulations continue to experience artificial disruption, it is 
generally assumed that this happens only for subhaloes below a mass 
resolution of 50 - 100 particles.  This notion is based on the fact 
that subhalo mass functions are typically converged down to this mass 
resolution limit \citep[e.g.,][]{Springel.etal.08, Onions.etal.12, 
  Knebe.etal.13, vdBosch.Jiang.16}. The general consensus therefore 
is that any disruption of subhaloes above this `resolution limit' 
must be physical in origin \citep[see][for a detailed
  discussion]{Diemand.Moore.Stadel.04}.

However, as we demonstrate in Paper II, state-of-the-art numerical
simulations of CDM structure formation still suffer from abundant {\it
  artificial} disruption well above this mass resolution limit, and 
it is therefore imperative that we revisit the issue of numerical 
overmerging. There are a number of
processes that may (potentially) give rise to numerical, artificial
disruption of dark matter substructure in numerical $N$-body
simulations. These are (i) evaporation resulting from two-body
relaxation of the subhalo, (ii) evaporation due to two-body encounters
with particles from the host halo, (iii) tidal heating due to
impulsive encounters with particles from the host halo, and (iv)
disruption due to issues with force softening.

Each of these processes have been discussed in more or less detail in
various previous studies, including \cite{Carlberg.94},
\cite{vKampen.95, vKampen.00a, vKampen.00b}, \cite{Moore.etal.96a}, and
\cite{Klypin.etal.99a}. For the sake of completeness, and in order to
correct and elucidate inconsistencies and errors in some of these
previous studies, we discuss each of these processes in detail in what
follows.

\subsection{Evaporation due to Two-Body Self-Relaxation}
\label{sec:evaporation}

In a gravitational $N$-body system, encounters (also called
`collisions') among the particles drive the system towards
equipartition of kinetic energy. The time-scale for this two-body
relaxation process is roughly
\begin{equation}\label{trelax}
t_{\rm relax} = {N \over 8 \, \ln\Lambda} \, t_{\rm cross}
\end{equation}
\citep[e.g.,][]{Binney.Tremaine.08}. Here $t_{\rm cross} \simeq
r/\sigma$ is the crossing time of the system of size $r$ and velocity
dispersion $\sigma$, and $\ln\Lambda = \ln(b_{\rm max}/b_{\rm min})$
is the Coulomb logarithm, with $b_{\rm max} \sim r$ and $b_{\rm min} =
2 G m_\rmp/\sigma^2$ the maximum and minimum impact parameters, and
$m_\rmp$ the particle mass. Using the virial theorem, one then finds
that $\ln\Lambda = \ln(N/2) \sim \ln N$.

Two-body relaxation drives the local velocity distribution of the
particles towards a Maxwellian, which has a tail that exceeds the
local escape speed. Hence, from time to time the gravitational
encounters can give enough energy to a particle so that it can
escape. This causes the $N$-body system to evaporate, which ultimately
leaves a final system with two particles on a Keplerian orbit. The
evaporation time scale is roughly 140 times the two-body relaxation
time \citep{Binney.Tremaine.08}, which for a system of point particles
implies
\begin{equation}\label{tevap}
t_{\rm evap} \simeq {15 \, N \over \ln N} \, t_{\rm cross}
\end{equation}
In $N$-body simulations of dark matter substructure, the number of
particles used is vastly smaller (typically by more than 50 orders of
magnitude) than the actual number of constituents of the physical
system being simulated. Hence, the relaxation and evaporation times in
the simulation are much, much smaller than they are in reality, which
in principle could give rise to artificial, excessive evaporation.

When considering the impact of two-body relaxation in numerical
$N$-body simulations, one has to modify the above estimate to take
account of force softening, which suppresses encounters with large
deflection angles. As shown in \cite{Farouki.Salpeter.82}, the net
effect of force softening is to boost the minimum impact parameter,
$b_{\rm min}$, such that the Coulomb logarithm becomes
\begin{equation}\label{Coulomb}
\Lambda = {\rm min} \left\{N,r/(4\varepsilon)\right\}\,,
\end{equation}
with $\varepsilon$ the softening length. Hence, softening increases
the two-body relaxation time (and thus also the evaporation time), but
by how much depends somewhat on the exact value of $\varepsilon$.
This is demonstrated in Fig.~\ref{fig:timescales}, where the green
lines plot $\log[t_{\rm evap}/t_{\rm cross}]$ as function of $\log[N]$
for an NFW subhalo with concentration parameter $c=10$.  The grey
shaded region marks time scales that are less than 30 crossing times,
which is a conservative upper limit on the age of the Universe. Hence,
time scales above this grey region are not expected to be of much
relevance. In the left-hand panel we adopt $\varepsilon/r_0 = 1.7
\times 10^{-2} \, (N/10^5)^{-0.23}$, where $r_0 = r_\rms/c$ is the
scale radius of the subhalo. As shown in \cite{Dehnen.01}, this
scaling for the (Plummer) softening length optimizes between
systematic force errors (which dominate when $\varepsilon$ is large)
and force noise (which dominates when $\varepsilon$ is
small)\footnote{This relation, which is taken from Table 2 in
  \cite{Dehnen.01} corresponds to a \cite{Hernquist.90} profile, which
  is similar, though, to an NFW profile.}. In the right-hand panel, we
adopt $\varepsilon/r_0 = 0.126 \, (N/10^5)^{-0.5}$, which corresponds
to the optimal softening length advocated by \cite{Power.etal.03},
based on a lower limit required to prevent discreteness effects. For
$N \lta 2000$, the Power \etal softening results in a much longer
evaporation time; the dramatic upturn in $\log[t_{\rm evap}/t_{\rm
    cross}]$ at small $N$ occurs when the softening length becomes
comparable to the size of the subhalo itself; obviously, in such an
extreme case, two-body relaxation is completely irrelevant, but at the
same time, such a simulation clearly is not able to capture the
relevant dynamics.  The softening advocated by \cite{Dehnen.01},
always has $\varepsilon/r_0 \ll 1$, which results in much smaller
evaporation times for small $N$. However, even for $N = 10$, the
evaporation time exceeds the Hubble time, and it therefore seems safe
to conclude that artificial evaporation of substructure due to
two-body relaxation should be of little concern \citep[see
  also][]{Moore.etal.96a}.

There is one important caveat, though: the above estimate for $t_{\rm
  relax}$ is based on the classical two-body treatment of
\cite{Chandrasekhar.43}, in which all collisions are assumed to be
independent. This ignores resonant effects and self-gravity
(`collective relaxation'), both of which can significantly boost
relaxation with respect to the two-body rate \citep{Weinberg.93}.  In
particular, as Weinberg demonstrates, the dominant contribution to
relaxation of a gravitational system comes from large scale modes,
with wavelengths comparable to the size of the system. As a
consequence, the treatment of force-softening has little to no effect
on the actual relaxation rate. This explains, among others, why
relaxation rates in expansion-type codes, such as the self-consistent
field code, reveals relaxation at a level that is similar to that of
tree-based codes \citep[][]{Hernquist.Ostriker.92}, and why the amount
of spurious (artificial) fragmentation apparent in warm dark matter
simulations is virtually independent of the softening length
\citep[][]{Power.etal.16}.  In Paper~II, we demonstrate that these
same large-scale fluctuations, in the presence of a tidal field, cause
a run-away instability in the mass evolution of dark matter subhaloes
in $N$-body simulations.

To summarize, we caution that although the two-body relaxation rate
(Eq.~[\ref{trelax}]) may exceed the Hubble time, the actual relaxation
time in numerical simulations may be significantly shorter, and the
overall impact of relaxation in $N$-body simulations remains a
contentious topic \citep[see e.g.,][and references therein]{Melott.07,
  Hahn.Angulo.16, Power.etal.16}.
  
\subsection{Evaporation due to Two-Body Collisions with Host Particles}
\label{sec:PStwobody}

In an attempt to explain the origin of numerical
`overmerging', \cite{Carlberg.94} suggested that subhaloes in
numerical simulations may artificially disrupt due to two-body
encounters between particles in the subhalo, and those in the host
halo. He argued that, since the host halo particles are so much hotter
(i.e., have higher velocities) than the subhalo particles, the former
will transfer kinetic energy to the latter, thereby `boiling' the
subhalo into dissolution. This idea was further promoted by
\cite{vKampen.95, vKampen.00a, vKampen.00b} who suggested that it is the
dominant numerical process to explain the overmerging of substructure
in cosmological $N$-body simulations. Unfortunately,
\cite{Carlberg.94} does not give a derivation of his expression for
the heating time scale, which in addition contains a typo, in that the
indices `c' and `h' in his Eq.~(11) are swapped. \cite{vKampen.95}
does present a correct derivation for the relaxation time, but then
incorrectly assumes that this is comparable to the {\it disruption}
time, thereby underestimating the latter by two orders of
magnitude. In order to set the record straight, we give a detailed
derivation in what follows.

Consider a dynamically {\it cold} system of size $r_\rmc$, consisting
of $N_\rmc$ particles of mass $m$ and with a velocity dispersion
$\sigma^2_\rmc \simeq G N_\rmc m / r_\rmc$, moving through a {\it hot}
system of size $r_\rmh$ consisting of $N_\rmh \gg N_\rmc$ particles of
identical mass $m$. The velocity dispersion of the hot system is
$\sigma^2_\rmh \simeq G N_\rmh m / r_\rmh$.  Each encounter between a
hot particle and a cold particle changes the velocity of the cold
particle such that
\begin{equation}\label{dv2enc}
(\Delta v^2)_{\rm enc} = {4 G^2 m^2 \over b^2 V^2}\,,
\end{equation}
\citep{Binney.Tremaine.08}. Here $b$ is the impact parameter of the
encounter, and $V$ is the relative speed between the particles. Since
the hot particles are moving much faster than the cold ones, we have
that $\langle V^2 \rangle = \sigma^2_\rmh + \sigma^2_\rmc \simeq
\sigma^2_\rmh$.  Per crossing through the {\it hot} system, each
particle in the subhalo has on average about $2 (N_\rmh/r^2_\rmh) \, b
\, \rmd b$ encounters with impact parameters in the range $b$, $b+\rmd
b$. Integrating over $b$, yields the total $\Delta v^2$ per crossing
time, which is
\begin{equation}
(\Delta v^2)_{\rm cross} = {8 \ln\Lambda_\rmh \over N_\rmh} \sigma^2_\rmh  
\end{equation}
Hence, the relaxation time for the cold system is
\begin{equation}\label{trelax1}
t^{\rm ch}_{\rm relax} \equiv {\sigma^2_\rmc \over (\Delta v^2)_{\rm
    cross}} t_{\rm cross,h} = {N_\rmh \over 8 \ln\Lambda_\rmh} \left({\sigma^2_\rmc
  \over \sigma^2_\rmh}\right) \, t_{\rm cross,h}\,,
\end{equation}
where the superscript `ch' indicates that this is two-body relaxation
of cold particles due to hot particles. Using that the densities of
collapsed haloes are independent of mass, we have that $t_{\rm
  cross,c} = t_{\rm cross,h}$, which allows us to rewrite
Eq.~(\ref{trelax1}) as
\begin{equation}\label{trelax2}
t^{\rm ch}_{\rm relax} = {\sigma_\rmh \over \sigma_\rmc} \, 
{\ln\Lambda_\rmc \over \ln\Lambda_\rmh} \, t^{\rm cc}_{\rm relax}
\end{equation}
where $t^{cc}_{\rm relax}$ is the two-body relaxation time of the cold
system in isolation, given by Eq.~(\ref{trelax}), and we have used
that $N_\rmh/N_\rms = (\sigma_\rmh/\sigma_\rmc)^3$, which follows from
the virial scaling relations for dark matter haloes, according to
which $\sigma \propto M^{1/3}$.

We thus see that the relaxation time due to encounters with the hot
particles is significantly {\it longer} than that due to encounters
with its own particles. This has its origin in the fact that
encounters with host halo particles have, on average, a larger impact
parameter and a higher relative velocity
(cf. Eq.~[\ref{dv2enc}]). This is partially compensated by the fact
that there are more host halo particles than subhalo particles, but
the combined effect is such that self-interactions are more important
than those between host and subhalo particles.

The relaxation time derived here is identical to that in
\cite{vKampen.95}. However, van Kampen then argued that the
evaporation time due to these particle-subhalo interactions is
identical to the relaxation time (i.e., no multiplication with a
factor 140 is required). Van Kampen incorrectly assumes that
relaxation heats the cold system to the `temperature' (velocity
dispersion) of the hot system. If this were true, then after a single
relaxation time a very large fraction of subhalo particles would
evaporate, such that the evaporation time scale is only slightly
longer than the relaxation time scale. But this is incorrect. After
all, the relaxation time is {\it defined} as the time required to heat
the velocity distribution of the cold system to a Maxwellian with a
velocity dispersion $\sigma_\rmc$, rather than $\sigma_\rmh$ (see
Eq.~[\ref{trelax1}]). Hence, one still requires of the order of 140
relaxation times for the cold system to evaporate, which implies that
the collisional heating due to interactions with host halo particles
is less important that that due to self-interactions.

\begin{figure*}
\includegraphics[width=\hdsize]{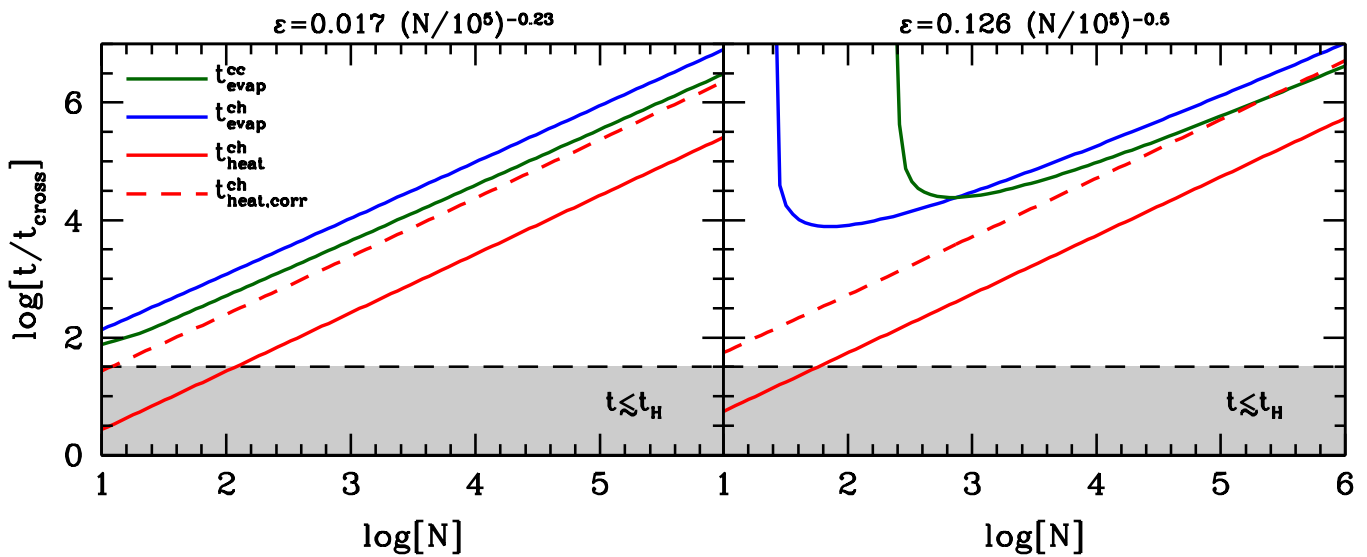}
\centering
\caption{Time scales, in units of the crossing time, for the
  disruption of subhaloes due to numerical effects as a function of
  the number of particles, $N$, used in simulating the subhalo (which
  is assumed to have a concentration parameter $c=10$). In particular,
  green, blue and red lines correspond to the times scales for
  evaporation due to two-body relaxation, evaporation due to
  collisional encounters with host halo particles, and disruption due
  to impulsive heating by host halo particles. For the latter, the red
  dashed line shows the proper, corrected time scale. See text for
  details. The grey-shaded region corresponds to $t/t_{\rm cross} \leq
  30$, which roughly corresponds to time scales shorter than a Hubble
  time. Left- and right-hand panels correspond to different
  assumptions regarding the softening length, as indicated.}
\label{fig:timescales}
\end{figure*}

This is illustrated in Fig.~\ref{fig:timescales}, where the blue lines
correspond to $t^{\rm ch}_{\rm evap}$ for a subhalo embedded in a host
halo with a mass 30 times larger than that of the subhalo.  Since
$t^{\rm ch}_{\rm evap}$ is roughly proportional to $(M_\rmh/M_\rmc)^{1/3}$,
and since most subhaloes have $M_\rmh/M_\rmc \gg 30$ this estimate for
$t^{\rm ch}_{\rm heat}$ may be considered a conservative lower limit.
Note that indeed $t^{\rm ch}_{\rm evap} > t^{\rm cc}_{\rm evap}$, except when
the softening length becomes comparable to the size of the subhalo
in question (which is not a meaningful situation to consider).

To summarize, contrary to claims by \cite{Carlberg.94} and
\cite{vKampen.95, vKampen.00a}, evaporation due to particle-subhalo
interactions is always {\it sub}-dominant to that due to
self-interactions among the subhalo particles. And since the latter is
always longer than the Hubble time, evaporation due to {\it
  collisional} encounters with host halo particles cannot be an
important numerical effect.

\subsection{Impulsive Heating due to encounters with host halo particles}
\label{sec:PStidal}

A subtly different numerical disruption mechanism is impulsive heating
of the subhalo due to encounters with the {\it artificially massive}
particles of the host halo. This is different from the two-body
particle-subhalo heating picture outlined above, which is a
collisional process. The heating mechanism considered here, and first
discussed in \cite{Moore.etal.96a}, considers the subhalo as a
collisionless system being tidally heated due to the high-speed,
impulsive encounters with the particles of the host halo.

Consider a particle from the hot host halo, which is modeled as a
Plummer sphere of mass $m$ and size $\varepsilon$ (the softening length), 
that has an impulsive encounter with a (cold) subhalo of mass 
$M_\rmc = N_\rmc \, m$ and size $r_\rmc$.  Let $b$ and ${\bf V} = V \, 
\hat{e}_z$ be the impact parameter and relative velocity (taken to be 
along the $z$-direction) of the encounter.

As long as the impact parameter $b > r_\rmc$ we can use the distant
tide approximation, according to which a subhalo particle increases
its kinetic energy following
\begin{equation}\label{dv2impulse}
\Delta v^2 = {4 G^2 m^2 \over V^2} \, {R^2 \over b^4}\,.
\end{equation}
Here $R^2 = x^2 + y^2$ is the cylindrical radius of the particle with
respect to the centre of the subhalo. For a penetrating encounter
with $b=0$ one finds that
\begin{equation}\label{dv2pene}
\Delta v^2 = {4 G^2 m^2 \over V^2} \, {R^2 \over (R^2 + \varepsilon^2)^2}
\end{equation}
\citep{Binney.Tremaine.87}. By smoothly interpolating between these
two cases, we obtain an expression for $\Delta v^2(R)$ that is valid
for any impact parameter:
\begin{equation}\label{dv2comb}
\Delta v^2(R) = {4 G^2 m^2 \over V^2} \, {R^2 \over b^4 + (R^2 + \varepsilon^2)^2}
\end{equation}
By far the largest increase of kinetic energy is due to encounters
with $b \sim 0$ and for particles with $R \sim \varepsilon$.  Using
the same approach as in \S\ref{sec:PStwobody}, we now integrate over
impact parameter, to obtain the $\Delta v^2(R)$ per crossing time
$t_{\rm cross,h}$:
\begin{eqnarray}\label{dv2crossheat}
(\Delta v^2)_{\rm cross}(R) & = & 32 \, {\sigma^2_\rmh \over N_\rmh} \, R^2 \, \int_0^{r_\rmh}
  {b \, \rmd b \over b^4 + (R^2 + \varepsilon^2)^2} \nonumber \\
 & \simeq & 8 \pi \, {\sigma^2_\rmh \over N_\rmh} \, 
{R^2 \over R^2 + \varepsilon^2}\,,
\end{eqnarray}
where we have used that $V^2 \simeq \sigma^2_\rmh$ and that $r_\rmh
\gg r_\rmc \geq R$. If we assume that the subhalo follows an NFW
density profile with scale parameter $r_0 = r_\rmc/c$, with $c$ the
subhalo's concentration parameter, then the total increase in energy,
per crossing time, due to impulsive shocks from high-speed encounters
with the host halo particles is
\begin{eqnarray}\label{dv2alt}
  (\Delta E)_{\rm cross} & = & {1 \over 2} \int_0^{r_\rmc} \Sigma_\rmc(R) \, 
(\Delta v^2)_{\rm cross}(R) \, 2 \pi R \, \rmd R \nonumber \\
  & = & 4 \pi \, {\vert E_\rmc \vert \over N_\rmc} \, 
  {\sigma_\rmc \over \sigma_\rmh}  \, {h_\rmc(\varepsilon/r_0) \over
    f_\rmE(c)}
\end{eqnarray}
where $E_\rmc$ is the subhalo's binding energy (Eq.~[\ref{Etotsub}]),
and
\begin{equation}\label{theat1}
  h_\rmc(y) = {1 \over f(c)} \, \int_0^c {x^3 \, g(x) \, \rmd x \over x^2 + y^2}
\end{equation}
with $g(x)$ given by Eq.~(\ref{gNFW}).

The function $h(y)$ transits from $\calO(1)$ for $y \ll 1$ to zero for
$y \gg 1$.  Hence, as long as the softening length used is small
compared to the scale radius of the subhalo, one can simply set
$h_\rmc(\varepsilon/r_\rms) = 1$ in the above expression for $(\Delta
E)_{\rm cross}$. This is immediately evident from
Eq.~(\ref{dv2crossheat}), which shows that every subhalo particle,
independent of its cylindrical radius $R$, experiences roughly the
same increase in kinetic energy per unit time. The only exception are
the particles with $R < \varepsilon$, for which $(\Delta v^2)_{\rm
  cross}(R) \propto R^2$. As long as $\varepsilon \ll r_0$, one
therefore does not make a large error by simply writing the {\it
  total} increase in kinetic energy of the subhalo, per crossing time,
as $(\Delta E)_{\rm cross} = {1 \over 2} N_\rmc \, m \, (\Delta
v^2)_{\rm cross}(R)$. This yields Eq.~(\ref{dv2alt}) with
$h_\rmc(\varepsilon/r_\rms) = 1$.

If we follow \cite{Moore.etal.96a} and \cite{vKampen.00a}, and define
the characteristic time scale for this heating process as $t^{\rm
  ch}_{\rm heat} \equiv (\vert E_\rmc \vert / (\Delta E)_{\rm cross})
\, t_{\rm cross}$, then
\begin{equation}\label{theatPS}
  t^{\rm ch}_{\rm heat}  = {2 \over \pi} \, \ln\Lambda_\rmh  \, {f_\rmE(c) \over h_\rmc(\varepsilon/r_0)} \, t^{\rm ch}_{\rm relax} = \zeta(c) \, \ln\Lambda_\rmh
 \,  t^{\rm ch}_{\rm relax} 
\end{equation}
where the second equality is only valid in the limit $\varepsilon \ll
r_0$, and $\zeta(c)$ is a function that depends weakly on the
concentration of the subhalo, increasing from $\sim 0.5$ for $c=5$ to
$\sim 1.3$ for $c=50$. Hence, in a typical $N$-body simulation with
sufficiently small softening length, the time scale for tidal heating
due to impulsive encounters with the host halo particles is roughly an
order of magnitude longer than the particle-subhalo two-body {\it
  relaxation} time, and thus roughly an order of magnitude shorter
than the corresponding {\it evaporation} time scale.

This is illustrated in the left-hand panel of
Fig.~\ref{fig:timescales}, where the red, solid line indicates $t^{\rm
  ch}_{\rm heat}$, computed using Eq.~(\ref{theatPS}), once again
assuming that $M_\rmh/M_\rmc = 30$. Note that this time scale falls in
the grey zone (i.e., becomes less than the Hubble time) for $N_\rmc
\lta 100$ particles. When using the softening length advocated by
\cite{Power.etal.03}, $t^{\rm ch}_{\rm heat}$ is somewhat larger for
small $N_\rmc$, but still of order the Hubble time for $N_\rmc \lta
100$. As face-value, this seems to suggest that particle-subhalo
heating might actually be an important mechanism causing artificial
disruption of subhaloes in numerical simulations. However, this
inference is based on the assumption that $t^{\rm ch}_{\rm heat}$
corresponds to the actual disruption time, which, as we now argue, is
incorrect.

As discussed in \S\ref{sec:applB} and \S\ref{sec:SubSubHeat}, the
problem is that the ratio between the {\it total} energy gain, $\Delta
E$, and the {\it total} binding energy, $|E_\rmc|$ is of little use to
gauge whether or not the system will disrupt. It is important to
account for how the energy $\Delta E$ is distributed over the
particles. Contrary to the case of tidal heating by a single
encounter, where the energy gain of particle $i$ is proportional to
the square of the particle's distance from the centre of the subhalo,
$(\Delta E)_i \propto r^2_i$, in the case considered here, every
subhalo particle roughly receives the same amount of energy (see
Eq.~[\ref{dEparticle}] below); the only exception are the particles
with $R < \varepsilon$, for which $(\Delta E)_i \propto R_i^2$.

In order to assess the impact of $(\Delta E)_{\rm cross}$ on the
system as a whole, we follow the approach of \S\ref{sec:applB} and
compare the individual $(\Delta E)_i$ to the individual binding
energies, $|E_i|$, and define the disruption time as the number of
crossing times required so that $(\Delta E)_i > |E_i|$ for the
majority of particles (i.e., 95 percent). To illustrate how this
modifies the inferred heating time scale, we proceed as follows. We
construct an $N$-body realization of an isotropic NFW subhalo with
concentration parameter $c=10$, using the method described in
\S\ref{sec:sims}. For each individual particle we compute the ratio
$\vert E_i \vert/(\Delta E)_i$.  Here $\vert E_i \vert = v_i^2/2 +
\Phi(r_i)$ is the specific binding energy of particle $i$, and
\begin{equation}\label{dEparticle}
(\Delta E)_i = {1 \over 2} (\Delta v^2)_i \simeq 4 \pi \,
  {\sigma^2_\rmh \over N_\rmh} \, {R^2 \over R^2 + \varepsilon^2}
\end{equation}
is the amount of specific kinetic energy added {\it per crossing time}
due to impulsive encounters with the host particles
(cf. Eq.~[\ref{dv2crossheat}]).  Note that we have assumed that the
$\vec{v}_i \cdot \Delta \vec{v}_i$ term (cf. Eq.~[\ref{dEpart}]),
summed over all encounters, adds to zero. This is motivated by the fact
that the particle is moving in between encounters. Hence, integrated 
over a crossing time, roughly every angle between $\vec{v}_i$ and 
$\Delta \vec{v}_i$ is equally represented.

For each particle, we interpret $\vert E_i \vert/(\Delta E)_i$ as the
number of crossing times required to unbind the particle. This ignores
the fact that the system will continuously re-virialize in order to
adjust to the newly acquired (impulsively added) energy and the
resulting mass loss, and what follows is therefore only a rough
estimate. We define the `corrected' time scale for disruption due to
particle-subhalo tidal heating as the number of crossing times
required for 95 percent of all subhalo particles to become unbound,
i.e., we find the 95-percentile of the distribution of $\vert E_i
\vert/(\Delta E)_i$ and interpret that as $t^{\rm ch}_{\rm heat,corr}
/ t_{\rm cross}$. The $t^{\rm ch}_{\rm heat,corr}$ thus obtained is
shown as dashed, red lines in Fig.~\ref{fig:timescales}. As long as
$\varepsilon \ll r_0$, this correction boosts $t^{\rm ch}_{\rm heat}$
by roughly an order of magnitude, making it very similar to the
two-body evaporation time, $t^{\rm cc}_{\rm evap}$. When the softening
length becomes comparable to the size of the subhalo, $t^{\rm ch}_{\rm
  heat,corr}$ becomes independent of $N_\rmc$, and many orders of
magnitude larger than the Hubble time.

We conclude that similar to two-body relaxation, artificial tidal
heating due to impulsive encounters with the overly massive particles
of the host halo cannot be responsible for artificial disruption of
subhaloes in numerical simulations.

\subsection{Force softening}
\label{sec:softening}
 
Despite many decades of simulation work, and numerous previous
studies, there is still no consensus as to the optimal choice for the
softening parameter in $N$-body simulations.  Numerous studies have
argued that minimizing the scatter in certain statistical descriptions
of the matter field (i.e., power spectrum or two-point correlation
function) among an ensemble of $N$-body realizations requires a
softening length $\varepsilon$ that is at least as large as the mean
inter-particle separation $l = L_{\rm box}/N^{1/3}$
\citep[e.g.,][]{Melott.etal.97, Splinter.etal.98, Romeo.etal.08,
  Joyce.etal.09}. Here $L_{\rm box}$ is the (periodic) box size of the
cosmological simulation and $N$ is the total number of particles
used. Most cosmological simulations, though, are run with much smaller
softening lengths, typically of the order of $\varepsilon/l \sim 0.01
- 0.02$. For example, the Millennium simulation
\citep{Springel.etal.05} has $\varepsilon/l \sim 0.022$, while the
MDPL and SMDPL simulations of \cite{Klypin.etal.16} adopt
$\varepsilon/l\sim 0.019$ and $\varepsilon/l\sim 0.014$,
respectively. Whether such simulations produce reliable results on
scales below the (comoving) mean inter-particle separation is still a
matter of fierce, ongoing debate \citep[see e.g.,][and references
  therein]{Knebe.etal.00, Melott.07, Joyce.etal.09, Benhaiem.etal.16,
  Power.etal.16}.

Several studies have focused on the optimal softening length for
simulating {\it individual} objects (i.e., individual dark matter
haloes or galaxies).  For example, \cite{Dehnen.01}, extending studies
by \cite{Merritt.96} and \cite{Athanassoula.etal.00}, suggests that
one should aim to minimize force errors. This can be achieved by
optimizing the softening length such that it simultaneously minimizes
the force bias (which increases with increasing $\varepsilon$) and the
force noise (which decreases with increasing $\varepsilon$). Based on
such considerations \cite{Dehnen.01} advocates that a collisionless
system with a \cite{Hernquist.90} density profile (which has the same
$r^{-1}$-cusp as a NFW profile) should be simulated with a Plummer
softening $\varepsilon/r_0 = 0.017 N_5^{-0.23}$ where $r_0$ is the
scale radius of the Hernquist (or NFW) profile and $N_5 = N/10^5$ with
$N$ the number of particles used to model the system in
question. \cite{Power.etal.03} use a different criterion to optimize
the softening length: by minimizing the impact of two-body effects on
the (central) density profiles of virialized dark matter haloes in
cosmological simulations, they advocate a much larger softening length
of $\varepsilon/r_0 = 0.126 \, (c/10) \, N_5^{-0.5}$, where $c$ is the
NFW halo concentration parameter. Note that both of these criteria
have the optimal softening length depend on the size of the system as
well as the number of particles that is used to represent the system.
If we adopt a concentration-mass relation of the form $c \propto
M^{-0.1}$ \citep[e.g.,][]{Dutton.Maccio.14}, and use that the virial
radii of dark matter haloes scale according to $r_{\rm vir} \propto
M^{1/3} \propto N^{1/3}$, we find that the optimal softening length
depends on halo mass according to $\varepsilon_{\rm opt} \propto
M^{\alpha}$, with $\alpha=+0.2$ and $-0.17$ for the criteria of Dehnen
and Power et al., respectively. This scaling is sufficiently weak so
that the softening is close to optimal for haloes spanning a
relatively large range in halo masses, as required for cosmological
simulations.

Of relevance for the discussion in this paper is how softening impacts
the dynamical evolution of dark matter substructure. To our knowledge,
there has been no concerted effort to investigate this. We fill this
gap in Paper II, with a detailed study of how softening impacts the
evolution and survivability of subhaloes. We demonstrate that whenever
the tidal field is strong (i.e., close to the centre of the host
halo), the survivability of substructure is extremely sensitive to the
softening used. Furthermore, as a subhalo experiences mass loss, both
its size and its number of particles change with time, which in turn
causes an evolution in the optimal softening length. As we demonstrate
in Paper~II, typically $\varepsilon_{\rm opt}$ drops by an order of
magnitude after the first peri-centric passage, after which it remains
fairly unchanged. Since simulations do not account for this drastic
drop in the optimal softening length, substructure is typically
evolved with a softening length that is too large, which results in
significant artificial subhalo disruption\footnote{The simulations
  used in {\it this} paper adopted a softening that was carefully
  calibrated using the detailed convergence studies described in
  Paper~II, and therefore do not suffer from this effect.}. Based on
the discussion in this section, and the results presented in Paper~II,
we therefore conclude that inadequate softening is the dominant driver
of subhalo disruption in numerical simulations.

\section{Summary \& Discussion}
\label{sec:concl}

Being able to accurately predict the abundance and demographics of
dark matter substructure is of paramount importance for many fields of
astrophysics: gravitational lensing, galaxy evolution, halo occupation
modeling, and even constraining the nature of the dark matter. Dark
matter substructure is subject to tidal stripping and heating, which
are highly non-linear processes and therefore best studied using
numerical $N$-body simulations. These reveal prevalent subhalo
disruption, with only $\sim 35$ percent of subhaloes accreted at $z=1$
surviving until $z=0$ \citep{Jiang.vdBosch.16}. Based on a hand-full
of convergence studies, it is generally assumed that most of the
disruption of subhaloes with more than 50-100 particles is physical,
rather then numerical. However, there is no consensus in the
literature as to what causes subhaloes to disrupt. In addition,
convergence is only a necessary, but not a sufficient condition, to
guarantee that the simulation results are physically correct.

All this warrants a detailed, comprehensive investigation into the
disruption of substructure in numerical simulations.  This is the
first paper in a series in which we use both idealized numerical
simulations and (semi)-analytical treatments to study under what
conditions dark matter substructure undergoes physical and/or
numerical disruption.  In this paper we have focused mainly on
analytical treatments of tidal heating and tidal stripping. We also
presented a crude analytical treatment of potential numerical
disruption mechanisms.  Our main conclusions are as follows:
\begin{itemize}
\item During their first orbit after accretion, subhaloes pass the
  center of their host halo at a median peri-centric distance of $\sim
  0.37 R_{\rm vir,h}$, while $\sim 7$ percent come within $0.1 R_{\rm
    vir,h }$.
\item Contrary to naive expectation, subhaloes that experience a tidal
  shock $\Delta E$ that exceeds the subhalo's binding energy,
  $|E_\rmb|$, do not necessarily undergo disruption. Rather, they
  experience mass loss. We have presented a fitting function
  (Eq.[\ref{fitfunc}]) that relates the amount of mass stripped to the
  ratio $\Delta E/|E_\rmb|$, which is valid for subhaloes with
  NFW density distributions. An NFW subhalo for which $\Delta
  E/|E_\rmb|=1$ ($100$) is stripped of only $\sim 20$ $(80)$ percent
  of its mass.
\item During first peri-centric passage, tidal heating increases the
  internal energy of the subhalo by a median $\Delta E \sim 2 |E_{\rm
    b}|$. This results in the subhalo loosing $\sim 30$ percent of its
  mass. In the rare case of a purely radial orbit, the subhalo
  experiences a tidal shock that may strip as much as 90 percent of
  its mass.
\item The impulse approximation, combined with the instantaneous mass
  loss approximation of \cite{Aguilar.White.85}, accurately predicts
  the amount of mass loss experienced by subhaloes on fairly radial
  orbits ($\eta \lta 0.2$).
\item Tidal heating due to high-speed (impulsive) encounters with
  other subhaloes, sometimes called `harassment', is negligible
  compared to the tidal heating associated with the peri-centric
  passage of the host halo.
\item Modeling tidal stripping in the non-impulsive regime is
  extremely difficult, mainly because the concept of tidal radius is
  ill-defined, and because we lack a theory to describe how a
  gravitational $N$-body system re-virializes as it undergoes
  stripping. Depending on which definition one adopts for the tidal
  radius, and what one assumes regarding the {\it rate} at which
  material outside of the (instantaneous) tidal radius is stripped,
  one can obtain dramatically different results, none of which
  adequately reproduce simulation results.
\item Instantaneously stripping matter from the outskirts of a NFW
  halo can leave a remnant with positive binding energy
  \citep[cf.][]{Hayashi.etal.03}. The remnant's binding energy depends
  strongly on the orbital anisotropy of the subhalo, with more
  radially anisotropic subhaloes having larger (i.e., more positive or
  less negative) binding energies for a given amount of stripping.
\item Instantaneously stripping matter from a NFW halo never leads to
  subhalo disruption, even if the remnant has positive binding energy.
\item Contrary to claims in the literature, two-body relaxation of
  subhaloes due to collisions between subhalo particles and (the more
  abundant) particles from the host halo is {\it less} efficient than
  that due to collisions among subhalo particles themselves. Neither
  is efficient enough, however, to cause significant evaporation of
  subhaloes in numerical simulations.
\item Tidal heating of subhaloes due to high-speed, impulsive
  encounters with host halo particles that are artificially massive
  (due to the limiting mass resolution of the simulation) cannot cause
  artificial disruption of substructure.
\end{itemize}

Based on these results, we conclude that it is extremely difficult to
physically disrupt CDM subhaloes (in the absence of baryonic effects;
see below). Tidal stripping alone never leads to complete disruption,
and tidal heating is not effective in the central regions of
subhaloes, which are adiabatically `shielded' \citep{Spitzer.87,
  Weinberg.94a, Weinberg.94b}.  In principle, multiple tidal shocks,
associated with multiple peri-centric passages, could ultimately cause
the subhalo to disrupt. We did not explore such a scenario in this
paper, since we lack a reliable analytical treatment for how subhaloes
re-virialize after tidally induced mass loss. However, we do
investigate the impact of multiple peri-centric passages in Papers II
and III, in which we use hundreds of idealized numerical
simulations. As we demonstrate in those papers, physical disruption in
CDM-based dark-matter-only simulations is extremely rare \citep[see
  also][]{Diemand.etal.07, Penarrubia.etal.10}. On the other hand, we
do find that the simulation results are extremely sensitive to the
choice of softening length. In particular, we demonstrate that
state-of-the-art cosmological simulations still suffer from serious
overmerging, largely as a consequence of inappropriate force
softening. In addition, we demonstrate that $N$-body simulations
suffer from a discreteness-driven run-away instability, that makes it
impossible to reliably trace the evolution of subhaloes in a strong
tidal field unless the subhalo has at least 1000 particles.

Numerical overmerging is a serious road-block for the many
astrophysical applications that require accurate characterization of
halo substructure.  In the coming decade a number of large galaxy
surveys, such as DESI, LSST, EUCLID, and WFIRST will provide
astrophysicists with an unprecedented wealth of data regarding galaxy
evolution and cosmology. In order to optimize the scientific impact of
these huge investments, it is prudent that galaxy clustering in these
surveys be interpreted in an unbiased, maximally informative manner. A
popular method to do so is to compare the observations to mock data,
obtained by populating dark matter (sub)haloes in dark-matter-only
simulations with `mock' galaxies, using methods such as subhalo
abundance matching (see \S\ref{sec:intro} for references). This entire
program faces a severe challenge if those simulations underpredict the
abundance of subhaloes due to artificial disruption. In principle one
can correct for this by including `orphans', i.e., mock galaxies
without an associated subhalo in the simulation
\citep[e.g.,][]{Kitzbichler.White.08, Moster.etal.13, Moster.etal.17,
  Guo.White.13}; however, unless it is known how many orphans to add,
and where, this seriously diminishes the information content of
small-scale clustering \citep[see][for a detailed
  discussion]{Campbell.etal.17}. Dark matter substructure is also an
important discriminator between different dark matter models (cold
vs. warm vs. self-interacting), and is important for translating a
potential, observed dark matter annihilation signal into a mass and
annihilation cross section of the dark matter particle. Unless we can
make accurate, and above all reliable, predictions regarding the
abundance and structure of dark matter subhaloes, we will forfeit one
of the main handles we have on learning about the nature of dark
matter.

We end by emphasizing that this work has entirely focused on dark
matter only, without taking account of potential baryonic effects.
There is a rapidly expanding literature on how baryons may impact the
abundance and demographics of dark matter
substructure \citep[e.g.,][]{Maccio.etal.06, Weinberg.etal.08,
  Dolag.etal.09, Arraki.etal.14, Brooks.Zolotov.14,
  Despali.Vegetti.16, Fiacconi.etal.16, Wetzel.etal.16,
  Garrison-Kimmel.etal.17}.  Our work, in no way, aims to undermine
the importance of baryons.  However, before we can make reliable
predictions for how baryonic processes modify substructure in the dark
sector, we first need to establish a better understanding of how tides
impact substructure, and develop tools and criteria to assess the
reliability of simulations (be it hydro or $N$-body). Hence, this work
is to be considered a necessary first step towards a more reliable
treatment of dark matter substructure (including satellite galaxies);
not the final answer.


\section*{Acknowledgments}

We are grateful to the referee, Chris Power, for an insightful referee
report, and to Andrew Hearin, Fangzhou Jiang, and Fred Rasio for
useful discussion.  FvdB is supported by the Klaus Tschira Foundation
and by the US National Science Foundation through grant AST 1516962,
and is grateful to the Munich Excellence Cluster for its
hospitality. Part of this work was performed at the Aspen Center for
Physics, which is supported by the National Science Foundation under
grant PHY-1066293, and at the Kavli Institute for Theoretical Physics,
which is supported by the National Science Foundation under grant
PHY-1125915.  OH and GO were supported by funding form the European
Research Council (ERC) under the European Union's Horizon 2020
research and innovation programme (grant agreement No. 679145, project
`COSMO-SIMS').


\bibliographystyle{mnras}
\bibliography{references_vdb}

\begin{thebibliography}{}
\makeatletter
\relax
\def\mn@urlcharsother{\let\do\@makeother \do\$\do\&\do\#\do\^\do\_\do\%\do\~}
\def\mn@doi{\begingroup\mn@urlcharsother \@ifnextchar [ {\mn@doi@}
  {\mn@doi@[]}}
\def\mn@doi@[#1]#2{\def\@tempa{#1}\ifx\@tempa\@empty \href
  {http://dx.doi.org/#2} {doi:#2}\else \href {http://dx.doi.org/#2} {#1}\fi
  \endgroup}
\def\mn@eprint#1#2{\mn@eprint@#1:#2::\@nil}
\def\mn@eprint@arXiv#1{\href {http://arxiv.org/abs/#1} {{\tt arXiv:#1}}}
\def\mn@eprint@dblp#1{\href {http://dblp.uni-trier.de/rec/bibtex/#1.xml}
  {dblp:#1}}
\def\mn@eprint@#1:#2:#3:#4\@nil{\def\@tempa {#1}\def\@tempb {#2}\def\@tempc
  {#3}\ifx \@tempc \@empty \let \@tempc \@tempb \let \@tempb \@tempa \fi \ifx
  \@tempb \@empty \def\@tempb {arXiv}\fi \@ifundefined
  {mn@eprint@\@tempb}{\@tempb:\@tempc}{\expandafter \expandafter \csname
  mn@eprint@\@tempb\endcsname \expandafter{\@tempc}}}

\bibitem[\protect\citeauthoryear{{Aguilar} \& {White}}{{Aguilar} \&
  {White}}{1985}]{Aguilar.White.85}
{Aguilar} L.~A.,  {White} S.~D.~M.,  1985, \mn@doi [\apj] {10.1086/163382},
  \href {http://adsabs.harvard.edu/abs/1985ApJ...295..374A} {295, 374}

\bibitem[\protect\citeauthoryear{{Arraki}, {Klypin}, {More}  \&
  {Trujillo-Gomez}}{{Arraki} et~al.}{2014}]{Arraki.etal.14}
{Arraki} K.~S.,  {Klypin} A.,  {More} S.,   {Trujillo-Gomez} S.,  2014, \mn@doi
  [\mnras] {10.1093/mnras/stt2279}, \href
  {http://adsabs.harvard.edu/abs/2014MNRAS.438.1466A} {438, 1466}

\bibitem[\protect\citeauthoryear{{Athanassoula}, {Fady}, {Lambert}  \&
  {Bosma}}{{Athanassoula} et~al.}{2000}]{Athanassoula.etal.00}
{Athanassoula} E.,  {Fady} E.,  {Lambert} J.~C.,   {Bosma} A.,  2000, \mn@doi
  [\mnras] {10.1046/j.1365-8711.2000.03316.x}, \href
  {http://adsabs.harvard.edu/abs/2000MNRAS.314..475A} {314, 475}

\bibitem[\protect\citeauthoryear{{Barnes} \& {Hut}}{{Barnes} \&
  {Hut}}{1986}]{Barnes.Hut.86}
{Barnes} J.,  {Hut} P.,  1986, \mn@doi [\nat] {10.1038/324446a0}, \href
  {http://adsabs.harvard.edu/abs/1986Natur.324..446B} {324, 446}

\bibitem[\protect\citeauthoryear{{Behroozi}, {Wechsler}  \& {Wu}}{{Behroozi}
  et~al.}{2013a}]{Behroozi.etal.13a}
{Behroozi} P.~S.,  {Wechsler} R.~H.,   {Wu} H.-Y.,  2013a, \mn@doi [\apj]
  {10.1088/0004-637X/762/2/109}, \href
  {http://adsabs.harvard.edu/abs/2013ApJ...762..109B} {762, 109}

\bibitem[\protect\citeauthoryear{{Behroozi}, {Wechsler}  \&
  {Conroy}}{{Behroozi} et~al.}{2013b}]{Behroozi.etal.13c}
{Behroozi} P.~S.,  {Wechsler} R.~H.,   {Conroy} C.,  2013b, \mn@doi [\apj]
  {10.1088/0004-637X/770/1/57}, \href
  {http://adsabs.harvard.edu/abs/2013ApJ...770...57B} {770, 57}

\bibitem[\protect\citeauthoryear{{Benhaiem}, {Joyce}  \& {Sylos
  Labini}}{{Benhaiem} et~al.}{2016}]{Benhaiem.etal.16}
{Benhaiem} D.,  {Joyce} M.,   {Sylos Labini} F.,  2016, preprint, \href
  {http://adsabs.harvard.edu/abs/2016arXiv160904580B} {} (\mn@eprint {arXiv}
  {1609.04580})

\bibitem[\protect\citeauthoryear{{Bergstr{\"o}m}, {Edsj{\"o}}, {Gondolo}  \&
  {Ullio}}{{Bergstr{\"o}m} et~al.}{1999}]{Bergstrom.etal.99}
{Bergstr{\"o}m} L.,  {Edsj{\"o}} J.,  {Gondolo} P.,   {Ullio} P.,  1999,
  \mn@doi [\prd] {10.1103/PhysRevD.59.043506}, \href
  {http://adsabs.harvard.edu/abs/1999PhRvD..59d3506B} {59, 043506}

\bibitem[\protect\citeauthoryear{{Binney} \& {Tremaine}}{{Binney} \&
  {Tremaine}}{1987}]{Binney.Tremaine.87}
{Binney} J.,  {Tremaine} S.,  1987, {Galactic dynamics}.
Princeton University Press

\bibitem[\protect\citeauthoryear{{Binney} \& {Tremaine}}{{Binney} \&
  {Tremaine}}{2008}]{Binney.Tremaine.08}
{Binney} J.,  {Tremaine} S.,  2008, {Galactic Dynamics: Second Edition}.
Princeton University Press

\bibitem[\protect\citeauthoryear{{Bose} et~al.,}{{Bose}
  et~al.}{2016}]{Bose.etal.16}
{Bose} S.,  et~al., 2016, preprint, \href
  {http://adsabs.harvard.edu/abs/2016arXiv160407409B} {} (\mn@eprint {arXiv}
  {1604.07409})

\bibitem[\protect\citeauthoryear{{Boylan-Kolchin}, {Springel}, {White}  \&
  {Jenkins}}{{Boylan-Kolchin} et~al.}{2010}]{Boylan-Kolchin.etal.10}
{Boylan-Kolchin} M.,  {Springel} V.,  {White} S.~D.~M.,   {Jenkins} A.,  2010,
  \mn@doi [\mnras] {10.1111/j.1365-2966.2010.16774.x}, \href
  {http://adsabs.harvard.edu/abs/2010MNRAS.406..896B} {406, 896}

\bibitem[\protect\citeauthoryear{{Boylan-Kolchin}, {Bullock}  \&
  {Kaplinghat}}{{Boylan-Kolchin} et~al.}{2011}]{Boylan-Kolchin.etal.11}
{Boylan-Kolchin} M.,  {Bullock} J.~S.,   {Kaplinghat} M.,  2011, \mn@doi
  [\mnras] {10.1111/j.1745-3933.2011.01074.x}, \href
  {http://adsabs.harvard.edu/abs/2011MNRAS.415L..40B} {415, L40}

\bibitem[\protect\citeauthoryear{{Brada{\v c}}, {Schneider}, {Steinmetz},
  {Lombardi}, {King}  \& {Porcas}}{{Brada{\v c}} et~al.}{2002}]{Bradac.02}
{Brada{\v c}} M.,  {Schneider} P.,  {Steinmetz} M.,  {Lombardi} M.,  {King}
  L.~J.,   {Porcas} R.,  2002, \mn@doi [\aap] {10.1051/0004-6361:20020559},
  \href {http://adsabs.harvard.edu/abs/2002A%26A...388..373B} {388, 373}

\bibitem[\protect\citeauthoryear{{Brainerd}, {Goldberg}  \& {Verner
  Villumsen}}{{Brainerd} et~al.}{1998}]{Brainerd.etal.98}
{Brainerd} T.~G.,  {Goldberg} D.~M.,   {Verner Villumsen} J.,  1998, \mn@doi
  [\apj] {10.1086/305917}, \href
  {http://adsabs.harvard.edu/abs/1998ApJ...502..505B} {502, 505}

\bibitem[\protect\citeauthoryear{{Brooks} \& {Zolotov}}{{Brooks} \&
  {Zolotov}}{2014}]{Brooks.Zolotov.14}
{Brooks} A.~M.,  {Zolotov} A.,  2014, \mn@doi [\apj]
  {10.1088/0004-637X/786/2/87}, \href
  {http://adsabs.harvard.edu/abs/2014ApJ...786...87B} {786, 87}

\bibitem[\protect\citeauthoryear{{Bryan} \& {Norman}}{{Bryan} \&
  {Norman}}{1998}]{Bryan.Norman.98}
{Bryan} G.~L.,  {Norman} M.~L.,  1998, \mn@doi [\apj] {10.1086/305262}, \href
  {http://adsabs.harvard.edu/abs/1998ApJ...495...80B} {495, 80}

\bibitem[\protect\citeauthoryear{{Bullock} \& {Boylan-Kolchin}}{{Bullock} \&
  {Boylan-Kolchin}}{2017}]{Bullock.BoylanKolchin.17}
{Bullock} J.~S.,  {Boylan-Kolchin} M.,  2017, preprint, \href
  {http://adsabs.harvard.edu/abs/2017arXiv170704256B} {} (\mn@eprint {arXiv}
  {1707.04256})

\bibitem[\protect\citeauthoryear{{Burkert}}{{Burkert}}{2000}]{Burkert.00}
{Burkert} A.,  2000, \mn@doi [\apjl] {10.1086/312674}, \href
  {http://adsabs.harvard.edu/abs/2000ApJ...534L.143B} {534, L143}

\bibitem[\protect\citeauthoryear{{Campbell}, {van den Bosch}, {Padmanabhan},
  {Mao}, {Zentner}, {Lange}, {Jiang}  \& {Villarreal}}{{Campbell}
  et~al.}{2017}]{Campbell.etal.17}
{Campbell} D.,  {van den Bosch} F.~C.,  {Padmanabhan} N.,  {Mao} Y.-Y.,
  {Zentner} A.~R.,  {Lange} J.~U.,  {Jiang} F.,   {Villarreal} A.,  2017,
  preprint, \href {http://adsabs.harvard.edu/abs/2017arXiv170506347C} {}
  (\mn@eprint {arXiv} {1705.06347})

\bibitem[\protect\citeauthoryear{{Carlberg}}{{Carlberg}}{1994}]{Carlberg.94}
{Carlberg} R.~G.,  1994, \mn@doi [\apj] {10.1086/174659}, \href
  {http://adsabs.harvard.edu/abs/1994ApJ...433..468C} {433, 468}

\bibitem[\protect\citeauthoryear{{Carlberg}}{{Carlberg}}{2009}]{Carlberg.09}
{Carlberg} R.~G.,  2009, \mn@doi [\apjl] {10.1088/0004-637X/705/2/L223}, \href
  {http://adsabs.harvard.edu/abs/2009ApJ...705L.223C} {705, L223}

\bibitem[\protect\citeauthoryear{{Cautun}, {Hellwing}, {van de Weygaert},
  {Frenk}, {Jones}  \& {Sawala}}{{Cautun} et~al.}{2014}]{Cautun.etal.14}
{Cautun} M.,  {Hellwing} W.~A.,  {van de Weygaert} R.,  {Frenk} C.~S.,  {Jones}
  B.~J.~T.,   {Sawala} T.,  2014, \mn@doi [\mnras] {10.1093/mnras/stu1829},
  \href {http://adsabs.harvard.edu/abs/2014MNRAS.445.1820C} {445, 1820}

\bibitem[\protect\citeauthoryear{{Chandrasekhar}}{{Chandrasekhar}}{1943}]{Chandrasekhar.43}
{Chandrasekhar} S.,  1943, \mn@doi [\apj] {10.1086/144517}, \href
  {http://adsabs.harvard.edu/abs/1943ApJ....97..255C} {97, 255}

\bibitem[\protect\citeauthoryear{{Col{\'{\i}}n}, {Avila-Reese},
  {Gonz{\'a}lez-Samaniego}  \& {Vel{\'a}zquez}}{{Col{\'{\i}}n}
  et~al.}{2015}]{Colin.etal.15}
{Col{\'{\i}}n} P.,  {Avila-Reese} V.,  {Gonz{\'a}lez-Samaniego} A.,
  {Vel{\'a}zquez} H.,  2015, \mn@doi [\apj] {10.1088/0004-637X/803/1/28}, \href
  {http://adsabs.harvard.edu/abs/2015ApJ...803...28C} {803, 28}

\bibitem[\protect\citeauthoryear{{Conroy}, {Wechsler}  \& {Kravtsov}}{{Conroy}
  et~al.}{2006}]{Conroy.etal.06}
{Conroy} C.,  {Wechsler} R.~H.,   {Kravtsov} A.~V.,  2006, \mn@doi [\apj]
  {10.1086/503602}, \href {http://adsabs.harvard.edu/abs/2006ApJ...647..201C}
  {647, 201}

\bibitem[\protect\citeauthoryear{{Dalal} \& {Kochanek}}{{Dalal} \&
  {Kochanek}}{2002}]{Dalal.Kochanek.02}
{Dalal} N.,  {Kochanek} C.~S.,  2002, \mn@doi [\apj] {10.1086/340303}, \href
  {http://adsabs.harvard.edu/abs/2002ApJ...572...25D} {572, 25}

\bibitem[\protect\citeauthoryear{{Dehnen}}{{Dehnen}}{2001}]{Dehnen.01}
{Dehnen} W.,  2001, \mn@doi [\mnras] {10.1046/j.1365-8711.2001.04237.x}, \href
  {http://adsabs.harvard.edu/abs/2001MNRAS.324..273D} {324, 273}

\bibitem[\protect\citeauthoryear{{Dekel}, {Devor}  \& {Hetzroni}}{{Dekel}
  et~al.}{2003}]{Dekel.etal.03}
{Dekel} A.,  {Devor} J.,   {Hetzroni} G.,  2003, \mn@doi [\mnras]
  {10.1046/j.1365-8711.2003.06432.x}, \href
  {http://adsabs.harvard.edu/abs/2003MNRAS.341..326D} {341, 326}

\bibitem[\protect\citeauthoryear{{Despali} \& {Vegetti}}{{Despali} \&
  {Vegetti}}{2016}]{Despali.Vegetti.16}
{Despali} G.,  {Vegetti} S.,  2016, preprint, \href
  {http://adsabs.harvard.edu/abs/2016arXiv160806938D} {} (\mn@eprint {arXiv}
  {1608.06938})

\bibitem[\protect\citeauthoryear{{Diemand}, {Moore}, {Stadel}  \&
  {Kazantzidis}}{{Diemand} et~al.}{2004a}]{Diemand.etal.04}
{Diemand} J.,  {Moore} B.,  {Stadel} J.,   {Kazantzidis} S.,  2004a, \mn@doi
  [\mnras] {10.1111/j.1365-2966.2004.07424.x}, \href
  {http://adsabs.harvard.edu/abs/2004MNRAS.348..977D} {348, 977}

\bibitem[\protect\citeauthoryear{{Diemand}, {Moore}  \& {Stadel}}{{Diemand}
  et~al.}{2004b}]{Diemand.Moore.Stadel.04}
{Diemand} J.,  {Moore} B.,   {Stadel} J.,  2004b, \mn@doi [\mnras]
  {10.1111/j.1365-2966.2004.07940.x}, \href
  {http://adsabs.harvard.edu/abs/2004MNRAS.352..535D} {352, 535}

\bibitem[\protect\citeauthoryear{{Diemand}, {Kuhlen}  \& {Madau}}{{Diemand}
  et~al.}{2007}]{Diemand.etal.07}
{Diemand} J.,  {Kuhlen} M.,   {Madau} P.,  2007, \mn@doi [\apj]
  {10.1086/520573}, \href {http://adsabs.harvard.edu/abs/2007ApJ...667..859D}
  {667, 859}

\bibitem[\protect\citeauthoryear{{Dolag}, {Borgani}, {Murante}  \&
  {Springel}}{{Dolag} et~al.}{2009}]{Dolag.etal.09}
{Dolag} K.,  {Borgani} S.,  {Murante} G.,   {Springel} V.,  2009, \mn@doi
  [\mnras] {10.1111/j.1365-2966.2009.15034.x}, \href
  {http://adsabs.harvard.edu/abs/2009MNRAS.399..497D} {399, 497}

\bibitem[\protect\citeauthoryear{{Drakos}, {Taylor}  \& {Benson}}{{Drakos}
  et~al.}{2017}]{Drakos.etal.17}
{Drakos} N.~E.,  {Taylor} J.~E.,   {Benson} A.~J.,  2017, \mn@doi [\mnras]
  {10.1093/mnras/stx652}, \href
  {http://adsabs.harvard.edu/abs/2017MNRAS.468.2345D} {468, 2345}

\bibitem[\protect\citeauthoryear{{Dutton} \& {Macci{\`o}}}{{Dutton} \&
  {Macci{\`o}}}{2014}]{Dutton.Maccio.14}
{Dutton} A.~A.,  {Macci{\`o}} A.~V.,  2014, \mn@doi [\mnras]
  {10.1093/mnras/stu742}, \href
  {http://adsabs.harvard.edu/abs/2014MNRAS.441.3359D} {441, 3359}

\bibitem[\protect\citeauthoryear{{Dutton} et~al.,}{{Dutton}
  et~al.}{2016}]{Dutton.etal.16}
{Dutton} A.~A.,  et~al., 2016, \mn@doi [\mnras] {10.1093/mnras/stw1537}, \href
  {http://adsabs.harvard.edu/abs/2016MNRAS.461.2658D} {461, 2658}

\bibitem[\protect\citeauthoryear{{Eddington}}{{Eddington}}{1916}]{Eddington.16}
{Eddington} A.~S.,  1916, \mn@doi [\mnras] {10.1093/mnras/76.7.572}, \href
  {http://adsabs.harvard.edu/abs/1916MNRAS..76..572E} {76, 572}

\bibitem[\protect\citeauthoryear{{Erkal}, {Belokurov}, {Bovy}  \&
  {Sanders}}{{Erkal} et~al.}{2016}]{Erkal.etal.16}
{Erkal} D.,  {Belokurov} V.,  {Bovy} J.,   {Sanders} J.~L.,  2016, \mn@doi
  [\mnras] {10.1093/mnras/stw1957}, \href
  {http://adsabs.harvard.edu/abs/2016MNRAS.463..102E} {463, 102}

\bibitem[\protect\citeauthoryear{{Faltenbacher} \& {Diemand}}{{Faltenbacher} \&
  {Diemand}}{2006}]{Faltenbacher.Diemand.06}
{Faltenbacher} A.,  {Diemand} J.,  2006, \mn@doi [\mnras]
  {10.1111/j.1365-2966.2006.10421.x}, \href
  {http://adsabs.harvard.edu/abs/2006MNRAS.369.1698F} {369, 1698}

\bibitem[\protect\citeauthoryear{{Farouki} \& {Salpeter}}{{Farouki} \&
  {Salpeter}}{1982}]{Farouki.Salpeter.82}
{Farouki} R.~T.,  {Salpeter} E.~E.,  1982, \mn@doi [\apj] {10.1086/159653},
  \href {http://adsabs.harvard.edu/abs/1982ApJ...253..512F} {253, 512}

\bibitem[\protect\citeauthoryear{{Fiacconi}, {Madau}, {Potter}  \&
  {Stadel}}{{Fiacconi} et~al.}{2016}]{Fiacconi.etal.16}
{Fiacconi} D.,  {Madau} P.,  {Potter} D.,   {Stadel} J.,  2016, \mn@doi [\apj]
  {10.3847/0004-637X/824/2/144}, \href
  {http://adsabs.harvard.edu/abs/2016ApJ...824..144F} {824, 144}

\bibitem[\protect\citeauthoryear{{Frenk}, {White}, {Davis}  \&
  {Efstathiou}}{{Frenk} et~al.}{1988}]{Frenk.etal.88}
{Frenk} C.~S.,  {White} S.~D.~M.,  {Davis} M.,   {Efstathiou} G.,  1988,
  \mn@doi [\apj] {10.1086/166213}, \href
  {http://adsabs.harvard.edu/abs/1988ApJ...327..507F} {327, 507}

\bibitem[\protect\citeauthoryear{{Gao}, {White}, {Jenkins}, {Stoehr}  \&
  {Springel}}{{Gao} et~al.}{2004}]{Gao.etal.04}
{Gao} L.,  {White} S.~D.~M.,  {Jenkins} A.,  {Stoehr} F.,   {Springel} V.,
  2004, \mn@doi [\mnras] {10.1111/j.1365-2966.2004.08360.x}, \href
  {http://adsabs.harvard.edu/abs/2004MNRAS.355..819G} {355, 819}

\bibitem[\protect\citeauthoryear{{Garrison-Kimmel}, {Boylan-Kolchin}, {Bullock}
   \& {Kirby}}{{Garrison-Kimmel} et~al.}{2014}]{Garrison-Kimmel.etal.14}
{Garrison-Kimmel} S.,  {Boylan-Kolchin} M.,  {Bullock} J.~S.,   {Kirby} E.~N.,
  2014, \mn@doi [\mnras] {10.1093/mnras/stu1477}, \href
  {http://adsabs.harvard.edu/abs/2014MNRAS.444..222G} {444, 222}

\bibitem[\protect\citeauthoryear{{Garrison-Kimmel} et~al.,}{{Garrison-Kimmel}
  et~al.}{2017}]{Garrison-Kimmel.etal.17}
{Garrison-Kimmel} S.,  et~al., 2017, preprint, \href
  {http://adsabs.harvard.edu/abs/2017arXiv170103792G} {} (\mn@eprint {arXiv}
  {1701.03792})

\bibitem[\protect\citeauthoryear{{Ghigna}, {Moore}, {Governato}, {Lake},
  {Quinn}  \& {Stadel}}{{Ghigna} et~al.}{1998}]{Ghigna.etal.98}
{Ghigna} S.,  {Moore} B.,  {Governato} F.,  {Lake} G.,  {Quinn} T.,   {Stadel}
  J.,  1998, \mn@doi [\mnras] {10.1046/j.1365-8711.1998.01918.x}, \href
  {http://adsabs.harvard.edu/abs/1998MNRAS.300..146G} {300, 146}

\bibitem[\protect\citeauthoryear{{Giocoli}, {Tormen}  \& {van den
  Bosch}}{{Giocoli} et~al.}{2008a}]{Giocoli.etal.08}
{Giocoli} C.,  {Tormen} G.,   {van den Bosch} F.~C.,  2008a, \mn@doi [\mnras]
  {10.1111/j.1365-2966.2008.13182.x}, \href
  {http://adsabs.harvard.edu/abs/2008MNRAS.386.2135G} {386, 2135}

\bibitem[\protect\citeauthoryear{{Giocoli}, {Pieri}  \& {Tormen}}{{Giocoli}
  et~al.}{2008b}]{Giocoli.etal.08b}
{Giocoli} C.,  {Pieri} L.,   {Tormen} G.,  2008b, \mn@doi [\mnras]
  {10.1111/j.1365-2966.2008.13283.x}, \href
  {http://adsabs.harvard.edu/abs/2008MNRAS.387..689G} {387, 689}

\bibitem[\protect\citeauthoryear{{Giocoli}, {Tormen}, {Sheth}  \& {van den
  Bosch}}{{Giocoli} et~al.}{2010}]{Giocoli.etal.10}
{Giocoli} C.,  {Tormen} G.,  {Sheth} R.~K.,   {van den Bosch} F.~C.,  2010,
  \mn@doi [\mnras] {10.1111/j.1365-2966.2010.16311.x}, \href
  {http://adsabs.harvard.edu/abs/2010MNRAS.404..502G} {404, 502}

\bibitem[\protect\citeauthoryear{{Gnedin} \& {Ostriker}}{{Gnedin} \&
  {Ostriker}}{1999}]{Gnedin.Ostriker.99}
{Gnedin} O.~Y.,  {Ostriker} J.~P.,  1999, \mn@doi [\apj] {10.1086/306864},
  \href {http://adsabs.harvard.edu/abs/1999ApJ...513..626G} {513, 626}

\bibitem[\protect\citeauthoryear{{Gnedin}, {Hernquist}  \& {Ostriker}}{{Gnedin}
  et~al.}{1999}]{Gnedin.etal.99}
{Gnedin} O.~Y.,  {Hernquist} L.,   {Ostriker} J.~P.,  1999, \mn@doi [\apj]
  {10.1086/306910}, \href {http://adsabs.harvard.edu/abs/1999ApJ...514..109G}
  {514, 109}

\bibitem[\protect\citeauthoryear{{Gonzalez-Casado}, {Mamon}  \&
  {Salvador-Sole}}{{Gonzalez-Casado} et~al.}{1994}]{Gonzales.etal.94}
{Gonzalez-Casado} G.,  {Mamon} G.~A.,   {Salvador-Sole} E.,  1994, \mn@doi
  [\apjl] {10.1086/187548}, \href
  {http://adsabs.harvard.edu/abs/1994ApJ...433L..61G} {433, L61}

\bibitem[\protect\citeauthoryear{{Griffen}, {Ji}, {Dooley}, {G{\'o}mez},
  {Vogelsberger}, {O'Shea}  \& {Frebel}}{{Griffen}
  et~al.}{2016}]{Griffen.etal.16}
{Griffen} B.~F.,  {Ji} A.~P.,  {Dooley} G.~A.,  {G{\'o}mez} F.~A.,
  {Vogelsberger} M.,  {O'Shea} B.~W.,   {Frebel} A.,  2016, \mn@doi [\apj]
  {10.3847/0004-637X/818/1/10}, \href
  {http://adsabs.harvard.edu/abs/2016ApJ...818...10G} {818, 10}

\bibitem[\protect\citeauthoryear{{Guo} \& {White}}{{Guo} \&
  {White}}{2013}]{Guo.White.13}
{Guo} Q.,  {White} S.,  2013, ArXiv:1303.3586, \href
  {http://adsabs.harvard.edu/abs/2013arXiv1303.3586G} {}

\bibitem[\protect\citeauthoryear{{Guo}, {White}, {Li}  \&
  {Boylan-Kolchin}}{{Guo} et~al.}{2010}]{Guo.etal.10}
{Guo} Q.,  {White} S.,  {Li} C.,   {Boylan-Kolchin} M.,  2010, \mn@doi [\mnras]
  {10.1111/j.1365-2966.2010.16341.x}, \href
  {http://adsabs.harvard.edu/abs/2010MNRAS.404.1111G} {404, 1111}

\bibitem[\protect\citeauthoryear{{Hahn} \& {Angulo}}{{Hahn} \&
  {Angulo}}{2016}]{Hahn.Angulo.16}
{Hahn} O.,  {Angulo} R.~E.,  2016, \mn@doi [\mnras] {10.1093/mnras/stv2304},
  \href {http://adsabs.harvard.edu/abs/2016MNRAS.455.1115H} {455, 1115}

\bibitem[\protect\citeauthoryear{{Han}, {Jing}, {Wang}  \& {Wang}}{{Han}
  et~al.}{2012}]{Han.etal.12}
{Han} J.,  {Jing} Y.~P.,  {Wang} H.,   {Wang} W.,  2012, \mn@doi [\mnras]
  {10.1111/j.1365-2966.2012.22111.x}, \href
  {http://adsabs.harvard.edu/abs/2012MNRAS.427.2437H} {427, 2437}

\bibitem[\protect\citeauthoryear{{Han}, {Cole}, {Frenk}  \& {Jing}}{{Han}
  et~al.}{2016}]{Han.etal.16}
{Han} J.,  {Cole} S.,  {Frenk} C.~S.,   {Jing} Y.,  2016, \mn@doi [\mnras]
  {10.1093/mnras/stv2900}, \href
  {http://adsabs.harvard.edu/abs/2016MNRAS.457.1208H} {457, 1208}

\bibitem[\protect\citeauthoryear{{Hayashi}, {Navarro}, {Taylor}, {Stadel}  \&
  {Quinn}}{{Hayashi} et~al.}{2003}]{Hayashi.etal.03}
{Hayashi} E.,  {Navarro} J.~F.,  {Taylor} J.~E.,  {Stadel} J.,   {Quinn} T.,
  2003, \mn@doi [\apj] {10.1086/345788}, \href
  {http://adsabs.harvard.edu/abs/2003ApJ...584..541H} {584, 541}

\bibitem[\protect\citeauthoryear{{Hearin}, {Zentner}, {Berlind}  \&
  {Newman}}{{Hearin} et~al.}{2013}]{Hearin.etal.13}
{Hearin} A.~P.,  {Zentner} A.~R.,  {Berlind} A.~A.,   {Newman} J.~A.,  2013,
  \mn@doi [\mnras] {10.1093/mnras/stt755}, \href
  {http://adsabs.harvard.edu/abs/2013MNRAS.433..659H} {433, 659}

\bibitem[\protect\citeauthoryear{{Hearin}, {Watson}  \& {van den
  Bosch}}{{Hearin} et~al.}{2015}]{Hearin.etal.15}
{Hearin} A.~P.,  {Watson} D.~F.,   {van den Bosch} F.~C.,  2015, \mn@doi
  [\mnras] {10.1093/mnras/stv1358}, \href
  {http://adsabs.harvard.edu/abs/2015MNRAS.452.1958H} {452, 1958}

\bibitem[\protect\citeauthoryear{{Hearin}, {Zentner}, {van den Bosch},
  {Campbell}  \& {Tollerud}}{{Hearin} et~al.}{2016}]{Hearin.etal.16}
{Hearin} A.~P.,  {Zentner} A.~R.,  {van den Bosch} F.~C.,  {Campbell} D.,
  {Tollerud} E.,  2016, \mn@doi [\mnras] {10.1093/mnras/stw840}, \href
  {http://adsabs.harvard.edu/abs/2016MNRAS.460.2552H} {460, 2552}

\bibitem[\protect\citeauthoryear{{Hernquist}}{{Hernquist}}{1990}]{Hernquist.90}
{Hernquist} L.,  1990, \mn@doi [\apj] {10.1086/168845}, \href
  {http://adsabs.harvard.edu/abs/1990ApJ...356..359H} {356, 359}

\bibitem[\protect\citeauthoryear{{Hernquist} \& {Ostriker}}{{Hernquist} \&
  {Ostriker}}{1992}]{Hernquist.Ostriker.92}
{Hernquist} L.,  {Ostriker} J.~P.,  1992, \mn@doi [\apj] {10.1086/171025},
  \href {http://adsabs.harvard.edu/abs/1992ApJ...386..375H} {386, 375}

\bibitem[\protect\citeauthoryear{{Jiang} \& {van den Bosch}}{{Jiang} \& {van
  den Bosch}}{2016a}]{Jiang.vdBosch.17}
{Jiang} F.,  {van den Bosch} F.~C.,  2016a, preprint, \href
  {http://adsabs.harvard.edu/abs/2016arXiv161002399J} {} (\mn@eprint {arXiv}
  {1610.02399})

\bibitem[\protect\citeauthoryear{{Jiang} \& {van den Bosch}}{{Jiang} \& {van
  den Bosch}}{2016b}]{Jiang.vdBosch.16}
{Jiang} F.,  {van den Bosch} F.~C.,  2016b, \mn@doi [\mnras]
  {10.1093/mnras/stw439}, \href
  {http://adsabs.harvard.edu/abs/2016MNRAS.458.2848J} {458, 2848}

\bibitem[\protect\citeauthoryear{{Jiang}, {Cole}, {Sawala}  \& {Frenk}}{{Jiang}
  et~al.}{2015}]{Jiang.etal.15}
{Jiang} L.,  {Cole} S.,  {Sawala} T.,   {Frenk} C.~S.,  2015, \mn@doi [\mnras]
  {10.1093/mnras/stv053}, \href
  {http://adsabs.harvard.edu/abs/2015MNRAS.448.1674J} {448, 1674}

\bibitem[\protect\citeauthoryear{{Joyce}, {Marcos}  \& {Baertschiger}}{{Joyce}
  et~al.}{2009}]{Joyce.etal.09}
{Joyce} M.,  {Marcos} B.,   {Baertschiger} T.,  2009, \mn@doi [\mnras]
  {10.1111/j.1365-2966.2008.14290.x}, \href
  {http://adsabs.harvard.edu/abs/2009MNRAS.394..751J} {394, 751}

\bibitem[\protect\citeauthoryear{{Kampakoglou} \& {Benson}}{{Kampakoglou} \&
  {Benson}}{2007}]{Kampakoglou.Benson.07}
{Kampakoglou} M.,  {Benson} A.~J.,  2007, \mn@doi [\mnras]
  {10.1111/j.1365-2966.2006.11223.x}, \href
  {http://adsabs.harvard.edu/abs/2007MNRAS.374..775K} {374, 775}

\bibitem[\protect\citeauthoryear{{Kang}, {Jing}, {Mo}  \& {B{\"o}rner}}{{Kang}
  et~al.}{2005}]{Kang.etal.05}
{Kang} X.,  {Jing} Y.~P.,  {Mo} H.~J.,   {B{\"o}rner} G.,  2005, \mn@doi [\apj]
  {10.1086/432493}, \href {http://adsabs.harvard.edu/abs/2005ApJ...631...21K}
  {631, 21}

\bibitem[\protect\citeauthoryear{{Keeton} \& {Moustakas}}{{Keeton} \&
  {Moustakas}}{2009}]{Keeton.Moustakas.09}
{Keeton} C.~R.,  {Moustakas} L.~A.,  2009, \mn@doi [\apj]
  {10.1088/0004-637X/699/2/1720}, \href
  {http://adsabs.harvard.edu/abs/2009ApJ...699.1720K} {699, 1720}

\bibitem[\protect\citeauthoryear{{King}}{{King}}{1962}]{King.62}
{King} I.,  1962, \mn@doi [\aj] {10.1086/108756}, \href
  {http://adsabs.harvard.edu/abs/1962AJ.....67..471K} {67, 471}

\bibitem[\protect\citeauthoryear{{Kitzbichler} \& {White}}{{Kitzbichler} \&
  {White}}{2008}]{Kitzbichler.White.08}
{Kitzbichler} M.~G.,  {White} S.~D.~M.,  2008, \mn@doi [\mnras]
  {10.1111/j.1365-2966.2008.13873.x}, \href
  {http://adsabs.harvard.edu/abs/2008MNRAS.391.1489K} {391, 1489}

\bibitem[\protect\citeauthoryear{{Klypin}, {Gottl{\"o}ber}, {Kravtsov}  \&
  {Khokhlov}}{{Klypin} et~al.}{1999a}]{Klypin.etal.99a}
{Klypin} A.,  {Gottl{\"o}ber} S.,  {Kravtsov} A.~V.,   {Khokhlov} A.~M.,
  1999a, \mn@doi [\apj] {10.1086/307122}, \href
  {http://adsabs.harvard.edu/abs/1999ApJ...516..530K} {516, 530}

\bibitem[\protect\citeauthoryear{{Klypin}, {Kravtsov}, {Valenzuela}  \&
  {Prada}}{{Klypin} et~al.}{1999b}]{Klypin.etal.99b}
{Klypin} A.,  {Kravtsov} A.~V.,  {Valenzuela} O.,   {Prada} F.,  1999b, \mn@doi
  [\apj] {10.1086/307643}, \href
  {http://adsabs.harvard.edu/abs/1999ApJ...522...82K} {522, 82}

\bibitem[\protect\citeauthoryear{{Klypin}, {Trujillo-Gomez}  \&
  {Primack}}{{Klypin} et~al.}{2011}]{Klypin.etal.11}
{Klypin} A.~A.,  {Trujillo-Gomez} S.,   {Primack} J.,  2011, \mn@doi [\apj]
  {10.1088/0004-637X/740/2/102}, \href
  {http://adsabs.harvard.edu/abs/2011ApJ...740..102K} {740, 102}

\bibitem[\protect\citeauthoryear{{Klypin}, {Prada}, {Yepes}, {He{\ss}}  \&
  {Gottl{\"o}ber}}{{Klypin} et~al.}{2015}]{Klypin.etal.15}
{Klypin} A.,  {Prada} F.,  {Yepes} G.,  {He{\ss}} S.,   {Gottl{\"o}ber} S.,
  2015, \mn@doi [\mnras] {10.1093/mnras/stu2685}, \href
  {http://adsabs.harvard.edu/abs/2015MNRAS.447.3693K} {447, 3693}

\bibitem[\protect\citeauthoryear{{Klypin}, {Yepes}, {Gottl{\"o}ber}, {Prada}
  \& {He{\ss}}}{{Klypin} et~al.}{2016}]{Klypin.etal.16}
{Klypin} A.,  {Yepes} G.,  {Gottl{\"o}ber} S.,  {Prada} F.,   {He{\ss}} S.,
  2016, \mn@doi [\mnras] {10.1093/mnras/stw248}, \href
  {http://adsabs.harvard.edu/abs/2016MNRAS.457.4340K} {457, 4340}

\bibitem[\protect\citeauthoryear{{Knebe}, {Kravtsov}, {Gottl{\"o}ber}  \&
  {Klypin}}{{Knebe} et~al.}{2000}]{Knebe.etal.00}
{Knebe} A.,  {Kravtsov} A.~V.,  {Gottl{\"o}ber} S.,   {Klypin} A.~A.,  2000,
  \mn@doi [\mnras] {10.1046/j.1365-8711.2000.03673.x}, \href
  {http://adsabs.harvard.edu/abs/2000MNRAS.317..630K} {317, 630}

\bibitem[\protect\citeauthoryear{{Knebe}, {Power}, {Gill}  \& {Gibson}}{{Knebe}
  et~al.}{2006}]{Knebe.etal.06}
{Knebe} A.,  {Power} C.,  {Gill} S.~P.~D.,   {Gibson} B.~K.,  2006, \mn@doi
  [\mnras] {10.1111/j.1365-2966.2006.10161.x}, \href
  {http://adsabs.harvard.edu/abs/2006MNRAS.368..741K} {368, 741}

\bibitem[\protect\citeauthoryear{{Knebe}, {Arnold}, {Power}  \&
  {Gibson}}{{Knebe} et~al.}{2008}]{Knebe.etal.08}
{Knebe} A.,  {Arnold} B.,  {Power} C.,   {Gibson} B.~K.,  2008, \mn@doi
  [\mnras] {10.1111/j.1365-2966.2008.13102.x}, \href
  {http://adsabs.harvard.edu/abs/2008MNRAS.386.1029K} {386, 1029}

\bibitem[\protect\citeauthoryear{{Knebe} et~al.,}{{Knebe}
  et~al.}{2013}]{Knebe.etal.13}
{Knebe} A.,  et~al., 2013, \mn@doi [\mnras] {10.1093/mnras/stt1403}, \href
  {http://adsabs.harvard.edu/abs/2013MNRAS.435.1618K} {435, 1618}

\bibitem[\protect\citeauthoryear{{Kravtsov}, {Klypin}  \&
  {Khokhlov}}{{Kravtsov} et~al.}{1997}]{Kravtsov.etal.97}
{Kravtsov} A.~V.,  {Klypin} A.~A.,   {Khokhlov} A.~M.,  1997, \mn@doi [\apjs]
  {10.1086/313015}, \href {http://adsabs.harvard.edu/abs/1997ApJS..111...73K}
  {111, 73}

\bibitem[\protect\citeauthoryear{{Kravtsov}, {Berlind}, {Wechsler}, {Klypin},
  {Gottl{\"o}ber}, {Allgood}  \& {Primack}}{{Kravtsov}
  et~al.}{2004}]{Kravtsov.etal.04}
{Kravtsov} A.~V.,  {Berlind} A.~A.,  {Wechsler} R.~H.,  {Klypin} A.~A.,
  {Gottl{\"o}ber} S.,  {Allgood} B.,   {Primack} J.~R.,  2004, \mn@doi [\apj]
  {10.1086/420959}, \href {http://adsabs.harvard.edu/abs/2004ApJ...609...35K}
  {609, 35}

\bibitem[\protect\citeauthoryear{{Kuhlen}, {Diemand}  \& {Madau}}{{Kuhlen}
  et~al.}{2008}]{Kuhlen.etal.08}
{Kuhlen} M.,  {Diemand} J.,   {Madau} P.,  2008, \mn@doi [\apj]
  {10.1086/590337}, \href {http://adsabs.harvard.edu/abs/2008ApJ...686..262K}
  {686, 262}

\bibitem[\protect\citeauthoryear{{Kuijken} \& {Dubinski}}{{Kuijken} \&
  {Dubinski}}{1994}]{Kuijken.Dubinski.94}
{Kuijken} K.,  {Dubinski} J.,  1994, \mn@doi [\mnras] {10.1093/mnras/269.1.13},
  \href {http://adsabs.harvard.edu/abs/1994MNRAS.269...13K} {269, 13}

\bibitem[\protect\citeauthoryear{{Lehmann}, {Mao}, {Becker}, {Skillman}  \&
  {Wechsler}}{{Lehmann} et~al.}{2015}]{Lehmann.etal.15}
{Lehmann} B.~V.,  {Mao} Y.-Y.,  {Becker} M.~R.,  {Skillman} S.~W.,   {Wechsler}
  R.~H.,  2015, preprint, \href
  {http://adsabs.harvard.edu/abs/2015arXiv151005651L} {} (\mn@eprint {arXiv}
  {1510.05651})

\bibitem[\protect\citeauthoryear{{{\L}okas}}{{{\L}okas}}{2009}]{Lokas.09}
{{\L}okas} E.~L.,  2009, \mn@doi [\mnras] {10.1111/j.1745-3933.2009.00620.x},
  \href {http://adsabs.harvard.edu/abs/2009MNRAS.394L.102L} {394, L102}

\bibitem[\protect\citeauthoryear{{Lovell}, {Frenk}, {Eke}, {Jenkins}, {Gao}  \&
  {Theuns}}{{Lovell} et~al.}{2014}]{Lovell.etal.14}
{Lovell} M.~R.,  {Frenk} C.~S.,  {Eke} V.~R.,  {Jenkins} A.,  {Gao} L.,
  {Theuns} T.,  2014, \mn@doi [\mnras] {10.1093/mnras/stt2431}, \href
  {http://adsabs.harvard.edu/abs/2014MNRAS.439..300L} {439, 300}

\bibitem[\protect\citeauthoryear{{Macci{\`o}}, {Moore}, {Stadel}  \&
  {Diemand}}{{Macci{\`o}} et~al.}{2006}]{Maccio.etal.06}
{Macci{\`o}} A.~V.,  {Moore} B.,  {Stadel} J.,   {Diemand} J.,  2006, \mn@doi
  [\mnras] {10.1111/j.1365-2966.2005.09976.x}, \href
  {http://adsabs.harvard.edu/abs/2006MNRAS.366.1529M} {366, 1529}

\bibitem[\protect\citeauthoryear{{Mateo}, {Olszewski}, {Pryor}, {Welch}  \&
  {Fischer}}{{Mateo} et~al.}{1993}]{Mateo.etal.93}
{Mateo} M.,  {Olszewski} E.~W.,  {Pryor} C.,  {Welch} D.~L.,   {Fischer} P.,
  1993, \mn@doi [\aj] {10.1086/116449}, \href
  {http://adsabs.harvard.edu/abs/1993AJ....105..510M} {105, 510}

\bibitem[\protect\citeauthoryear{{Melott}}{{Melott}}{2007}]{Melott.07}
{Melott} A.~L.,  2007, preprint, \href
  {http://adsabs.harvard.edu/abs/2007arXiv0709.0745M} {} (\mn@eprint {arXiv}
  {0709.0745})

\bibitem[\protect\citeauthoryear{{Melott}, {Shandarin}, {Splinter}  \&
  {Suto}}{{Melott} et~al.}{1997}]{Melott.etal.97}
{Melott} A.~L.,  {Shandarin} S.~F.,  {Splinter} R.~J.,   {Suto} Y.,  1997,
  \mn@doi [\apjl] {10.1086/310590}, \href
  {http://adsabs.harvard.edu/abs/1997ApJ...479L..79M} {479, L79}

\bibitem[\protect\citeauthoryear{{Merritt}}{{Merritt}}{1996}]{Merritt.96}
{Merritt} D.,  1996, \mn@doi [\aj] {10.1086/117980}, \href
  {http://adsabs.harvard.edu/abs/1996AJ....111.2462M} {111, 2462}

\bibitem[\protect\citeauthoryear{{Metcalf} \& {Madau}}{{Metcalf} \&
  {Madau}}{2001}]{Metcalf.Madau.01}
{Metcalf} R.~B.,  {Madau} P.,  2001, \mn@doi [\apj] {10.1086/323695}, \href
  {http://adsabs.harvard.edu/abs/2001ApJ...563....9M} {563, 9}

\bibitem[\protect\citeauthoryear{{Mo}, {Mao}  \& {White}}{{Mo}
  et~al.}{1998}]{Mo.etal.98}
{Mo} H.~J.,  {Mao} S.,   {White} S.~D.~M.,  1998, \mn@doi [\mnras]
  {10.1046/j.1365-8711.1998.01227.x}, \href
  {http://adsabs.harvard.edu/abs/1998MNRAS.295..319M} {295, 319}

\bibitem[\protect\citeauthoryear{{Mo}, {van den Bosch}  \& {White}}{{Mo}
  et~al.}{2010}]{MBW10}
{Mo} H.,  {van den Bosch} F.~C.,   {White} S.,  2010, {Galaxy Formation and
  Evolution}.
Cambridge University Press

\bibitem[\protect\citeauthoryear{{Molin{\'e}}, {S{\'a}nchez-Conde},
  {Palomares-Ruiz}  \& {Prada}}{{Molin{\'e}} et~al.}{2016}]{Moline.etal.16}
{Molin{\'e}} {\'A}.,  {S{\'a}nchez-Conde} M.~A.,  {Palomares-Ruiz} S.,
  {Prada} F.,  2016, preprint, \href
  {http://adsabs.harvard.edu/abs/2016arXiv160304057M} {} (\mn@eprint {arXiv}
  {1603.04057})

\bibitem[\protect\citeauthoryear{{Moore}, {Katz}, {Lake}, {Dressler}  \&
  {Oemler}}{{Moore} et~al.}{1996a}]{Moore.etal.96b}
{Moore} B.,  {Katz} N.,  {Lake} G.,  {Dressler} A.,   {Oemler} A.,  1996a,
  \mn@doi [\nat] {10.1038/379613a0}, \href
  {http://adsabs.harvard.edu/abs/1996Natur.379..613M} {379, 613}

\bibitem[\protect\citeauthoryear{{Moore}, {Katz}  \& {Lake}}{{Moore}
  et~al.}{1996b}]{Moore.etal.96a}
{Moore} B.,  {Katz} N.,   {Lake} G.,  1996b, \mn@doi [\apj] {10.1086/176745},
  \href {http://adsabs.harvard.edu/abs/1996ApJ...457..455M} {457, 455}

\bibitem[\protect\citeauthoryear{{Moore}, {Governato}, {Quinn}, {Stadel}  \&
  {Lake}}{{Moore} et~al.}{1998}]{Moore.etal.98b}
{Moore} B.,  {Governato} F.,  {Quinn} T.,  {Stadel} J.,   {Lake} G.,  1998,
  \mn@doi [\apjl] {10.1086/311333}, \href
  {http://adsabs.harvard.edu/abs/1998ApJ...499L...5M} {499, L5}

\bibitem[\protect\citeauthoryear{{Moore}, {Ghigna}, {Governato}, {Lake},
  {Quinn}, {Stadel}  \& {Tozzi}}{{Moore} et~al.}{1999}]{Moore.etal.99}
{Moore} B.,  {Ghigna} S.,  {Governato} F.,  {Lake} G.,  {Quinn} T.,  {Stadel}
  J.,   {Tozzi} P.,  1999, \mn@doi [\apjl] {10.1086/312287}, \href
  {http://adsabs.harvard.edu/abs/1999ApJ...524L..19M} {524, L19}

\bibitem[\protect\citeauthoryear{{Moster}, {Somerville}, {Maulbetsch}, {van den
  Bosch}, {Macci{\`o}}, {Naab}  \& {Oser}}{{Moster}
  et~al.}{2010}]{Moster.etal.10}
{Moster} B.~P.,  {Somerville} R.~S.,  {Maulbetsch} C.,  {van den Bosch} F.~C.,
  {Macci{\`o}} A.~V.,  {Naab} T.,   {Oser} L.,  2010, \mn@doi [\apj]
  {10.1088/0004-637X/710/2/903}, \href
  {http://adsabs.harvard.edu/abs/2010ApJ...710..903M} {710, 903}

\bibitem[\protect\citeauthoryear{{Moster}, {Naab}  \& {White}}{{Moster}
  et~al.}{2013}]{Moster.etal.13}
{Moster} B.~P.,  {Naab} T.,   {White} S.~D.~M.,  2013, \mn@doi [\mnras]
  {10.1093/mnras/sts261}, \href
  {http://adsabs.harvard.edu/abs/2013MNRAS.428.3121M} {428, 3121}

\bibitem[\protect\citeauthoryear{{Moster}, {Naab}  \& {White}}{{Moster}
  et~al.}{2017}]{Moster.etal.17}
{Moster} B.~P.,  {Naab} T.,   {White} S.~D.~M.,  2017, preprint, \href
  {http://adsabs.harvard.edu/abs/2017arXiv170505373M} {} (\mn@eprint {arXiv}
  {1705.05373})

\bibitem[\protect\citeauthoryear{{Navarro}, {Frenk}  \& {White}}{{Navarro}
  et~al.}{1997}]{Navarro.etal.97}
{Navarro} J.~F.,  {Frenk} C.~S.,   {White} S.~D.~M.,  1997, \mn@doi [\apj]
  {10.1086/304888}, \href {http://adsabs.harvard.edu/abs/1997ApJ...490..493N}
  {490, 493}

\bibitem[\protect\citeauthoryear{{Navarro} et~al.,}{{Navarro}
  et~al.}{2010}]{Navarro.etal.10}
{Navarro} J.~F.,  et~al., 2010, \mn@doi [\mnras]
  {10.1111/j.1365-2966.2009.15878.x}, \href
  {http://adsabs.harvard.edu/abs/2010MNRAS.402...21N} {402, 21}

\bibitem[\protect\citeauthoryear{{Ogiya} \& {Burkert}}{{Ogiya} \&
  {Burkert}}{2015}]{Ogiya.Burkert.15}
{Ogiya} G.,  {Burkert} A.,  2015, \mn@doi [\mnras] {10.1093/mnras/stu2283},
  \href {http://adsabs.harvard.edu/abs/2015MNRAS.446.2363O} {446, 2363}

\bibitem[\protect\citeauthoryear{{Ogiya} \& {Burkert}}{{Ogiya} \&
  {Burkert}}{2016}]{Ogiya.Burkert.16}
{Ogiya} G.,  {Burkert} A.,  2016, \mn@doi [\mnras] {10.1093/mnras/stw091},
  \href {http://adsabs.harvard.edu/abs/2016MNRAS.457.2164O} {457, 2164}

\bibitem[\protect\citeauthoryear{{Ogiya} \& {Mori}}{{Ogiya} \&
  {Mori}}{2014}]{Ogiya.Mori.14}
{Ogiya} G.,  {Mori} M.,  2014, \mn@doi [\apj] {10.1088/0004-637X/793/1/46},
  \href {http://adsabs.harvard.edu/abs/2014ApJ...793...46O} {793, 46}

\bibitem[\protect\citeauthoryear{{Ogiya}, {Mori}, {Miki}, {Boku}  \&
  {Nakasato}}{{Ogiya} et~al.}{2013}]{Ogiya.etal.13}
{Ogiya} G.,  {Mori} M.,  {Miki} Y.,  {Boku} T.,   {Nakasato} N.,  2013, in
  Journal of Physics Conference Series. p. 012014,
  \mn@doi{10.1088/1742-6596/454/1/012014}

\bibitem[\protect\citeauthoryear{{Oguri} \& {Lee}}{{Oguri} \&
  {Lee}}{2004}]{Oguri.Lee.04}
{Oguri} M.,  {Lee} J.,  2004, \mn@doi [\mnras]
  {10.1111/j.1365-2966.2004.08304.x}, \href
  {http://adsabs.harvard.edu/abs/2004MNRAS.355..120O} {355, 120}

\bibitem[\protect\citeauthoryear{{Onions} et~al.,}{{Onions}
  et~al.}{2012}]{Onions.etal.12}
{Onions} J.,  et~al., 2012, \mn@doi [\mnras]
  {10.1111/j.1365-2966.2012.20947.x}, \href
  {http://adsabs.harvard.edu/abs/2012MNRAS.423.1200O} {423, 1200}

\bibitem[\protect\citeauthoryear{{Pe{\~n}arrubia} \& {Benson}}{{Pe{\~n}arrubia}
  \& {Benson}}{2005}]{Penarrubia.Benson.05}
{Pe{\~n}arrubia} J.,  {Benson} A.~J.,  2005, \mn@doi [\mnras]
  {10.1111/j.1365-2966.2005.09633.x}, \href
  {http://adsabs.harvard.edu/abs/2005MNRAS.364..977P} {364, 977}

\bibitem[\protect\citeauthoryear{{Pe{\~n}arrubia}, {Benson}, {Walker},
  {Gilmore}, {McConnachie}  \& {Mayer}}{{Pe{\~n}arrubia}
  et~al.}{2010}]{Penarrubia.etal.10}
{Pe{\~n}arrubia} J.,  {Benson} A.~J.,  {Walker} M.~G.,  {Gilmore} G.,
  {McConnachie} A.~W.,   {Mayer} L.,  2010, \mn@doi [\mnras]
  {10.1111/j.1365-2966.2010.16762.x}, \href
  {http://adsabs.harvard.edu/abs/2010MNRAS.406.1290P} {406, 1290}

\bibitem[\protect\citeauthoryear{{Pieri}, {Bertone}  \& {Branchini}}{{Pieri}
  et~al.}{2008}]{Pieri.etal.08}
{Pieri} L.,  {Bertone} G.,   {Branchini} E.,  2008, \mn@doi [\mnras]
  {10.1111/j.1365-2966.2007.12828.x}, \href
  {http://adsabs.harvard.edu/abs/2008MNRAS.384.1627P} {384, 1627}

\bibitem[\protect\citeauthoryear{{Power}, {Navarro}, {Jenkins}, {Frenk},
  {White}, {Springel}, {Stadel}  \& {Quinn}}{{Power}
  et~al.}{2003}]{Power.etal.03}
{Power} C.,  {Navarro} J.~F.,  {Jenkins} A.,  {Frenk} C.~S.,  {White} S.~D.~M.,
   {Springel} V.,  {Stadel} J.,   {Quinn} T.,  2003, \mn@doi [\mnras]
  {10.1046/j.1365-8711.2003.05925.x}, \href
  {http://adsabs.harvard.edu/abs/2003MNRAS.338...14P} {338, 14}

\bibitem[\protect\citeauthoryear{{Power}, {Robotham}, {Obreschkow}, {Hobbs}  \&
  {Lewis}}{{Power} et~al.}{2016}]{Power.etal.16}
{Power} C.,  {Robotham} A.~S.~G.,  {Obreschkow} D.,  {Hobbs} A.,   {Lewis}
  G.~F.,  2016, \mn@doi [\mnras] {10.1093/mnras/stw1644}, \href
  {http://adsabs.harvard.edu/abs/2016MNRAS.462..474P} {462, 474}

\bibitem[\protect\citeauthoryear{{Press}, {Teukolsky}, {Vetterling}  \&
  {Flannery}}{{Press} et~al.}{1992}]{Press.etal.92}
{Press} W.~H.,  {Teukolsky} S.~A.,  {Vetterling} W.~T.,   {Flannery} B.~P.,
  1992, {Numerical recipes in FORTRAN. The art of scientific computing}.
Cambridge University Press

\bibitem[\protect\citeauthoryear{{Pullen}, {Benson}  \& {Moustakas}}{{Pullen}
  et~al.}{2014}]{Pullen.etal.14}
{Pullen} A.~R.,  {Benson} A.~J.,   {Moustakas} L.~A.,  2014, \mn@doi [\apj]
  {10.1088/0004-637X/792/1/24}, \href
  {http://adsabs.harvard.edu/abs/2014ApJ...792...24P} {792, 24}

\bibitem[\protect\citeauthoryear{{Read}, {Wilkinson}, {Evans}, {Gilmore}  \&
  {Kleyna}}{{Read} et~al.}{2006}]{Read.etal.06a}
{Read} J.~I.,  {Wilkinson} M.~I.,  {Evans} N.~W.,  {Gilmore} G.,   {Kleyna}
  J.~T.,  2006, \mn@doi [\mnras] {10.1111/j.1365-2966.2005.09861.x}, \href
  {http://adsabs.harvard.edu/abs/2006MNRAS.366..429R} {366, 429}

\bibitem[\protect\citeauthoryear{{Reddick}, {Tinker}, {Wechsler}  \&
  {Lu}}{{Reddick} et~al.}{2014}]{Reddick.etal.14}
{Reddick} R.~M.,  {Tinker} J.~L.,  {Wechsler} R.~H.,   {Lu} Y.,  2014, \mn@doi
  [\apj] {10.1088/0004-637X/783/2/118}, \href
  {http://adsabs.harvard.edu/abs/2014ApJ...783..118R} {783, 118}

\bibitem[\protect\citeauthoryear{{Rocha}, {Peter}, {Bullock}, {Kaplinghat},
  {Garrison-Kimmel}, {O{\~n}orbe}  \& {Moustakas}}{{Rocha}
  et~al.}{2013}]{Rocha.etal.13}
{Rocha} M.,  {Peter} A.~H.~G.,  {Bullock} J.~S.,  {Kaplinghat} M.,
  {Garrison-Kimmel} S.,  {O{\~n}orbe} J.,   {Moustakas} L.~A.,  2013, \mn@doi
  [\mnras] {10.1093/mnras/sts514}, \href
  {http://adsabs.harvard.edu/abs/2013MNRAS.430...81R} {430, 81}

\bibitem[\protect\citeauthoryear{{Romeo}, {Agertz}, {Moore}  \&
  {Stadel}}{{Romeo} et~al.}{2008}]{Romeo.etal.08}
{Romeo} A.~B.,  {Agertz} O.,  {Moore} B.,   {Stadel} J.,  2008, \mn@doi [\apj]
  {10.1086/591236}, \href {http://adsabs.harvard.edu/abs/2008ApJ...686....1R}
  {686, 1}

\bibitem[\protect\citeauthoryear{{Sanders}, {Bovy}  \& {Erkal}}{{Sanders}
  et~al.}{2016}]{Sanders.etal.16}
{Sanders} J.~L.,  {Bovy} J.,   {Erkal} D.,  2016, \mn@doi [\mnras]
  {10.1093/mnras/stw232}, \href
  {http://adsabs.harvard.edu/abs/2016MNRAS.457.3817S} {457, 3817}

\bibitem[\protect\citeauthoryear{{Spitzer}}{{Spitzer}}{1958}]{Spitzer.58}
{Spitzer} Jr. L.,  1958, \mn@doi [\apj] {10.1086/146435}, \href
  {http://adsabs.harvard.edu/abs/1958ApJ...127...17S} {127, 17}

\bibitem[\protect\citeauthoryear{{Spitzer}}{{Spitzer}}{1987}]{Spitzer.87}
{Spitzer} L.,  1987, {Dynamical evolution of globular clusters}.
Princeton University Press

\bibitem[\protect\citeauthoryear{{Splinter}, {Melott}, {Shandarin}  \&
  {Suto}}{{Splinter} et~al.}{1998}]{Splinter.etal.98}
{Splinter} R.~J.,  {Melott} A.~L.,  {Shandarin} S.~F.,   {Suto} Y.,  1998,
  \mn@doi [\apj] {10.1086/305450}, \href
  {http://adsabs.harvard.edu/abs/1998ApJ...497...38S} {497, 38}

\bibitem[\protect\citeauthoryear{{Springel}, {White}, {Tormen}  \&
  {Kauffmann}}{{Springel} et~al.}{2001}]{Springel.etal.01}
{Springel} V.,  {White} S.~D.~M.,  {Tormen} G.,   {Kauffmann} G.,  2001,
  \mn@doi [\mnras] {10.1046/j.1365-8711.2001.04912.x}, \href
  {http://adsabs.harvard.edu/abs/2001MNRAS.328..726S} {328, 726}

\bibitem[\protect\citeauthoryear{{Springel} et~al.,}{{Springel}
  et~al.}{2005}]{Springel.etal.05}
{Springel} V.,  et~al., 2005, \mn@doi [\nat] {10.1038/nature03597}, \href
  {http://adsabs.harvard.edu/abs/2005Natur.435..629S} {435, 629}

\bibitem[\protect\citeauthoryear{{Springel} et~al.,}{{Springel}
  et~al.}{2008}]{Springel.etal.08}
{Springel} V.,  et~al., 2008, \mn@doi [\mnras]
  {10.1111/j.1365-2966.2008.14066.x}, \href
  {http://adsabs.harvard.edu/abs/2008MNRAS.391.1685S} {391, 1685}

\bibitem[\protect\citeauthoryear{{Strigari}, {Koushiappas}, {Bullock}  \&
  {Kaplinghat}}{{Strigari} et~al.}{2007}]{Strigari.etal.07}
{Strigari} L.~E.,  {Koushiappas} S.~M.,  {Bullock} J.~S.,   {Kaplinghat} M.,
  2007, \mn@doi [\prd] {10.1103/PhysRevD.75.083526}, \href
  {http://adsabs.harvard.edu/abs/2007PhRvD..75h3526S} {75, 083526}

\bibitem[\protect\citeauthoryear{{Taffoni}, {Mayer}, {Colpi}  \&
  {Governato}}{{Taffoni} et~al.}{2003}]{Taffoni.etal.03}
{Taffoni} G.,  {Mayer} L.,  {Colpi} M.,   {Governato} F.,  2003, \mn@doi
  [\mnras] {10.1046/j.1365-8711.2003.06395.x}, \href
  {http://adsabs.harvard.edu/abs/2003MNRAS.341..434T} {341, 434}

\bibitem[\protect\citeauthoryear{{Taylor} \& {Babul}}{{Taylor} \&
  {Babul}}{2001}]{Taylor.Babul.01}
{Taylor} J.~E.,  {Babul} A.,  2001, \mn@doi [\apj] {10.1086/322276}, \href
  {http://adsabs.harvard.edu/abs/2001ApJ...559..716T} {559, 716}

\bibitem[\protect\citeauthoryear{{Taylor} \& {Babul}}{{Taylor} \&
  {Babul}}{2004}]{Taylor.Babul.04}
{Taylor} J.~E.,  {Babul} A.,  2004, \mn@doi [\mnras]
  {10.1111/j.1365-2966.2004.07395.x}, \href
  {http://adsabs.harvard.edu/abs/2004MNRAS.348..811T} {348, 811}

\bibitem[\protect\citeauthoryear{{Tollet}, {Cattaneo}, {Mamon}, {Moutard}  \&
  {van den Bosch}}{{Tollet} et~al.}{2017}]{Tollet.etal.17}
{Tollet} {\'E}.,  {Cattaneo} A.,  {Mamon} G.,  {Moutard} T.,   {van den Bosch}
  F.,  2017, preprint, \href
  {http://adsabs.harvard.edu/abs/2017arXiv170706264T} {} (\mn@eprint {arXiv}
  {1707.06264})

\bibitem[\protect\citeauthoryear{{Tormen}}{{Tormen}}{1997}]{Tormen.97}
{Tormen} G.,  1997, \mn@doi [\mnras] {10.1093/mnras/290.3.411}, \href
  {http://adsabs.harvard.edu/abs/1997MNRAS.290..411T} {290, 411}

\bibitem[\protect\citeauthoryear{{Tormen}, {Bouchet}  \& {White}}{{Tormen}
  et~al.}{1997}]{Tormen.etal.97}
{Tormen} G.,  {Bouchet} F.~R.,   {White} S.~D.~M.,  1997, \mn@doi [\mnras]
  {10.1093/mnras/286.4.865}, \href
  {http://adsabs.harvard.edu/abs/1997MNRAS.286..865T} {286, 865}

\bibitem[\protect\citeauthoryear{{Tormen}, {Diaferio}  \& {Syer}}{{Tormen}
  et~al.}{1998}]{Tormen.etal.98}
{Tormen} G.,  {Diaferio} A.,   {Syer} D.,  1998, \mn@doi [\mnras]
  {10.1046/j.1365-8711.1998.01775.x}, \href
  {http://adsabs.harvard.edu/abs/1998MNRAS.299..728T} {299, 728}

\bibitem[\protect\citeauthoryear{{Trujillo-Gomez}, {Klypin}, {Primack}  \&
  {Romanowsky}}{{Trujillo-Gomez} et~al.}{2011}]{Trujillo-Gomez.etal.11}
{Trujillo-Gomez} S.,  {Klypin} A.,  {Primack} J.,   {Romanowsky} A.~J.,  2011,
  \mn@doi [\apj] {10.1088/0004-637X/742/1/16}, \href
  {http://adsabs.harvard.edu/abs/2011ApJ...742...16T} {742, 16}

\bibitem[\protect\citeauthoryear{{Vale} \& {Ostriker}}{{Vale} \&
  {Ostriker}}{2004}]{Vale.Ostriker.04}
{Vale} A.,  {Ostriker} J.~P.,  2004, \mn@doi [\mnras]
  {10.1111/j.1365-2966.2004.08059.x}, \href
  {http://adsabs.harvard.edu/abs/2004MNRAS.353..189V} {353, 189}

\bibitem[\protect\citeauthoryear{{Vegetti} \& {Koopmans}}{{Vegetti} \&
  {Koopmans}}{2009}]{Vegetti.Koopmans.09}
{Vegetti} S.,  {Koopmans} L.~V.~E.,  2009, \mn@doi [\mnras]
  {10.1111/j.1365-2966.2009.15559.x}, \href
  {http://adsabs.harvard.edu/abs/2009MNRAS.400.1583V} {400, 1583}

\bibitem[\protect\citeauthoryear{{Vegetti}, {Koopmans}, {Auger}, {Treu}  \&
  {Bolton}}{{Vegetti} et~al.}{2014}]{Vegetti.etal.14}
{Vegetti} S.,  {Koopmans} L.~V.~E.,  {Auger} M.~W.,  {Treu} T.,   {Bolton}
  A.~S.,  2014, \mn@doi [\mnras] {10.1093/mnras/stu943}, \href
  {http://adsabs.harvard.edu/abs/2014MNRAS.442.2017V} {442, 2017}

\bibitem[\protect\citeauthoryear{{Vogelsberger}, {Zavala}  \&
  {Loeb}}{{Vogelsberger} et~al.}{2012}]{Vogelsberger.etal.12}
{Vogelsberger} M.,  {Zavala} J.,   {Loeb} A.,  2012, preprint, \href
  {http://adsabs.harvard.edu/abs/2012arXiv1201.5892V} {} (\mn@eprint {arXiv}
  {1201.5892})

\bibitem[\protect\citeauthoryear{{Walker}, {Mateo}, {Olszewski},
  {Pe{\~n}arrubia}, {Wyn Evans}  \& {Gilmore}}{{Walker}
  et~al.}{2009}]{Walker.etal.09}
{Walker} M.~G.,  {Mateo} M.,  {Olszewski} E.~W.,  {Pe{\~n}arrubia} J.,  {Wyn
  Evans} N.,   {Gilmore} G.,  2009, \mn@doi [\apj]
  {10.1088/0004-637X/704/2/1274}, \href
  {http://adsabs.harvard.edu/abs/2009ApJ...704.1274W} {704, 1274}

\bibitem[\protect\citeauthoryear{{Weinberg}}{{Weinberg}}{1993}]{Weinberg.93}
{Weinberg} M.~D.,  1993, \mn@doi [\apj] {10.1086/172773}, \href
  {http://adsabs.harvard.edu/abs/1993ApJ...410..543W} {410, 543}

\bibitem[\protect\citeauthoryear{{Weinberg}}{{Weinberg}}{1994a}]{Weinberg.94a}
{Weinberg} M.~D.,  1994a, \mn@doi [\aj] {10.1086/117161}, \href
  {http://adsabs.harvard.edu/abs/1994AJ....108.1398W} {108, 1398}

\bibitem[\protect\citeauthoryear{{Weinberg}}{{Weinberg}}{1994b}]{Weinberg.94b}
{Weinberg} M.~D.,  1994b, \mn@doi [\aj] {10.1086/117162}, \href
  {http://adsabs.harvard.edu/abs/1994AJ....108.1403W} {108, 1403}

\bibitem[\protect\citeauthoryear{{Weinberg}}{{Weinberg}}{1997}]{Weinberg.97}
{Weinberg} M.~D.,  1997, \mn@doi [\apj] {10.1086/303828}, \href
  {http://adsabs.harvard.edu/abs/1997ApJ...478..435W} {478, 435}

\bibitem[\protect\citeauthoryear{{Weinberg}, {Colombi}, {Dav{\'e}}  \&
  {Katz}}{{Weinberg} et~al.}{2008}]{Weinberg.etal.08}
{Weinberg} D.~H.,  {Colombi} S.,  {Dav{\'e}} R.,   {Katz} N.,  2008, \mn@doi
  [\apj] {10.1086/524646}, \href
  {http://adsabs.harvard.edu/abs/2008ApJ...678....6W} {678, 6}

\bibitem[\protect\citeauthoryear{{Wetzel}}{{Wetzel}}{2011}]{Wetzel.11}
{Wetzel} A.~R.,  2011, \mn@doi [\mnras] {10.1111/j.1365-2966.2010.17877.x},
  \href {http://adsabs.harvard.edu/abs/2011MNRAS.412...49W} {412, 49}

\bibitem[\protect\citeauthoryear{{Wetzel}, {Hopkins}, {Kim},
  {Faucher-Gigu{\`e}re}, {Kere{\v s}}  \& {Quataert}}{{Wetzel}
  et~al.}{2016}]{Wetzel.etal.16}
{Wetzel} A.~R.,  {Hopkins} P.~F.,  {Kim} J.-h.,  {Faucher-Gigu{\`e}re} C.-A.,
  {Kere{\v s}} D.,   {Quataert} E.,  2016, \mn@doi [\apjl]
  {10.3847/2041-8205/827/2/L23}, \href
  {http://adsabs.harvard.edu/abs/2016ApJ...827L..23W} {827, L23}

\bibitem[\protect\citeauthoryear{{Widrow}}{{Widrow}}{2000}]{Widrow.00}
{Widrow} L.~M.,  2000, \mn@doi [\apjs] {10.1086/317367}, \href
  {http://adsabs.harvard.edu/abs/2000ApJS..131...39W} {131, 39}

\bibitem[\protect\citeauthoryear{{Wu}, {Hahn}, {Evrard}, {Wechsler}  \&
  {Dolag}}{{Wu} et~al.}{2013}]{Wu.etal.13}
{Wu} H.-Y.,  {Hahn} O.,  {Evrard} A.~E.,  {Wechsler} R.~H.,   {Dolag} K.,
  2013, \mn@doi [\mnras] {10.1093/mnras/stt1582}, \href
  {http://adsabs.harvard.edu/abs/2013MNRAS.436..460W} {436, 460}

\bibitem[\protect\citeauthoryear{{Zentner} \& {Bullock}}{{Zentner} \&
  {Bullock}}{2003}]{Zentner.Bullock.03}
{Zentner} A.~R.,  {Bullock} J.~S.,  2003, \mn@doi [\apj] {10.1086/378797},
  \href {http://adsabs.harvard.edu/abs/2003ApJ...598...49Z} {598, 49}

\bibitem[\protect\citeauthoryear{{Zentner}, {Berlind}, {Bullock}, {Kravtsov}
  \& {Wechsler}}{{Zentner} et~al.}{2005}]{Zentner.etal.05}
{Zentner} A.~R.,  {Berlind} A.~A.,  {Bullock} J.~S.,  {Kravtsov} A.~V.,
  {Wechsler} R.~H.,  2005, \mn@doi [\apj] {10.1086/428898}, \href
  {http://adsabs.harvard.edu/abs/2005ApJ...624..505Z} {624, 505}

\bibitem[\protect\citeauthoryear{{Zentner}, {Hearin}  \& {van den
  Bosch}}{{Zentner} et~al.}{2014}]{Zentner.etal.14}
{Zentner} A.~R.,  {Hearin} A.~P.,   {van den Bosch} F.~C.,  2014, \mn@doi
  [\mnras] {10.1093/mnras/stu1383}, \href
  {http://adsabs.harvard.edu/abs/2014MNRAS.443.3044Z} {443, 3044}

\bibitem[\protect\citeauthoryear{{Zentner}, {Hearin}, {van den Bosch}, {Lange}
  \& {Villarreal}}{{Zentner} et~al.}{2016}]{Zentner.etal.16}
{Zentner} A.~R.,  {Hearin} A.,  {van den Bosch} F.~C.,  {Lange} J.~U.,
  {Villarreal} A.,  2016, preprint, \href
  {http://adsabs.harvard.edu/abs/2016arXiv160607817Z} {} (\mn@eprint {arXiv}
  {1606.07817})

\bibitem[\protect\citeauthoryear{{Zhao}}{{Zhao}}{2004}]{Zhao.04}
{Zhao} H.,  2004, \mn@doi [\mnras] {10.1111/j.1365-2966.2004.07835.x}, \href
  {http://adsabs.harvard.edu/abs/2004MNRAS.351..891Z} {351, 891}

\bibitem[\protect\citeauthoryear{{Zolotov} et~al.,}{{Zolotov}
  et~al.}{2012}]{Zolotov.etal.12}
{Zolotov} A.,  et~al., 2012, \mn@doi [\apj] {10.1088/0004-637X/761/1/71}, \href
  {http://adsabs.harvard.edu/abs/2012ApJ...761...71Z} {761, 71}

\bibitem[\protect\citeauthoryear{{van Kampen}}{{van Kampen}}{1995}]{vKampen.95}
{van Kampen} E.,  1995, \mn@doi [\mnras] {10.1093/mnras/273.2.295}, \href
  {http://adsabs.harvard.edu/abs/1995MNRAS.273..295V} {273, 295}

\bibitem[\protect\citeauthoryear{{van Kampen}}{{van
  Kampen}}{2000a}]{vKampen.00a}
{van Kampen} E.,  2000a, ArXiv Astrophysics e-prints, \href
  {http://adsabs.harvard.edu/abs/2000astro.ph..2027V} {}

\bibitem[\protect\citeauthoryear{{van Kampen}}{{van
  Kampen}}{2000b}]{vKampen.00b}
{van Kampen} E.,  2000b, ArXiv Astrophysics e-prints, \href
  {http://adsabs.harvard.edu/abs/2000astro.ph..8453V} {}

\bibitem[\protect\citeauthoryear{{van den Bosch}}{{van den
  Bosch}}{2017}]{vdBosch.17}
{van den Bosch} F.~C.,  2017, \mn@doi [\mnras] {10.1093/mnras/stx520}, \href
  {http://adsabs.harvard.edu/abs/2017MNRAS.468..885V} {468, 885}

\bibitem[\protect\citeauthoryear{{van den Bosch} \& {Jiang}}{{van den Bosch} \&
  {Jiang}}{2016}]{vdBosch.Jiang.16}
{van den Bosch} F.~C.,  {Jiang} F.,  2016, \mn@doi [\mnras]
  {10.1093/mnras/stw440}, \href
  {http://adsabs.harvard.edu/abs/2016MNRAS.458.2870V} {458, 2870}

\bibitem[\protect\citeauthoryear{{van den Bosch}, {Norberg}, {Mo}  \&
  {Yang}}{{van den Bosch} et~al.}{2004}]{vdBosch.etal.04}
{van den Bosch} F.~C.,  {Norberg} P.,  {Mo} H.~J.,   {Yang} X.,  2004, \mn@doi
  [\mnras] {10.1111/j.1365-2966.2004.08021.x}, \href
  {http://adsabs.harvard.edu/abs/2004MNRAS.352.1302V} {352, 1302}

\bibitem[\protect\citeauthoryear{{van den Bosch}, {Tormen}  \& {Giocoli}}{{van
  den Bosch} et~al.}{2005}]{vdBosch.etal.05}
{van den Bosch} F.~C.,  {Tormen} G.,   {Giocoli} C.,  2005, \mn@doi [\mnras]
  {10.1111/j.1365-2966.2005.08964.x}, \href
  {http://adsabs.harvard.edu/abs/2005MNRAS.359.1029V} {359, 1029}

\bibitem[\protect\citeauthoryear{{van den Bosch}, {Jiang}, {Campbell}  \&
  {Behroozi}}{{van den Bosch} et~al.}{2016}]{vdBosch.etal.16}
{van den Bosch} F.~C.,  {Jiang} F.,  {Campbell} D.,   {Behroozi} P.,  2016,
  \mn@doi [\mnras] {10.1093/mnras/stv2338}, \href
  {http://adsabs.harvard.edu/abs/2016MNRAS.455..158V} {455, 158}

\makeatother
\end{thebibliography}


\appendix

\section{Computing Bound Fractions of $N$-body Haloes}
\label{App:fbound}

This appendix describes the method we use to compute the bound
fraction of an $N$-body (sub)halo. We consider a particle $i$ to be
bound if
\begin{equation}\label{Ebound}
E_i \equiv {1 \over 2} m_i v^2_i - \sum_{j\ne i}
{G \, m_i \, m_j \, \calQ_j \over (\vert \br_j - \br_i \vert^2 + \varepsilon^2)^{1/2}} < 0\,.
\end{equation}
Here $\varepsilon$ is the softening length, and $\calQ_j$ is 
equal to 1 (0) if particle $j$ is bound
(unbound). Hence, in order to be able to compute $E_i$, one first
needs to know which are the bound particles. Obviously, this can 
only be solved using an iterative scheme.  In our analysis we 
proceed as follows:
\begin{enumerate}
\item We begin by making an initial guess for the boundness,
  $\calQ_i$, of each particle.  At $t=0$ (ICs) we assume that each
  particle is bound ($\calQ_i = 1 \,\,\, \forall i=1,...,N$), while at later
  times we assume that $\calQ_i$ is the same as in the previous simulation
  output.
\item For each particle we compute $E_i$ using
  Eq.~(\ref{Ebound}).  We update $\calQ_i$ accordingly and compute the
  new, bound fraction $\fbound = {1 \over N_{\rm tot}} \sum_i
  \calQ_i$.
\item Compute the centre-of-mass position and velocity of the halo,
  $\br_{\rm com}$ and $\bv_{\rm com}$, as the average position and
  velocity of the $N = {\rm MAX}(N_{\rm min},f_{\rm com} \, N_{\rm bound})$ 
  particles that are most bound. Here $N_{\rm bound} = \fbound\, N_{\rm tot}$, 
  while $N_{\rm min}$ and $f_{\rm com}$ are free parameters.
\item Update $\bv_i$ for each particle using the new
  centre-of-mass velocity.
\item Go back to (ii) and iterate until the changes in $\br_{\rm com}$
  and $\bv_{\rm com}$ are smaller than $10^{-4} \rvir$ and $10^{-4}
  \Vvir$, respectively. This typically requires 3-10 iterations.
\end{enumerate}
When computing the gravitational potential term of Eq.~(\ref{Ebound})
we use a Barnes \& Hut octree with the same opening angle,
$\theta$, as in the simulation. We also adopt the same
softening. Generally, smaller values of $f_{\rm com}$ results in a
more noisy time-evolution of $\br_{\rm com}$ and $\bv_{\rm com}$,
while for larger values of $f_{\rm com}$ it is more difficult to trace
$\fbound(t)$ when it becomes small. We obtain stable results for
$N_{\rm min} = 10$ (we never use fewer than 10 particles to determine
the centre-of-mass properties\footnote{Except when $N_{\rm bound}<10$, 
in which case we adopt $N=N_{\rm bound}$.}) and $f_{\rm com} = 0.05$
(centre-of-mass properties are determined using the 5 percent most
bound particles), which are the parameters we use throughout. 
\begin{figure*}
\includegraphics[width=\hdsize]{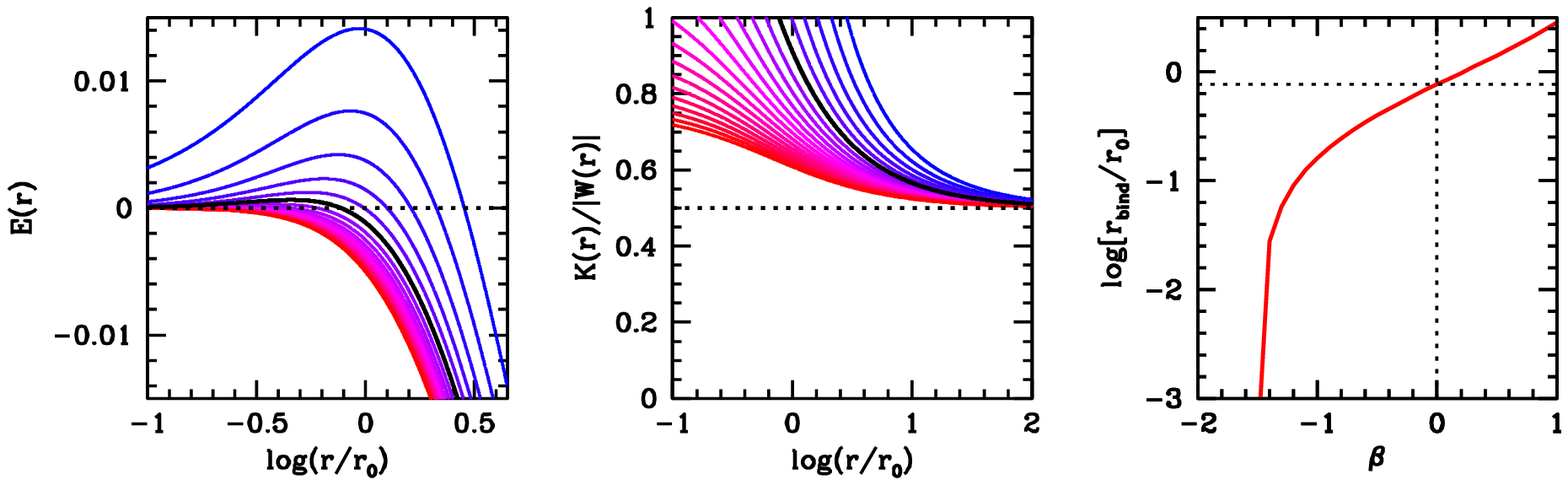}
\centering
\caption{Energy structure of spherical, instantaneously truncated NFW
  haloes.  The left panel plots the total binding energy, $E(r)$, as a
  function of the truncation radius, $r$, in units of the NFW scale
  radius, $r_0$. The middle panel shows the corresponding virial
  ratio, $K(r)/|W(r)|$. Different colors correspond to different
  values of the anisotropy parameter; from blue to red $\beta = 1.0,
  0.8,...,-3.0$.  The panel on the right shows $\log[r_{\rm
      crit}/r_0]$ as function of $\beta$, where $r_{\rm crit}$ is
  defined as the truncation radius below which the remnant has
  positive binding energy. $r_{\rm crit}\simeq 0.77 r_0$ for an
  isotropic NFW halo ($\beta = 0$), becomes $\sim 2.84 r_0$ for an NFW
  halo with purely radial orbits ($\beta=1$), and is absent (i.e., the
  remnant always has negative binding energy) whenever $\beta \lta
  -1.5$.}
\label{fig:rbind}
\end{figure*}

\section{The Energy Structure of Dark Matter Haloes}
\label{App:NFWener}

Consider a spherical dark matter halo with density distribution
$\rho(r)$ and enclosed mass profile
\begin{equation}
M(r) = 4\pi \int_0^r \rho(r') \, r'^2 \, \rmd r'
\end{equation}
The total energy of the dark matter particles inside radius $r$ is
given by $E(r) = K(r) + W(r)$, where $K$ and $W$ are the corresponding
kinetic and potential energies, respectively. For a spherical system,
\begin{equation}\label{Wpot_generic}
W(r) = 2 \pi \int_0^r \rho(r') \, \Phi(r') \, r'^2 \rmd r'\,.
\end{equation}
Using that the gravitational potential
\begin{equation}\label{potential}
\Phi(r) = -4 \pi G \left[{1 \over r} \int_0^r \rho(r') \, r'^2 \, \rmd r'
+ \int_r^{\infty} \rho(r') \, r' \, \rmd r' \right]
\end{equation}
(Binney \& Tremaine 2008), and using integration by parts, this can be
written as
\begin{eqnarray}\label{Epot}
W(r) & = & -4 \pi G \int_0^r \rho(r') \, M(r') \, r' \, \rmd r'
- \nonumber \\
& & 2 \pi G \, M(r) \, \int_r^{\infty} \rho(r') \, r' \, \rmd r'\,.
\end{eqnarray}
The second term describes the contribution to the gravitational
potential energy due to matter located at radii larger than $r$, and
vanishes if $r \rightarrow \infty$. In what follows, we compute the
energy of a spherical halo whose mass is instantaneously truncated
(e.g., due to tides) at radius $r_\rmt$. In that case, $\rho(r) = 0$
for $r > r_\rmt$ and $W(r_\rmt)$ is simply given by the first term
of Eq.~(\ref{Epot}).

The Jeans equation for hydrostatic equilibrium in a spherical stellar
system is
\begin{equation}\label{Jeans}
{\rmd(\rho \sigma^2_r) \over \rmd r} + 2 {\beta \rho \sigma^2_r \over r}
= -\rho {\rmd \Phi \over \rmd r}\,,
\end{equation}
where $\beta \equiv 1 - \sigma^2_{\theta}/\sigma^2_r$ describes the
velocity dispersion anisotropy (here assumed constant). Haloes with
$\beta = 0$ are isotropic, haloes with $\beta < 0$ are tangentially
anisotropic, and haloes with $0 < \beta \leq 1$ are radially
anisotropic. The solution of Eq.~(\ref{Jeans}) is
\begin{equation}
\sigma^2_r(r) = {r^{-2\beta} \over \rho(r)} \int_r^{\infty} \rho(r') M(r')
r'^{2\beta-2} \, \rmd r'
\end{equation}
where we have used that $\rmd \Phi/\rmd r = G M(r)/r^2$.  The kinetic
energy of particles inside radius $r_\rmt$ is given by
\begin{equation}\label{Kkin}
K(r_\rmt) = 2 \pi \int_0^{r_\rmt} \rho(r) \, \langle v^2 \rangle(r) \, r^2 \rmd r\,.
\end{equation}
If we assume that the halo has no net streaming motion, so that
$\sigma^2_{\phi} = \sigma^2_{\theta}$, then $\langle v^2 \rangle(r) =
[1+2(1-\beta)] \sigma^2_r(r)$ and we have that
\begin{eqnarray}\label{Ekin}
K(r_\rmt) & = & 2 \pi G \, [1 + 2(1-\beta)] \int_0^{r_\rmt} \rmd r \, r^{2-2\beta} 
\nonumber \\
& & \int_{r}^{\infty} \rho(r') \, M(r') \, r'^{2\beta-2} \, \rmd r'\,.
\end{eqnarray}

The left-hand panel of Fig.~\ref{fig:rbind} plots the total binding
energy $E(r_\rmt) = K(r_\rmt) + W(r_\rmt)$ of a spherical NFW halo
that is instantaneously truncated at radius $r_\rmt$.  Different lines
correspond to different values of the anisotropy parameter $\beta$, as
described in the figure caption.  Note that these results are independent
of halo concentration.  The middle panel shows the corresponding
virial ratio $K(r_\rmt)/|W(r_\rmt)|$, which asymptotes to 0.5
(corresponding to virial equilibrium) in the limit $r_\rmt \rightarrow
\infty$. Note that $K(r_\rmt)/|W(r_\rmt)| > 1$ corresponds to
$E(r_\rmt) > 0$, and thus a positive, total binding energy. Finally,
the right-hand panel plots $\log[r_{\rm crit}/r_0]$ as function of
$\beta$, where $r_{\rm crit}$ is defined such that $E(r_\rmt) > 0$ for
$r_\rmt < r_{\rm crit}$. For an isotropic NFW profile ($\beta=0$), we
have that $r_{\rm crit} = 0.77 r_0$, in agreement with
\cite{Hayashi.etal.03}. For a maximally, radially anisotropic NFW
halo, $r_{\rm crit} = 2.84 r_0$, while $r_{\rm crit} = 0$ if $\beta
\lta -1.6$. In this case the tidally truncated remnant will always
have negative, total binding energy, independent of $r_\rmt$. To see
this, recall that the limit $\beta \rightarrow -\infty$ corresponds to
a halo in which all particles are on circular orbits, so that $\langle
v^2 \rangle(r) = v^2_{\rm circ}(r) = GM(r)/r$. In that case,
Eq.~(\ref{Kkin}) reduces to
\begin{equation}\label{Kkinmod}
  K(r_\rmt) = 2 \pi G \int_0^{r_\rmt} \rho(r) \, M(r) \, r \, \rmd r =
  -{W(r_\rmt) \over 2}\,.
\end{equation}
Hence, we have that $K(r_\rmt)/|W(r_\rmt)| = {1 \over 2}$, and thus
$E(r_\rmt) = -K(r_\rmt) < 0$.

\label{lastpage}

\end{document}